\documentclass[aps,pra,reprint,groupedaddress,nofootinbib,longbibliography,showpacs]{revtex4-1}
\usepackage{graphicx}
\usepackage{latexsym,amssymb,amsmath}
\usepackage{bm}
\usepackage[caption=false]{subfig}
\usepackage{xcolor}
\usepackage{mathrsfs}
\usepackage[colorlinks=true,
urlcolor=blue,
linkcolor=blue,
citecolor=blue
]{hyperref}
\begin{document}
\title{Topological photonic states in one-dimensional dimerized ultracold atomic chains}

\author{B. X. Wang}
\author{C. Y. Zhao}
\email{changying.zhao@sjtu.edu.cn}

\affiliation{Institute of Engineering Thermophysics, Shanghai Jiao Tong University,
	Shanghai, 200240, China}
\date{\today}

\begin{abstract}
We study the topological optical states in one-dimensional (1D) dimerized ultracold atomic chains, as an extension of the Su-Schrieffer-Heeger (SSH) model. By taking the fully retarded near-field and far-field dipole-dipole interactions into account, we describe the system by an effective non-Hermitian Hamiltonian, vastly different from the Hermitian Hamiltonian of the conventional SSH model. We analytically calculate the complex bandstructures for infinitely long chains, and show that the topological invariant, i.e., the complex Zak phase, is still quantized and becomes nontrivial when the dimerization parameter $\beta>0.5$, despite the broken chiral symmetry and non-Hermiticity. We have verified the validity of the bulk-boundary correspondence for this non-Hermitian system by further analyzing the eigenstate distributions along with their inverse participation ratios (IPRs) for finite chains, where topologically protected edge states are unambiguously identified. We also reveal that such topological edge states are robust under symmetry-breaking disorders. For transverse eigenstates, we further discover the increase of localization length of topological edge states with the increase of lattice period due to the presence of strong far-field dipole-dipole interactions. Moreover, the ultra-strong scattering cross section and ultra-narrow linewidth of a single cold atom allow us to observe in more detail about topological states than in conventional systems, such as the frequency shift with respect to the single-atom resonance and the largely tunable bandgap. We envisage these topological photonic states can provide an efficient interface between light and matter. 
\end{abstract}
\maketitle
\section{Introduction}
The area of topological photonics attracts growing attention in the last decade \cite{luNPhoton2014,khanikaevNPhoton2017,ozawa2018topological}. Specially designed topological photonic systems are able create topologically protected optical states, which can propagate unidirectionally without any backscattering processes, even in the presence of disorder and impurities \cite{luNPhoton2014,khanikaevNPhoton2017,ozawa2018topological}. Such systems show promising applications as novel photonic devices, including unidirectional waveguides \cite{poliNComms2015}, optical isolators \cite{el-GanainyOL2015} as well topological lasers \cite{partoPRL2018}, etc. One-dimensional (1D) and quasi-1D topological photonic systems, for instance, plasmonic nanoparticle chains \cite{lingOE2015,downingPRB2017}, lattices of optical waveguides \cite{longhiOL2013,blanco-RedondoPRL2016} and arrays of dielectric resonators \cite{poliNComms2015} have received intense investigation due to their simplicity of fabrication and capabilities of guiding and confining light.  Among them, the simplest type of one-dimensional photonic systems with remarkable topological features is based on a photonic analogy of the Su-Schrieffer-Heeger (SSH) model describing dimerized chains \cite{asboth2016short,longhiOL2013,poliNComms2015,lingOE2015,dengOL2016,blanco-RedondoPRL2016,meierNC2016,downingPRB2017,partoPRL2018}, which was originally proposed in the electronic context. 

In fact, such topological photonic systems can be also realized through ultracold atoms \cite{yelinPRL20172,perczelPRA2017}. This is inspired by the recent works on the topological quantum optics in 2D atomic arrays in optical lattices \cite{yelinPRL20172}. Nowadays, the techniques of optical lattices \cite{blochNaturephys2005} and optical tweezers \cite{lesterPRL2015} are able to accurately manipulate the positions of ultracold atoms to create artificial atomic lattices \cite{zollerNaturephys2016}. Utilizing the superfluid-Mott insulator transition for bosons, periodic atomic arrays with a single atom per lattice site can be fabricated using optical lattices \cite{greinerNature2002,shersonNature2010,bakrNature2011,bakrScience2010,weitenbergNature2011}. Recently, it is even possible to assemble 1D and 2D arrays of ultracold atoms with near-unity perfect filling factor in a large scale (up to 50 atoms) in an atom-by-atom level \cite{endresScience2016,lahayeScience2016}. Regarding the very high resonant scattering cross section at the dipole transition ($\sim\lambda_0^2$ where $\lambda_0$ is the wavelength of driving light) of a single ultracold atom \cite{guerin2017light}, such ultracold atom arrays offer people an excellent platform for achieving strong light-matter interaction. Furthermore, the prominent collective behaviors of photon scattering by cold atomic arrays, can be harnessed to create artificial ultrocold atom-based photonic crystals and metamaterials \cite{jenkinsPRA2012}, offering opportunities for novel miniaturized and efficient nanophotonic devices like perfect reflectors \cite{bettlesPRL2016,yelinPRL2017} and polarizers \cite{wangOE2017}. The ultra-narrow radiative linewidth of ultracold atoms \cite{Schilder2016} compared to conventional optical scatterers also provide finer observations about topological states than conventional systems. Moreover, if nonlinear effects in atoms are included at high-intensity excitation, such dimerized chains can be applied to study the many-body physics of interacting photons, offering an ideal playground with both controllable topology and nonlinearity \cite{changNaturephys2007, yelinPRL2017,yelinPRL20172}. Therefore, in this paper, we propose the one-dimensional dimerized ultracold atomic chain to study the topological edge and interface states of light, as a first step towards this goal. 

Another motivation of this work is to understand the topological physics of non-Hermitian systems that is developing rapidly. Different from the Hermitian Hamiltonian of the conventional SSH model, the effective Hamiltonians of light-matter interaction associated with the photonic analog of the SSH model using cold atoms and plasmonic nanoparticles are non-Hermitian, which, in general, implies complex eigenvalues with imaginary parts describing the interaction with environment \cite{yelinPRL20172,downing2018topological,pocockArxiv2017}. Actually, the non-Hermiticity of the effective Hamiltonian becomes more prominent when the retardation effect and far-field dipole-dipole interactions are taken into account \cite{downing2018topological,pocockArxiv2017}. Such non-Hermiticity may have relevant consequences on the topological phenomena in these systems. Recently, there has been a growing interest on the topological properties of non-Hermitian Hamiltonians \cite{garrisonPLA1988,rudnerPRL2009,huPRB2011,esakiPRB2011,liangPRA2013,schomerusOL2013,leePRL2016,lingSR2016,leykamPRL2017,jinPRA2017,weimannNaturemat2017,lieuPRB2018,yucePRA2018,xiongJPC2018,shenPRL2018,yao2018edge,yinPRA2018,alvarez2018topologicalreview,dangel2018topological,kunst2018biorthogonal,gong2018topological,kawabata2018nonhermitian}. Nevertheless, there are several crucial open questions yet to be carefully answered in this field, to name a few, including: (1) Can appropriate topological invariants be defined in non-Hermitian systems to describe the topology of bandstructures \cite{esakiPRB2011,liangPRA2013,leykamPRL2017,ozawa2018topological,alvarez2018topologicalreview,shenPRL2018}? (2) Is the bulk-boundary correspondence still valid in non-Hermitian systems \cite{ozawa2018topological,alvarez2018topologicalreview,xiongJPC2018}? (3) Are there stable and nontrivial edge states in the interface of topologically different non-Hermitian systems \cite{weimannNaturemat2017,alvarez2018topologicalreview}? Our system is naturally positioned in this context, and in this paper, we aim to answer these questions, at least partially, on the basis of the cold atomic realization of the SSH model for photons. 
	
On the other hand, unlike the conventional SSH model based on the nearest-neighbor approximation \cite{asboth2016short}, the consideration of retardation effect and far-field dipole-dipole interactions, which is especially necessary when the lattice period is comparable with or even larger than the wavelength \cite{pocockArxiv2017,downing2018topological}, will lead to the breaking of chiral symmetry, which is regarded as a necessary condition to realize topological protection for the edge states in the SSH model \cite{ryuNJP2010}. As a result, a thorough investigation on the consequences of the simultaneous presence of chiral-symmetry breaking and non-Hermiticity should be carried out. In addition, the presence of far-field dipole-dipole interactions, especially in the transverse eigenstates, which introduce very strong long-range hoppings of photons, can possibly change the entire topological landscape, should be examined carefully \cite{pocockArxiv2017}.
	
In this paper, by taking both near-field and far-field dipole-dipole interactions into account in the effective Hamiltonian describing light-matter interaction,  we analytically calculate the bandstructures of infinitely long 1D dimerized cold atomic systems. It is shown that a topological invariant, i.e., the complex Zak phase, can be well-defined to classify the topology of the Bloch bands. By carrying out open-boundary-condition calculations, we find that topologically protected states emerge at the edges of topologically nontrival dimerized chains as well as at the interface between chains with different complex Zak phases, implying the validity of the bulk-boundary correspondence in our system. We further verify that these conclusions are valid for both longitudinal and transverse eigenstates. We also reveal such topological photonic states are robust under symmetry-breaking disorders. For transverse eigenstates, we further discover the increase of localization length of topological edge states with the increase of lattice period due to the presence of strong far-field, long-range dipole-dipole interactions. The topologically protected edge states in dimerized ultracold atomic chains can provide an efficient interface for studying the topological states of light and matter. And we also expect our study can provide implications for the study of the topological properties of non-Hermitian Hamiltonians.

\section{Model}\label{model}
\begin{figure}[htbp]
	\centering
	\subfloat{
		\includegraphics[width=0.8\linewidth]{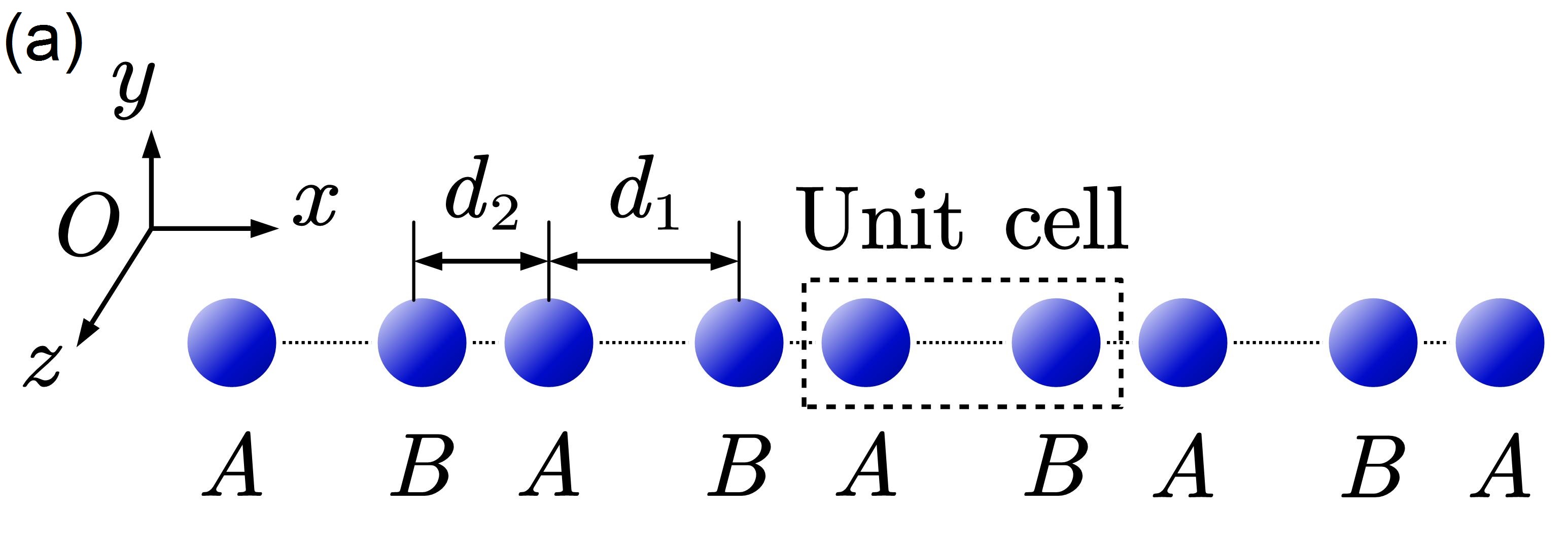}\label{schematic}
	}
	\hspace{0.01in}
	\flushleft
	\subfloat{
		\includegraphics[width=0.46\linewidth]{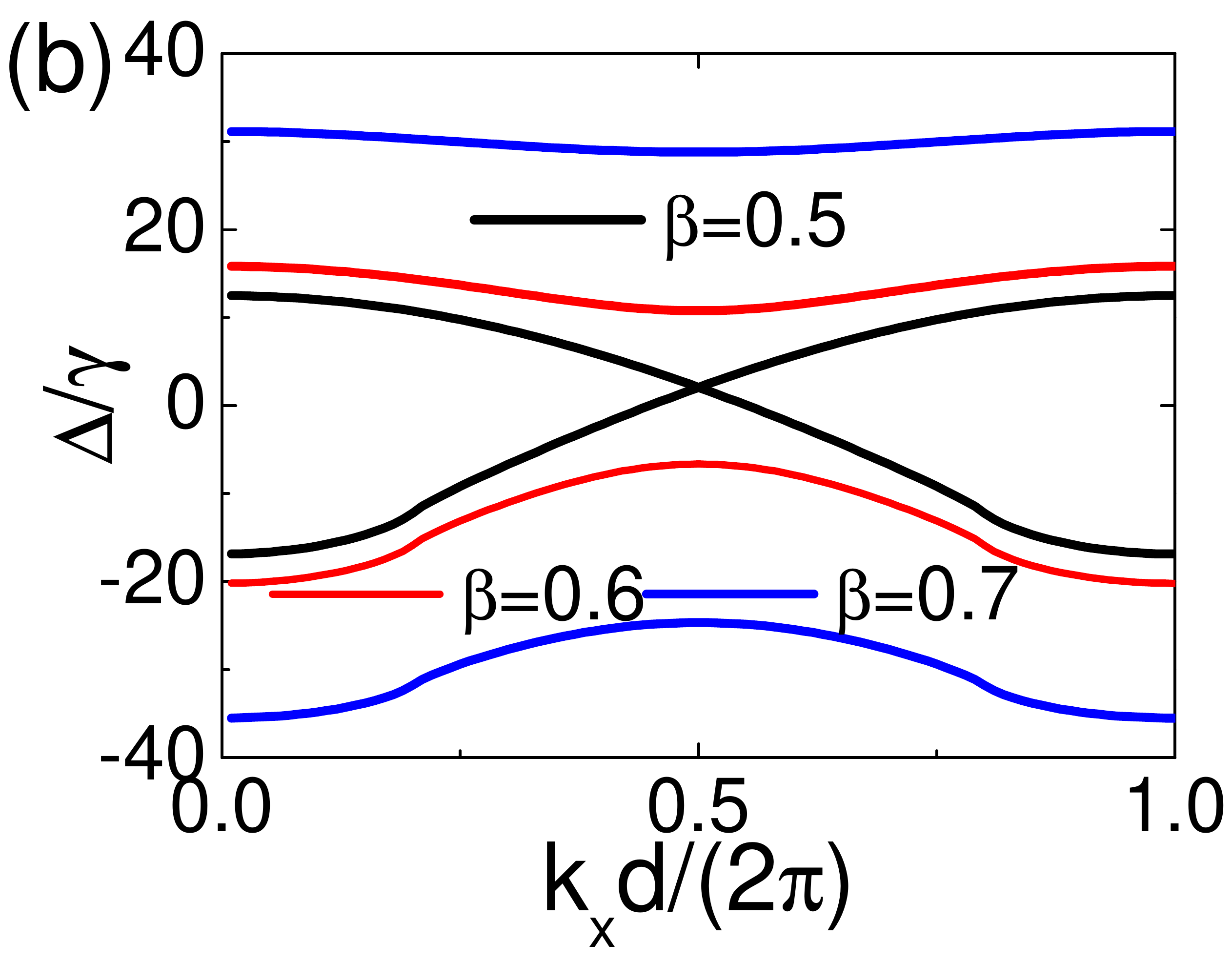}\label{bandstructure02}
	}
	\hspace{0.01in}
	\subfloat{
		\includegraphics[width=0.45\linewidth]{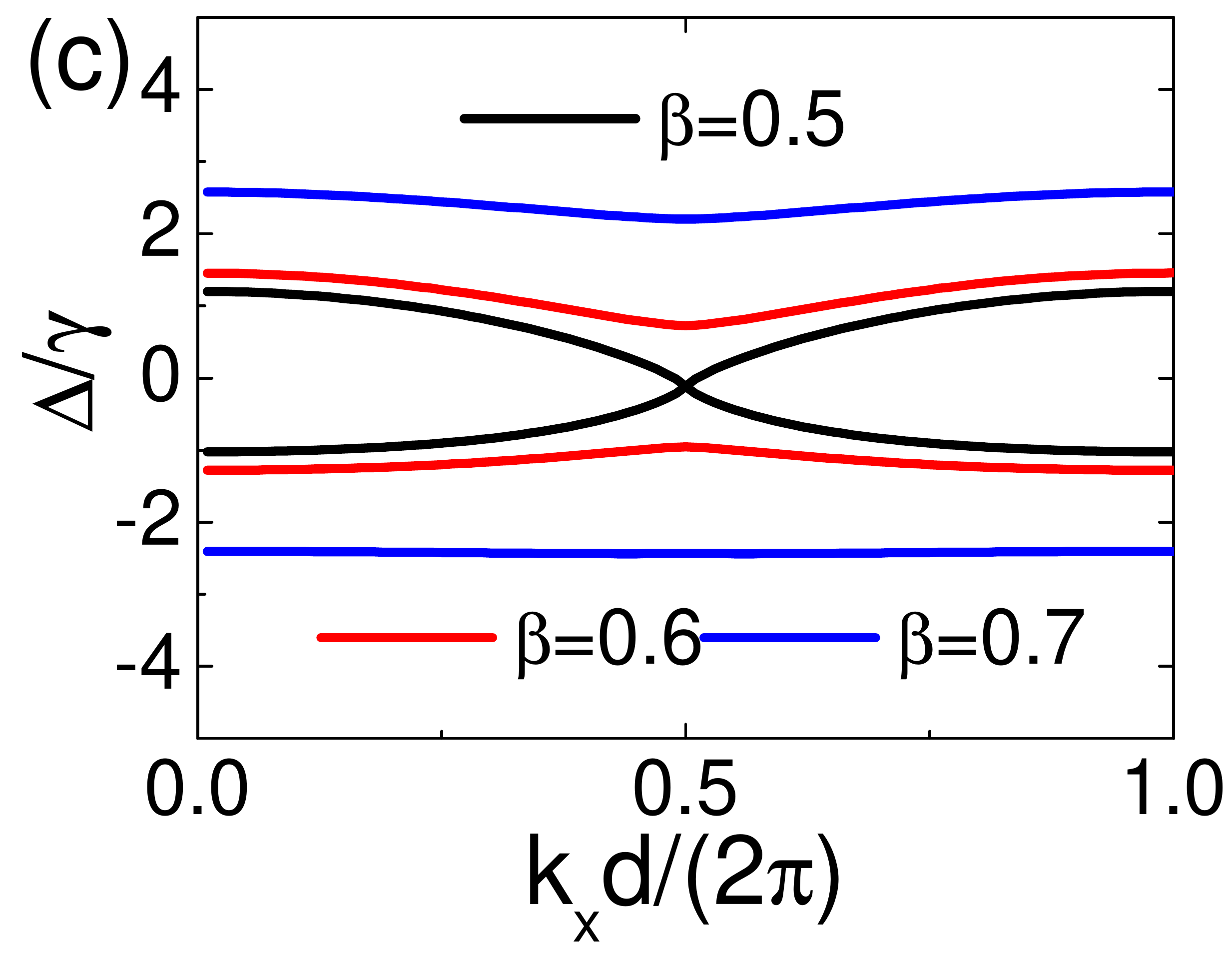}\label{bandstructure05}
	}
	\hspace{0.01in}
	\subfloat{
	\includegraphics[width=0.46\linewidth]{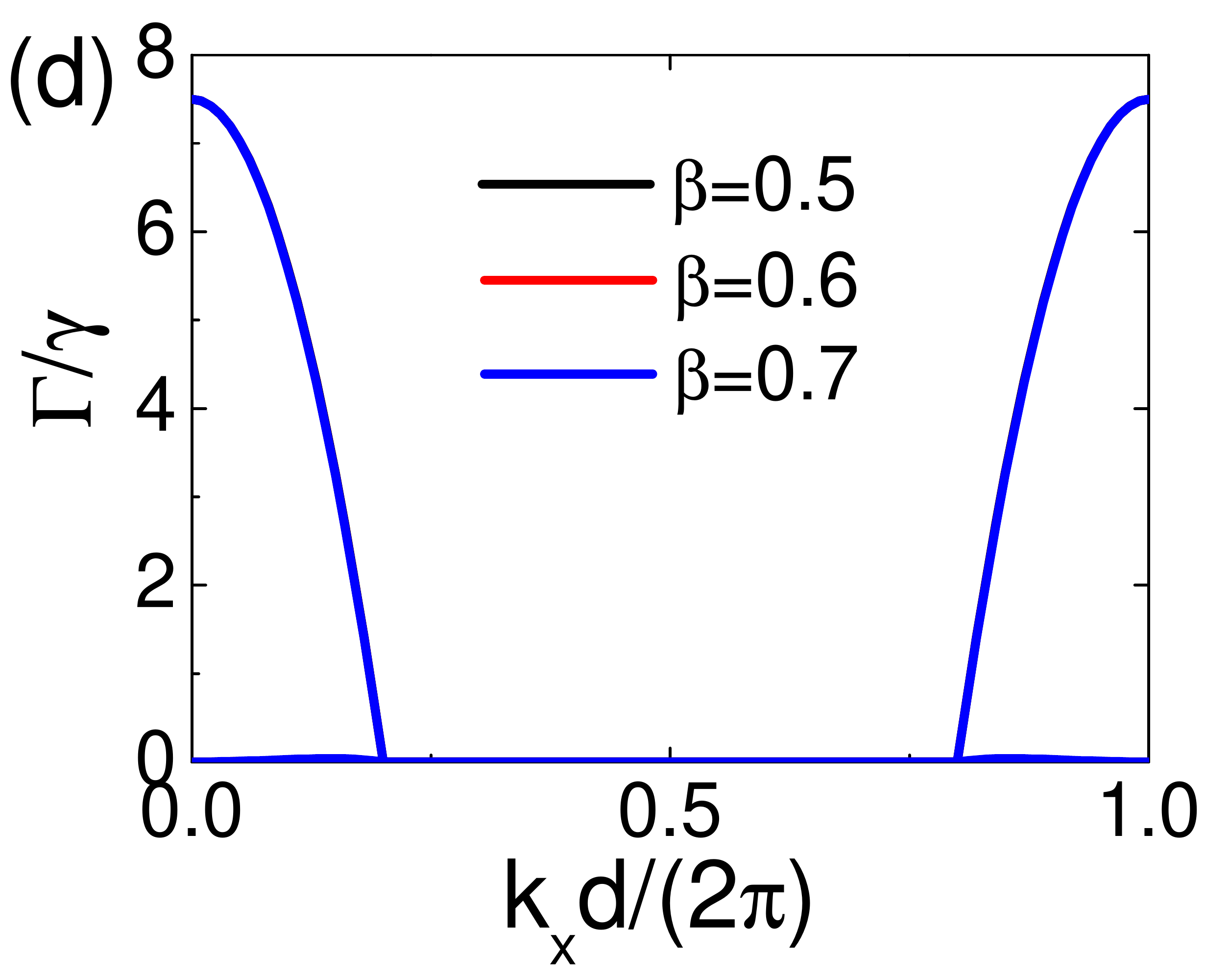}\label{imagband02}
}
\hspace{0.01in}
\subfloat{
	\includegraphics[width=0.46\linewidth]{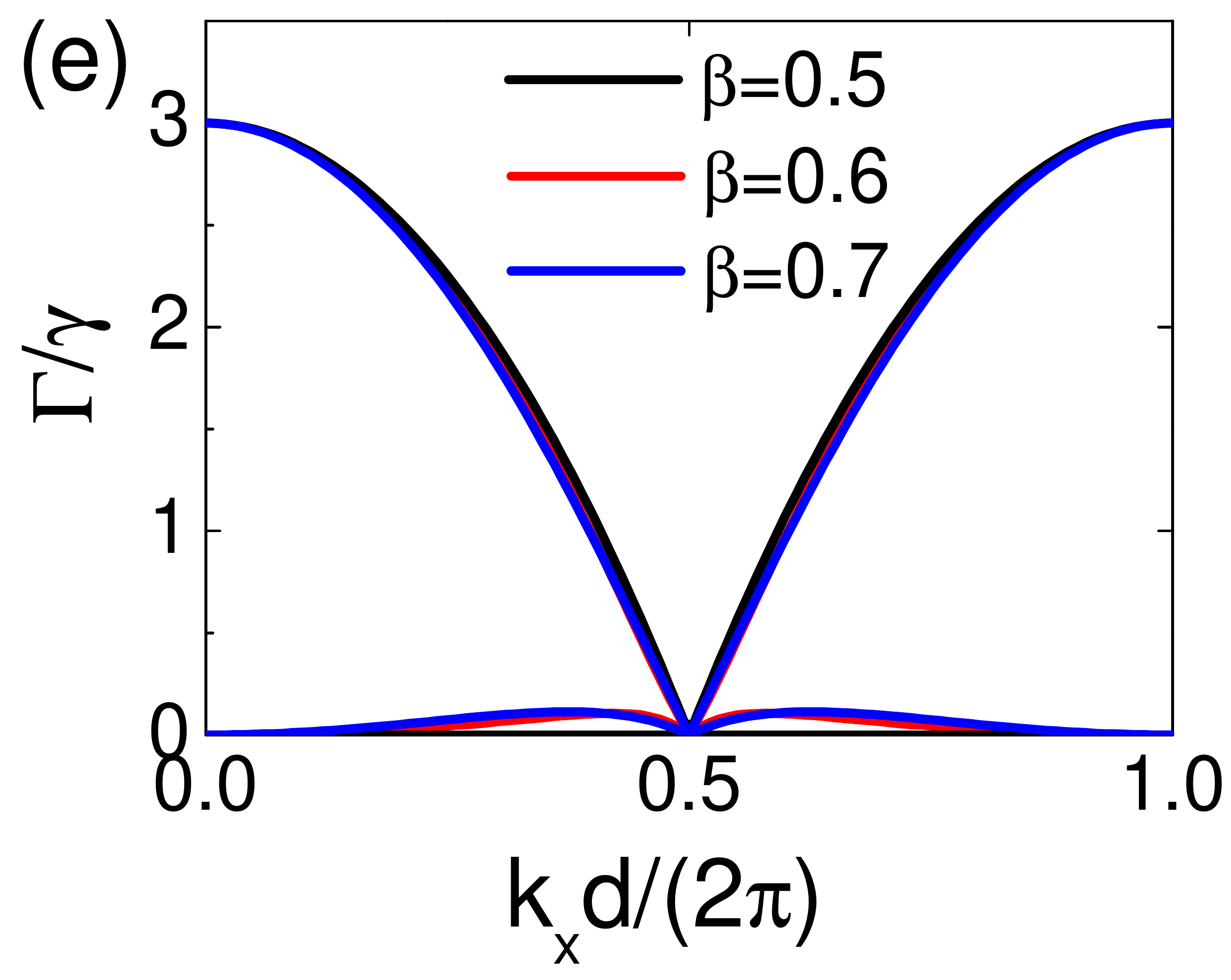}\label{imagband05}
}
	\caption{(a) Schematic of the dimerized ultracold atomic chain. The atoms are identical with alternate atom spacings represented by $d_1$ and $d_2$, and the two sublattices are denoted by $A$ and $B$ respectively. A unit cell is identified by the dashed-line rectangle. (b) Real parts of the bandstructures for longitudinal eigenstates of a dimerized chain with $d=0.2\lambda_0$ for different dimerization parameters $\beta$. Here $\lambda_0=\omega_0/c$ is the wavelength at the single-atom resonance. (c) The same as that in (b) but with $d=0.5\lambda_0$. (d) Imaginary parts of the bandstructures for longitudinal eigenstates of a dimerized chain with $d=0.2\lambda_0$ for different dimerization parameters $\beta$. (e) The same as that in (d) but with $d=0.5\lambda_0$.}\label{figcda}
\end{figure}

The schematic of a dimerized, ultracold, and two-level atomic chain is shown in Fig.\ref{schematic}, where the dimerization is introduced by using inequivalent spacings, i.e., $d_1\neq d_2$, between two sublattices, denoted by $A$ and $B$ respectively. The overall period is then $d=d_1+d_2$.  Here we first assume these atoms are perfectly trapped in a perfect Mott insulator state in the optical lattice \cite{antezzaPRA2009,yelinPRL20172}. Such dimerization leads to different ``hopping" rates of photons in different directions, mimicking the SSH model for electron hopping.  However, in the presence of near-field and far-field dipole-dipole interactions, the physical picture is much more complicated than nearest-neighbor ``hopping" in the conventional SSH model.

The two-level atom, for simplicity, has three degenerate excited states denoted by $|e_{\alpha}\rangle$ polarized along different directions, where $\alpha=x,y,z$ stands for the Cartesian coordinates, with a ground state denoted by $|g\rangle$. By applying the single excitation approximation (which is valid for sufficiently weakly excited system) \cite{kaiserJMO2011,kaiserFP2012,guerinPRL2016}, we can work in the subspace spanned by the ground states $|G\rangle\equiv|g...g\rangle$ and the single excited states $|i\rangle\equiv|g...e_i...g\rangle$ of the atoms \cite{kaiserJMO2011,kaiserFP2012,guerinPRL2016}. Moreover, by adiabatically eliminating the photonic degrees of freedom in the reservoir (i.e., the quantized electromagnetic field), we obtain the effective Hamiltonian describing light-atom interaction in the absence of the external field as \cite{antezzaPRL2009,antezzaPRA2009,kaiserJMO2011,kaiserFP2012,guerinPRL2016,perczelPRA2017,yelinPRL2017,yelinPRL20172} 
\begin{equation}\label{Hamiltonian}
\begin{split}
\mathcal{H}&=\hbar\sum_{i=1}^N\sum_{\alpha=x,y,z}(\omega_0-i\gamma/2)|e_{i,\alpha}\rangle\langle e_{i,\beta}|\\&+\frac{3\pi\hbar\gamma c}{\omega_0}\sum_{i=1,i\neq j}\sum_{\alpha,\beta=x,y,z}G_{\alpha\beta}(\mathbf{r}_j,\mathbf{r}_i)|e_{i,\alpha}\rangle\langle e_{j,\beta}|,
\end{split}
\end{equation}
which acts on the single excited states of the atoms. Here $\hbar$ is the Planck's constant, $\omega_0$ is angular frequency of the dipole transition from $|g\rangle$ to $|e\rangle$ in a single atom in free space with a radiative linewidth of $\gamma$, and $c$ is the speed of light in vacuum. $G_{\alpha\beta}(\mathbf{r}_j,\mathbf{r}_i)$ is the free-space dyadic Green's function describing the propagation of field emitting from the $i$-th atom to $j$-th atom, where $\mathbf{r}_j$ and $\mathbf{r}_i$ indicate their positions \cite{svidzinskyPRA2010,guerinPRL2016}:
\begin{equation}
\begin{split}
G_{\alpha\beta}(\mathbf{r}_j,\mathbf{r}_i)&=-\frac{\exp{(ikr)}}{4\pi r}\Big[\Big(1+\frac{i}{kr}-\frac{1}{(kr)^2}\Big)\delta_{\alpha\beta}\\&+\Big(-1-\frac{3i}{kr}+\frac{3}{(kr)^2}\Big)\hat{r}_{\alpha}\hat{r}_{\beta}\Big]
\end{split}
\end{equation}
where $k=\omega_0/c$ is the wavenumber in vacuum, $r=|\mathbf{r}|$, $\mathbf{r}=\mathbf{r}_j-\mathbf{r}_i$, $\hat{r}_\alpha=r_\alpha/r$, and $\delta_{\alpha\beta}$ is the Kronecker delta. For 1D atomic chains, the collective eigenstates can be regarded as transverse (all atoms are excited to the $|e_y\rangle$ or $|e_z\rangle$ states) or longitudinal (all atoms are excited to the $|e_x\rangle$) according to the polarization directions of the atoms \cite{weberPRB2004}. Here we first focus on the longitudinal states and discuss transverse states later because the longitudinal states couple with the light line very weakly \cite{lingSR2016,pocockArxiv2017,downing2018topological}. By invoking the Bloch theorem, for an infinite periodic dimerized lattice, we construct the longitudinal eigenstate with a wavenumber $k_x$ along the $x$-axis as a linear combination of the single-excited states \cite{kaiserJMO2011,kaiserFP2012,guerinPRL2016,yelinPRL20172,yelinPRL2017,perczelPRA2017}:
\begin{equation}\label{eigenstate}
|\psi_{k_x}\rangle=\sum_{n=-\infty}^{\infty}\exp(ik_xnd)[p_{A,k_x}|e_{2n-1,x}\rangle+p_{B,k_x}|e_{2n,x}\rangle],
\end{equation}
where $n$ denotes the $n$-th unit cell, $|e_{2n-1,x}\rangle$ and $|e_{2n,x}\rangle$ stand for the single excited states of the A-type atom and B-type atom, respectively, and $p_{A,k_x}$ and $p_{B,k_x}$ are corresponding expansion coefficients depending on $k_x$. Therefore, we have the following non-Hermitian eigenstate problem:
\begin{equation}\label{eigeneq}
\mathcal{H}|\psi_{k_x}\rangle=\hbar E_{k_x}|\psi_{k_x}\rangle,
\end{equation}
where $E_{k_x}$ is the complex eigenfrequency (energy) of the Bloch eigenstate described by $E_{k_x}=\omega_{k_x}-i\Gamma_{k_x}/2$, in which $\omega_{k_x}$ is the angular frequency and $\Gamma_{k_x}$ is the radiative linewidth of the eigenstate. Inserting Eqs.(\ref{Hamiltonian}) and (\ref{eigenstate}) into Eq.(\ref{eigeneq}), we can straightforwardly obtain the following equation:
\begin{widetext}
\begin{equation}\label{blocheqlong}
-\frac{3\pi \gamma c}{\omega_0}\left(\begin{matrix}
\sum_{n\neq 0}G_{xx}(nd)\exp{(ik_xnd)} & \sum G_{xx}(nd+d_1)\exp{(ik_xnd)}\\
\sum G_{xx}(nd-d_1)\exp{(ik_xnd)} &
\sum_{n\neq 0}G_{xx}(nd)\exp{(ik_xnd)}
\end{matrix}\right)\left(\begin{matrix}p_{A,k_x}\\p_{B,k_x}\end{matrix}\right)=(E_{k_x}-\omega_0 + i \gamma/2)\left(\begin{matrix}p_{A,k_x}\\p_{B,k_x}\end{matrix}\right),
\end{equation}
\end{widetext}
where 
\begin{equation}\label{gxxeq}
G_{xx}(x)=-2\Big[\frac{i}{k|x|}-\frac{1}{(kx)^2}\Big]\frac{\exp{(ik|x|)}}{4\pi |x|},
\end{equation}
Therefore we can obtain the following eigenvalue equation
\begin{equation}\label{eigenvalue_long}
\left(\begin{matrix}
a_{11}^{L}(k_x) & a_{12}^{L}(k_x)\\
a_{21}^{L}(k_x) & a_{22}^{L}(k_x)
\end{matrix}\right)\left(\begin{matrix}p_{A,k_x}\\p_{B,k_x}\end{matrix}\right)=\frac{E_{k_x}-\omega_0 + i \gamma/2}{-3\pi\gamma}\left(\begin{matrix}p_{A,k_x}\\p_{B,k_x}\end{matrix}\right),
\end{equation}
The matrix in the left hand side (LHS) of Eq.(\ref{eigenvalue_long}) can be regarded as the effective Hamiltonian in the reciprocal domain, denoted by $H(k_x)$. Using the polylogrithm (or Jonqui\'ere's function) defined as $Li_s(z)=\sum_{n=1}^\infty z^n/n^s$, we can straightforwardly obtain the diagonal elements as

\begin{equation}
\begin{split}
a_{11}^{L}(k_x)=a_{22}^{L}(k_x)&=-i\frac{Li_2(z^+)+Li_2(z^-)}{2\pi k^2d^2}\\&+\frac{Li_3(z^+)+Li_3(z^-)}{2\pi k^3d^3},
\end{split}
\end{equation}
and invoking the Lerch transcendent defined as $\Phi(z,s,a)=\sum_{n=0}^\infty z^n/(n+a)^s$, we get the off-diagonal elements
\begin{equation}\label{a12Leq}
\begin{split}
&a_{12}^{L}(k_x)=\Big[-i\frac{\Phi(z^+,2,\beta)}{2\pi k^2d^2}+\frac{\Phi(z^+,3,\beta)}{2\pi k^3d^3}\Big]\exp{(ik\beta d)}\\&+\Big[-i\frac{\Phi(z^-,2,1-\beta)}{2\pi k^2d^2}+\frac{\Phi(z^-,3,1-\beta)}{2\pi k^3d^3}\Big]z^-\exp{(-ik\beta d)},
\end{split}
\end{equation}
\begin{equation}\label{a21Leq}
\begin{split}
&a_{21}^{L}(k_x)=\Big[-i\frac{\Phi(z^+,2,1-\beta)}{2\pi k^2d^2}+\frac{\Phi(z^+,3,1-\beta)}{2\pi k^3d^3}\Big]z^+\\&\times\exp{(-ik\beta d)}+\Big[-i\frac{\Phi(z^-,2,\beta)}{2\pi k^2d^2}+\frac{\Phi(z^-,3,\beta)}{2\pi k^3d^3}\Big]\exp{(ik\beta d)},
\end{split}
\end{equation}
where $\beta=d_1/d$ is the dimerization parameter, $z^+=\exp{(i(k+k_x)d)}$ and $z^-=\exp{(i(k-k_x)d)}$ \cite{pocockArxiv2017}. Here the matrix elements are not unique, depending on the choice of unit cell \cite{atalaNaturephys2013}, which are chosen such that it fulfills $a_{ij}^L(k_x)=a_{ij}^L(k_x+2\pi/d)$, i.e., using the periodic gauge \cite{lingOE2015,ozawa2018topological}. Therefore, the calculated dispersion relations in real and imaginary spaces are:

\begin{equation}\label{deltaeq}
\frac{\Delta}{\gamma}=3\pi\mathrm{Re}\left[-a_{11}^{L}(k_x)\pm\sqrt{a_{12}^{L}(k_x)}\sqrt{a_{21}^{L}(k_x)}\right],
\end{equation} 
\begin{equation}\label{gammaeq}
\frac{\Gamma}{\gamma}=1+6\pi\mathrm{Im}\left[a_{11}^{L}(k_x)\mp\sqrt{a_{12}^{L}(k_x)}\sqrt{a_{21}^{L}(k_x)}\right],
\end{equation} 
respectively, where $\Delta=\mathrm{Re}E_{k_x}-\omega_0$ stands for the detuning of the eigenstate, and $\Gamma=-2\mathrm{Im}E_{k_x}$ is the corresponding radiative linewidth (decay rate). It can be inferred that for the nearest-neighbor approximation, the upper and lower bands are symmetric with respect to $\Delta=0$ (i.e., the single atom resonance frequency) \cite{zhangPRB2018}, while the intercell dipole-dipole couplings are introduced, the bandstructures are not symmetric any more. The calculated bulk (Bloch) bandstructures for $\beta=0.5, 0.6, 0.7$ are shown in Figs.\ref{bandstructure02} and \ref{imagband02} for $d=0.2\lambda_0$, and Figs.\ref{bandstructure02} and \ref{imagband05} show the real and imaginary parts of the spectrum for $d=0.5\lambda_0$. The bandstructures for $\beta=0.3$ and $\beta=0.4$ are the same as those for $\beta=0.7$ and $\beta=0.6$, respectively. It is observed that for $\beta\neq0.5$, bandgaps in the real space are opened in both cases, while a smaller period leads to a wider bandgap. This is due to that the near-field dipole-dipole interactions between atoms can give rise to very strong frequency shifts \cite{Schilder2016,wangOE2017}. On the other hand, the imaginary parts of the bandstructures are ungapped.

\section{Topological invariant and midgap states}
\subsection{Topological invariant}\label{topoinvariant}
Previously, much attention concerning the topological properties of non-Hermitian systems was paid to the study of exceptional points (EPs), which largely determine the exotic topological phenomena of such systems \cite{fengNaturephton2017,elganainyNaturephys2018}. Exceptional points are singularities of energy spectrum of non-Hermitian Hamiltonians where the eigenvalues and eigenvectors (eigen wave functions) coalesce \cite{heissJPA2012}. Various winding numbers respect to the EPs are then defined to characterize non-Hermitian systems \cite{leePRL2016,leykamPRL2017,yinPRA2018} are defined to depict the topology of non-Hermitian bandstructures. Recently, more general discussions were carried out to classify the topological properties of non-Hermitian systems in analogy with the Altland-Zirnbauer (AZ) classification \cite{schnyderPRB2008} for Hermitian systems \cite{gong2018topological}. On the other hand, for non-Hermitian systems without band degeneracies (i.e., EPs), which means the different bands are separable, the complex Zak phase \cite{garrisonPLA1988,liangPRA2013,lieuPRB2018,partoPRL2018,dangel2018topological} and the (first) Chern number \cite{shenPRL2018,kawabata2018nonhermitian} are used to characterize the topological properties of one-dimensional and two-dimensional systems, respectively. Nevertheless, for non-Hermitian systems, the left-eigenvectors are different from the right-eigenvectors, and both of them should be exploited in the calculation of topological invariants. 	
	
More rigorously, here a ``separable" band with a band number $n$ means that for any $m\neq n$ in the entire bandstructure, the energy $E_{m,\mathbf{k}}\neq E_{n,\mathbf{k}}$ for all $\mathbf{k}$ in the complex plane. Moreover, if the energy $E_{m,\mathbf{k}'}\neq E_{n,\mathbf{k}}$ for all $\mathbf{k}$ and $\mathbf{k}'$ in the complex plane, the band $n$ is called ``isolated" or ``gapped" \cite{shenPRL2018}. According to Eqs.(\ref{deltaeq}-\ref{gammaeq}), since $a_{12}(k_x)a_{21}(k_x)\neq0$ is always valid in our case when $\beta\neq0.5$, the bulk bandstructures are always separable. Actually, in the complex plane, the cases investigated in this study are always gapped except for  $\beta\neq0.5$ (which can be seen in the examples exhibited below). As a consequence, in this paper, we expect that for the present non-Hermitian system the complex Zak phase is quantized and can describe the topological phase transition, where the transition point is the gap closing point, i.e., $\beta=0.5$ \cite{lieuPRB2018}. 

Specifically, since the non-Hermitian effective Hamiltonian in Eq.(\ref{eigenvalue_long}) has no chiral symmetry, according to the conventional AZ classification for Hermitian systems, this system seems to be topologically trivial \cite{schnyderPRB2008,asboth2016short}. However, a close scrutiny of the effective Hamiltonian tells us that although the frequencies of eigenstates are affected by the diagonal terms (chiral-symmetry breaking terms), the eigenvectors of eigenstates are still the same as those of the chirally symmetric counterpart  of the Hamiltonian, i.e., 
\begin{equation}
\tilde{H}(k_x)=\left(\begin{matrix}0 & a_{12}^{L}(k_x)\\
a_{21}^{L}(k_x) & 0\end{matrix}\right),
\end{equation}
which obviously fulfills the chiral symmetry condition $\sigma_z\tilde{H}(k_x)\sigma_z=-\tilde{H}(k_x)$. Here $\sigma_i$ with $i=x,y,z$ refers to Pauli matrices. Such a property can be viewed as a trivial chiral-symmetry breaking as pointed out by Pocock et al \cite{pocockArxiv2017}. In that sense, the complex Zak phase preserves the feature in a chirally symmetric system, and thus is still quantized and can be used to determine the topology of bulk bandstructure, as recently discussed by Lieu \cite{lieuPRB2018}. Note this quantization does not refer to the inversion symmetry, unlike the real Zak phase, which is defined solely based on right eigenvectors and requires the inversion symmetry to be quantized \cite{lieuPRB2018}. Nevertheless, here the effective Hamiltonian still obeys this symmetry, i.e., $\sigma_xH(k_x)\sigma_x=H(-k_x)$ over the center of the lattice. That is why some authors argued that the inversion symmetry quantizes the Berry phase \cite{downing2018topological}.

\begin{figure}[htbp]
	\centering
	\flushleft
	\subfloat{
		\includegraphics[width=0.46\linewidth]{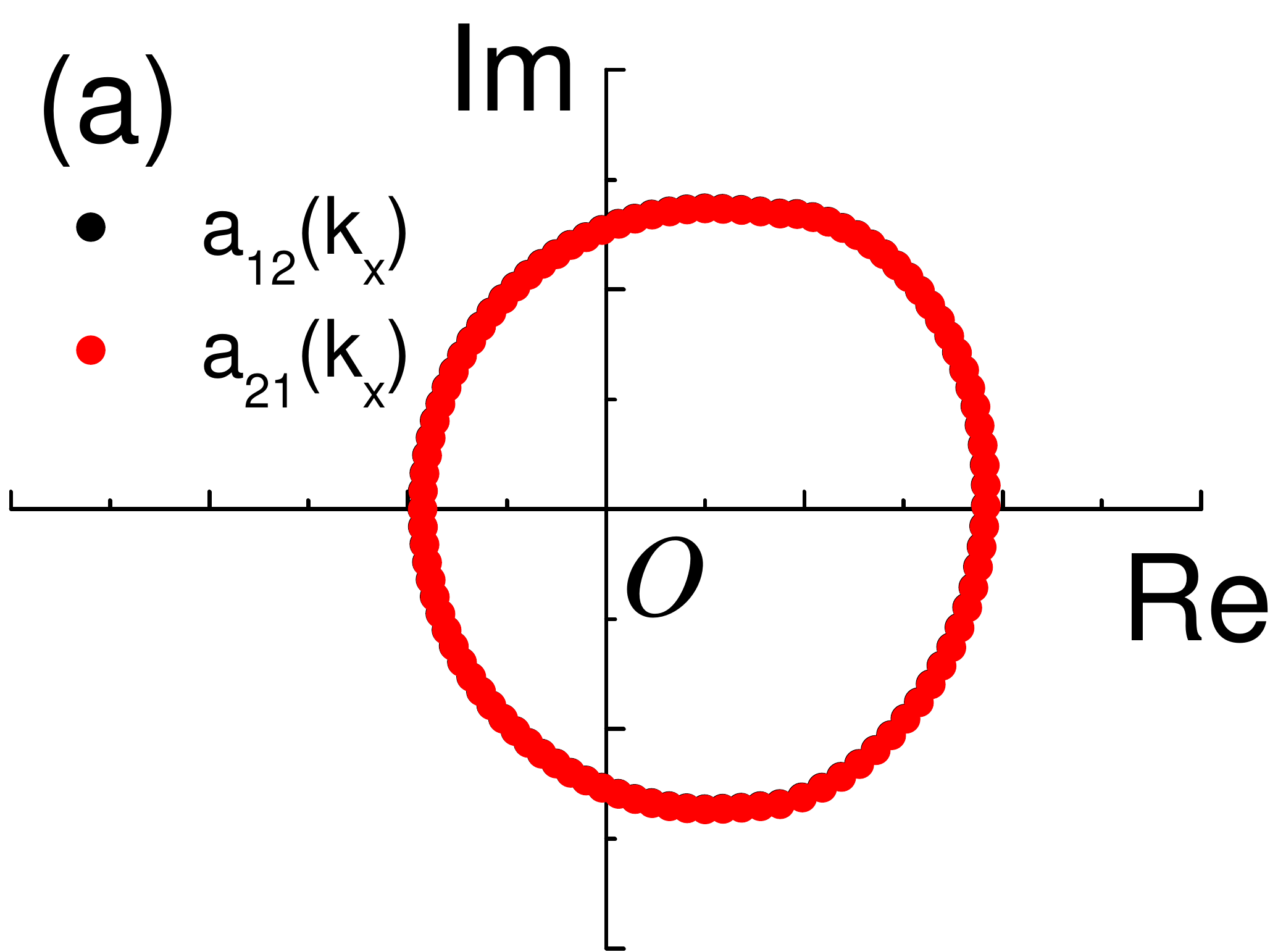}\label{vorticitybeta06d02}
	}
	\hspace{0.01in}
	\subfloat{
		\includegraphics[width=0.46\linewidth]{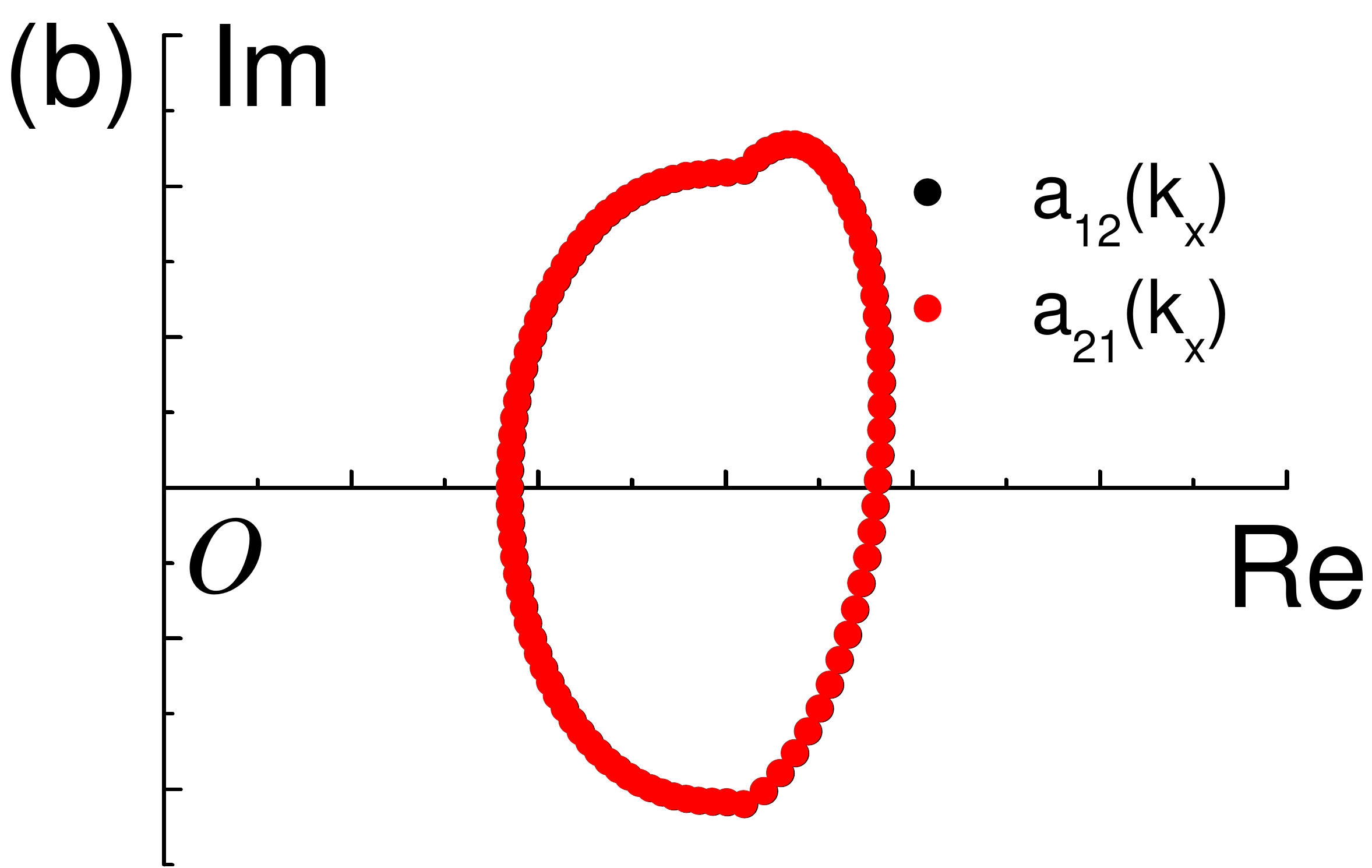}\label{vorticitybeta04d02}
	}
	\hspace{0.01in}
	\subfloat{
	\includegraphics[width=0.46\linewidth]{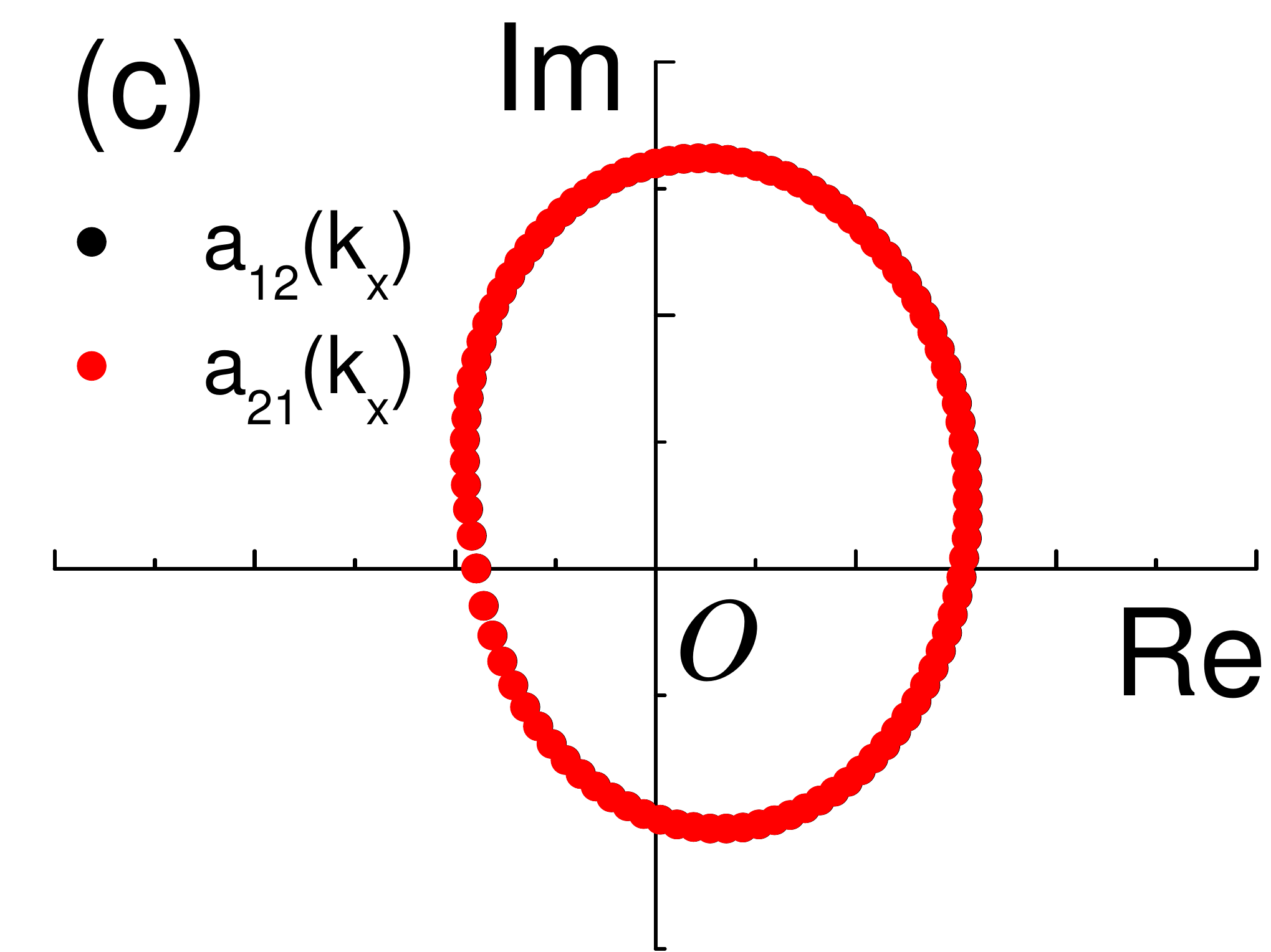}\label{vorticitybeta06d05}
}
\hspace{0.01in}
\subfloat{
	\includegraphics[width=0.46\linewidth]{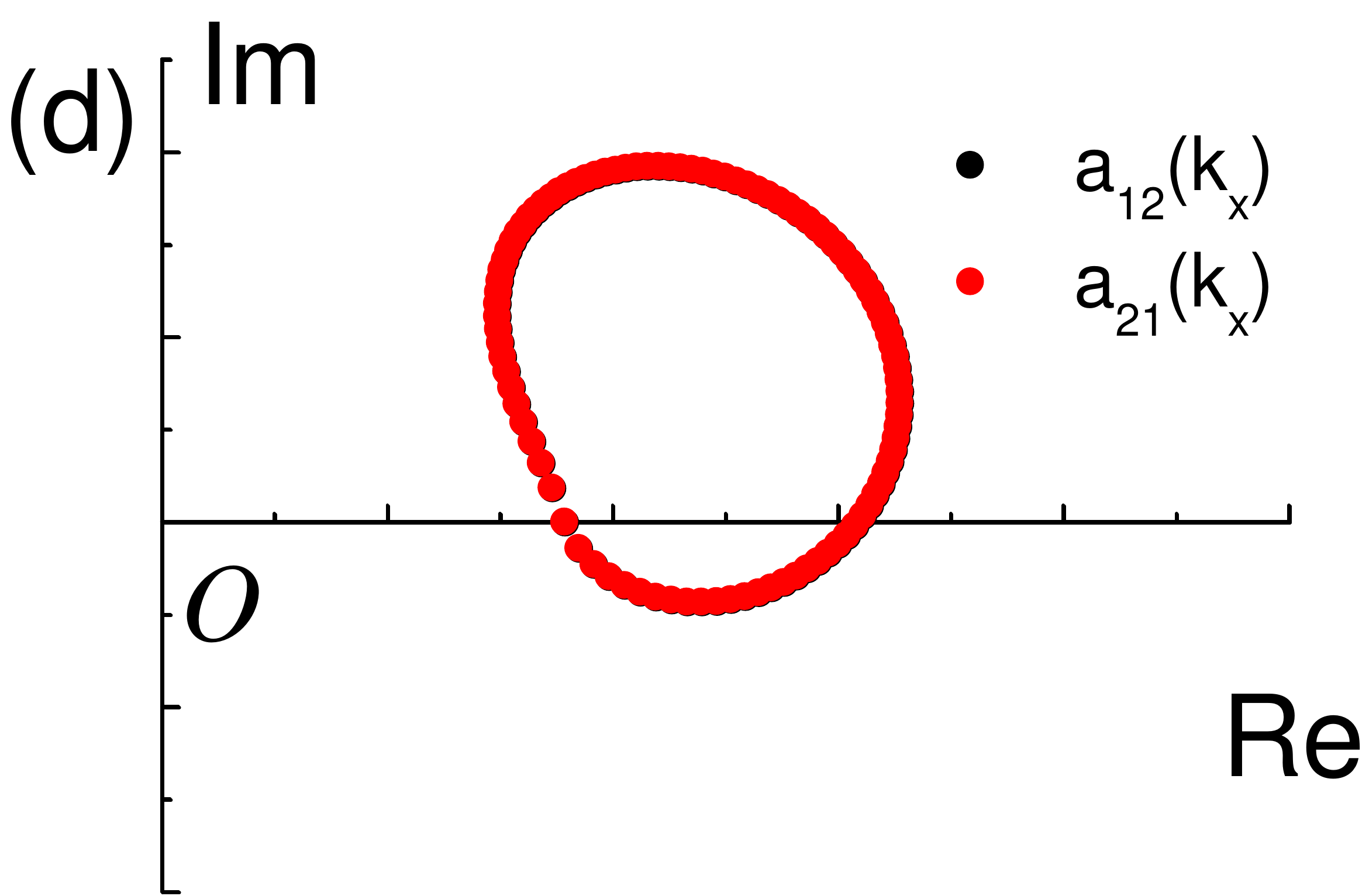}\label{vorticitybeta04d05}
}
	\caption{The winding of $a_{12}(k_x)$ and $a_{21}(k_x)$ over the origin in the complex plane. The lattice structures are (a) $d=0.2\lambda_0$, $\beta=0.6$. (b) $d=0.2\lambda_0$, $\beta=0.4$. (c) $d=0.5\lambda_0$,$\beta=0.6$. (d) $d=0.5\lambda_0$, $\beta=0.4$.}\label{windingnumber}
	
\end{figure}

In 1D non-Hermitian systems, the complex Zak phase should be reformulated according to the orthogonality of left and right eigenstates (also called biorthogonality) \cite{esakiPRB2011,schomerusOL2013,lingSR2016,weimannNaturemat2017,dingPRB2015,jinPRA2017,lieuPRB2018,shenPRL2018,partoPRL2018,yucePLA2015,wagnerAP2017,yucePRA2018,lieuPRB2018,alvarezPRB2018} as 
\begin{equation}
\theta_B=i\int_\mathrm{BZ}\sum_{j=A,B}\langle p_{j,k_x}^{L}|i\nabla_{k_x}|p^R_{j,k_x}\rangle,
\end{equation}
where $|p_{k_x}^{R}\rangle$ and $|p_{k_x}^{L}\rangle$ denote right and left eigenstates of $H(k_x)$. The left eigenstate satisfies $H^\dag(k_x)|p_{k_x}^{L}\rangle=E_{k_x}^*|p_{k_x}^{L}\rangle$. These normalized (to make $\langle p_{k_x}^{L}|p_{k_x}^{R}\rangle=1$) eigenvectors of the eigenstates are calculated as
\begin{equation}
|p_{k_x}^{L}\rangle=\frac{1}{\sqrt{2}}\left(\begin{matrix}\mp\frac{\sqrt{a_{21}^{L,*}(k_x)}}{\sqrt{a_{12}^{L,*}(k_x)}}\\1\end{matrix}\right),
\end{equation}
\begin{equation}
|p_{k_x}^{R}\rangle=\frac{1}{\sqrt{2}}\left(\begin{matrix}\mp\frac{\sqrt{a_{12}^L(k_x)}}{\sqrt{a_{21}^L(k_x)}}\\1\end{matrix}\right).
\end{equation}
Therefore the complex Zak phase is
\begin{equation}\label{cberryphase}
\begin{split}
\theta_\mathrm{B}&=\int_\mathrm{BZ}dk_x\mathcal{A}(k_x)\\&=i\int_{-\pi/d}^{\pi/d}\Big[p_{A,k_x}^{L,*}\frac{\partial p_{A,k_x}^R}{\partial k_x}+p_{B,k_x}^{L,*}\frac{\partial p_{B,k_x}^R}{\partial k_x}\Big]dk_x\\&=\frac{\arg[a_{21}(k_x)]-\arg[a_{12}(k_x)]+i\ln(\frac{|a_{12}(k_x)|}{|a_{21}(k_x)|}}{4}\Big|_{-\pi/d}^{\pi/d}.
\end{split}
\end{equation}
According to Eq.(\ref{cberryphase}), the real part of $\theta_B$ is simply half the difference of the winding numbers of $a_{21}(k_x)$ and $a_{12}(k_x)$ encircling the origin multiplied by $\pi$. Since $|a_{12}(-\pi/d)|=|a_{21}(-\pi/d)|=|a_{12}(\pi/d)|=|a_{21}(\pi/d)|$ from Eqs.(\ref{a12Leq}-\ref{a21Leq}), the imaginary part of $\theta_\mathrm{B}$ is exactly zero. Therefore, the complex Zak phase is a real quantity. In Fig.\ref{windingnumber}, we show the winding of $a_{12}(k_x)$ and $a_{21}(k_x)$ over the origin in the complex
plane, for the cases of $\beta=0.4$ and $\beta=0.6$ with lattice constant $d=0.2\lambda_0$ and $d=0.5\lambda_0$. Since the winding directions of $a_{12}(k_x)$ and $a_{21}(k_x)$ are always opposite because $a_{12}(k_x)=a_{21}(-k_x)$, the winding numbers of $a_{12}(k_x)$ and $a_{21}(k_x)$ are +1 and -1 respectively when the dimerization parameter is $\beta=0.6$, while when $\beta=0.4$, the winding numbers are both zero for the cases of $d=0.2\lambda_0$ and $d=0.5\lambda_0$. As a consequence, the complex Zak phase for $\beta=0.6$ is $\pi$ and is $0$ for $\beta=0.4$, and the topological phase transition point is the gap closing point, i.e., $\beta=0.5$. 
%Here the complex Zak phase is either $\pi$ or $0$ and therefore results in a $\mathbb{Z}_2$ topological insulator \cite{miert2DM2017}. 
The quantized complex Zak phase gives rise to the total winding number $\mathcal{W}=\theta_\mathrm{B}/\pi$, which determines the number of edge states in either boundary of a finite 1D chain if the bulk-boundary correspondence is valid, which will be discussed in the next subsection. Note in the present case, the next-nearest-neighbor or high-order couplings cannot provide an increase in the winding number due to the nature of dipole-dipole interactions. However, if these couplings are of different nature and become strong enough, higher winding numbers are also possible, for example, the cases discussed by  Yin et al \cite{yinPRA2018} and P{\'e}rez-Gonz{\'a}lez et al \cite{perezArxiv2018}, although in different physical contexts.

\subsection{Bulk-boundary correspondence and midgap states}
In Hermitian systems, according to the bulk-boundary correspondence, the topological invariant (Berry phase) determines the existence of edge states \cite{luNPhoton2014,khanikaevNPhoton2017,rhimPRB2017,ozawa2018topological}. The existence of nontrivial topological invariant predicts topologically protected edge states at the boundary between the chain and the topologically trivial vacuum. However, for some non-Hermitian Hamiltonians in one dimension, it was recently found that the conventional bulk-boundary correspondence becomes invalid \cite{yao2018edge,kunst2018biorthogonal,xiongJPC2018}. This breakdown is because the appreciable difference between the Bloch bandstructures under the periodic boundary condition (PBC) and the bulk bandstructures calculated using the open boundary condition (OBC).  Consequently, some modified topological invariants were proposed, for instance, the so-called biorthogonal polarization \cite{kunst2018biorthogonal}, and the modified winding number defined with respect to a deformed Brillouin zone \cite{yao2018edge}, to correctly characterize the topology of the bulk bandstructures of finite 1D non-Hermitian systems. Nevertheless, in the current non-Hermitian system, we show that the previously derived complex Zak phase is able to characterize these edge states, and the bulk-boundary correspondence is still valid.  

Let us first consider the longitudinal eigenstates of a finite-size chain with $N=100$ identical atoms (50 dimers) with an open boundary. Its bandstructure can be determined numerically by calculating the eigenstates of the Hamiltonian in Eq.(\ref{Hamiltonian}) with respect to the wave function $|\psi\rangle=\sum_{j=1}^{N}p_{j}|e_{j,x}\rangle$, where $p_{i}$ is the expansion coefficient denoting the probability amplitude of each excited state $|e_{j,x}\rangle$ \cite{weberPRB2004,guerinPRL2016,kaiserFP2012,kaiserJMO2011}, whose physical significance in classical electrodynamics is the (normalized) dipole moment of the $j$-th atom. This equation specifies a set of solutions in the form $E=\omega-i\Gamma/2$ ($\Gamma>0$) in the lower complex plane denoting to the eigenstates of the dimerized chain. The same as in Sec.\ref{model}, $\omega$ amounts to the angular frequency of an eigenstate while $\Gamma$ refers to its radiative linewidth (decay rate), where the corresponding right eigenvector $|\mathbf{p}^R\rangle=[p_1p_2...p_j...p_N]$ then indicates the dipole moment distribution of an eigenstate in a classical interpretation. The wavenumber for an eigenstate can be determined by \cite{weberPRB2004,pocockArxiv2017}
\begin{equation}
k_x\Big(\frac{2\pi}{d}\Big)^{-1}=\frac{(N-2)n+1}{N(N-1)},
\end{equation}
where $n$ is the mode number of an eigenstate, which is 1 plus the number of times of sign changes of $\mathrm{Re}(p_j)$ along the chain for that eigenstate \cite{weberPRB2004,pocockArxiv2017}. To demonstrate the degree of state localization in space, we further analyze the inverse participation ratio (IPR) of an eigenstate from its eigenvector as \cite{wangOL2018}
\begin{equation}
\mathrm{IPR}=\frac{\sum_{j=1}^{N}|p_j|^4}{[\sum_{j=1}^{N}|p_j|^2]^2}.
\end{equation}
The IPR can be used to indicate the spatial confinement of different eigenstates \cite{Skipetrov2014,wangOL2018}. For instance, for an IPR approaches $1/M$, where $M$ is an integer, the corresponding eigenstate involves the excitation of $M$ atoms \cite{Skipetrov2014,wangOL2018}. For a highly localized topological edge/interface state, its IPR should be substantially larger compared to those of bulk states \cite{wangOL2018}. 
\begin{figure}[htbp]
	\centering
	\flushleft
	\subfloat{
	\includegraphics[width=0.46\linewidth]{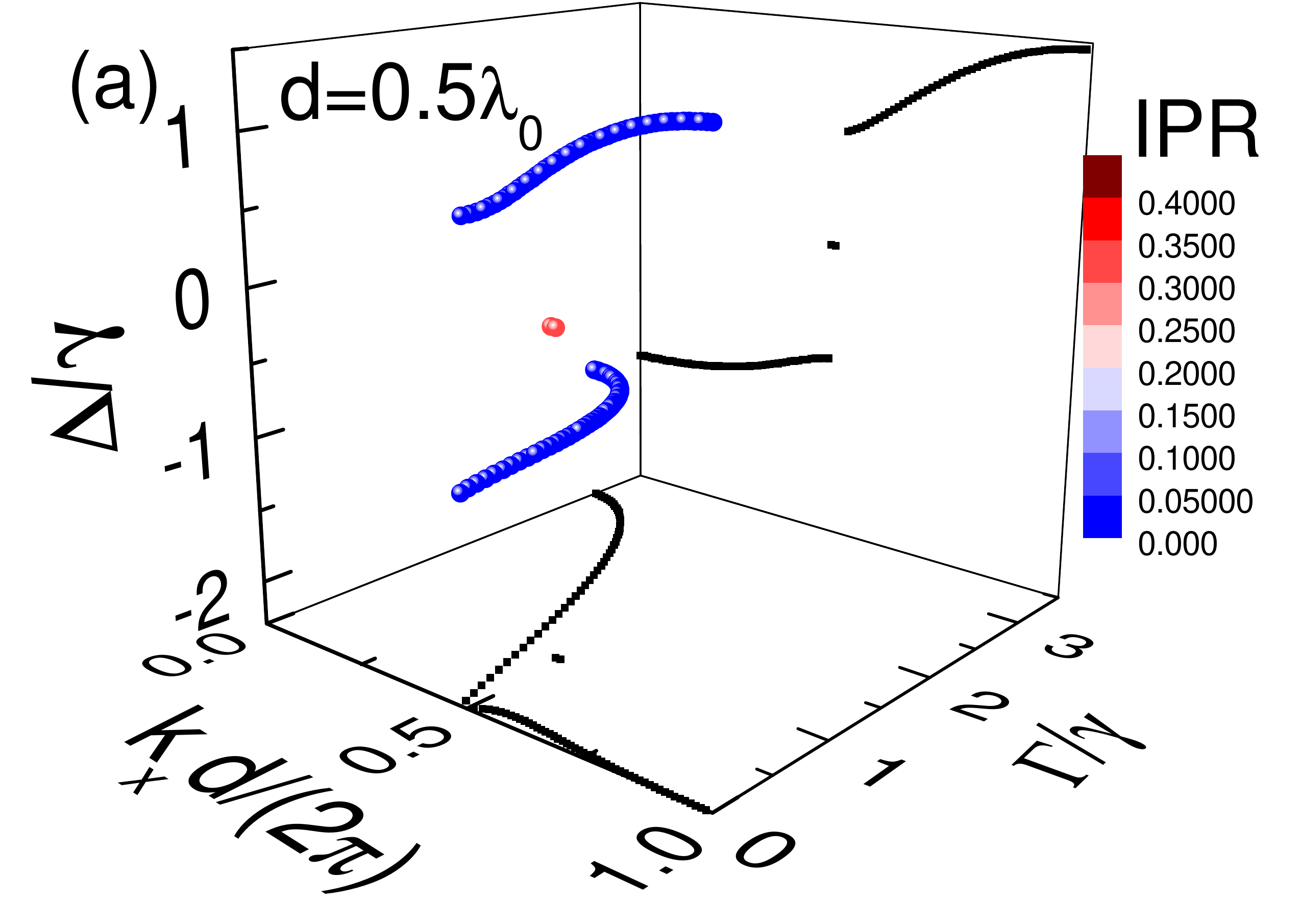}\label{beta06longband05}
}
	\hspace{0.01in}
	\subfloat{
	\includegraphics[width=0.46\linewidth]{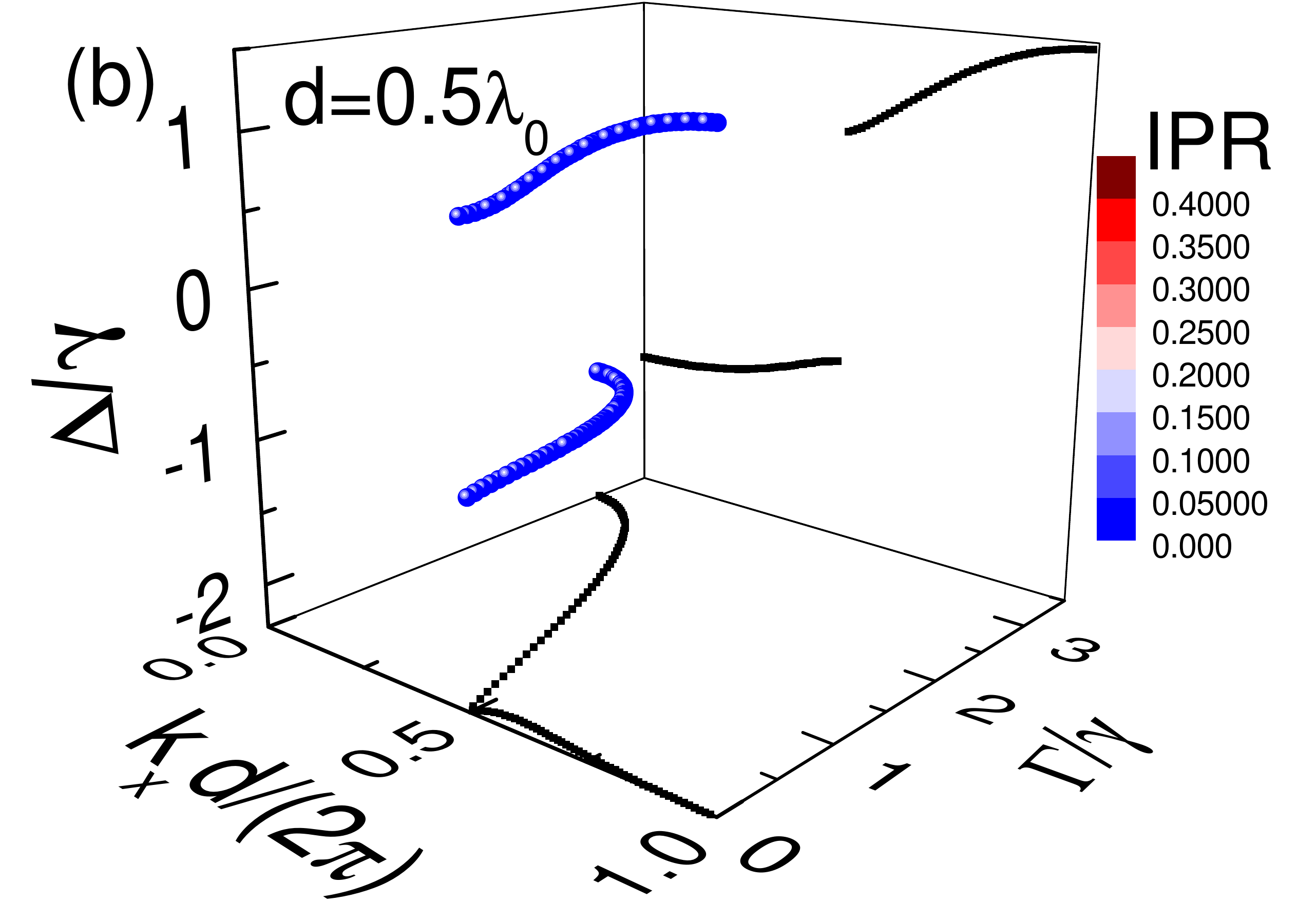}\label{beta04longband05}
}
	\hspace{0.01in}
\subfloat{
	\includegraphics[width=0.46\linewidth]{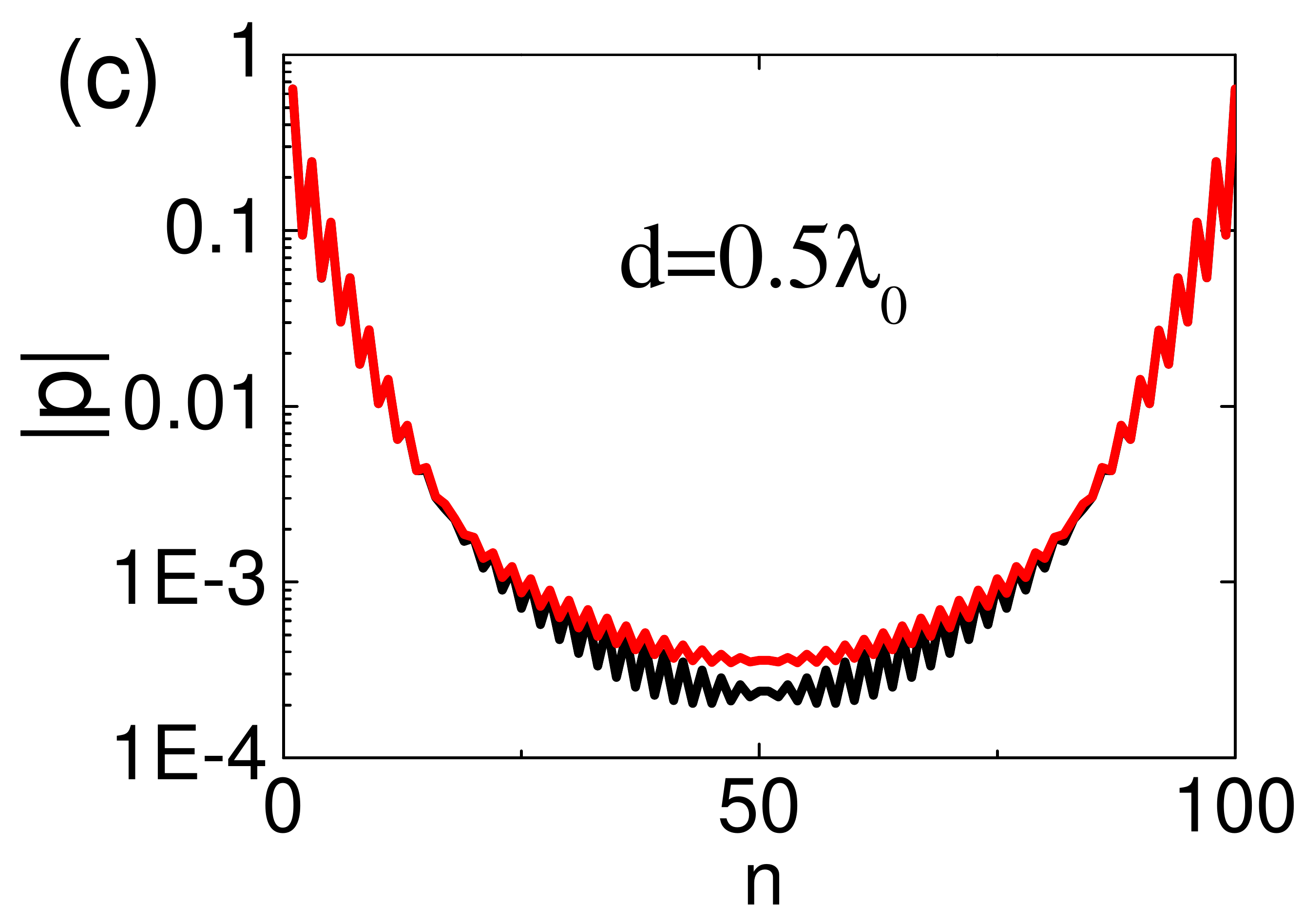}\label{midgapmode}
}
	\hspace{0.01in}
\subfloat{
	\includegraphics[width=0.46\linewidth]{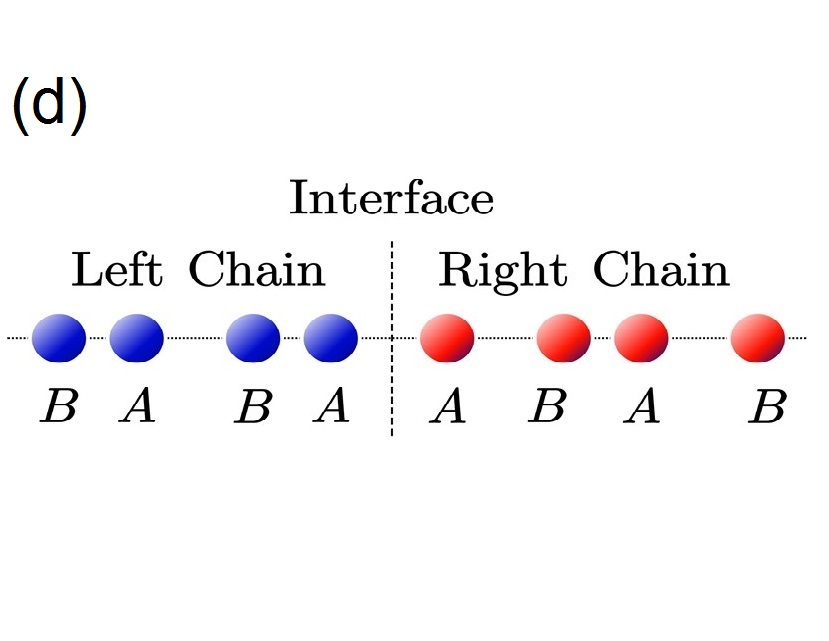}\label{interfaceschematic}
}
	\hspace{0.01in}
	\subfloat{
	\includegraphics[width=0.46\linewidth]{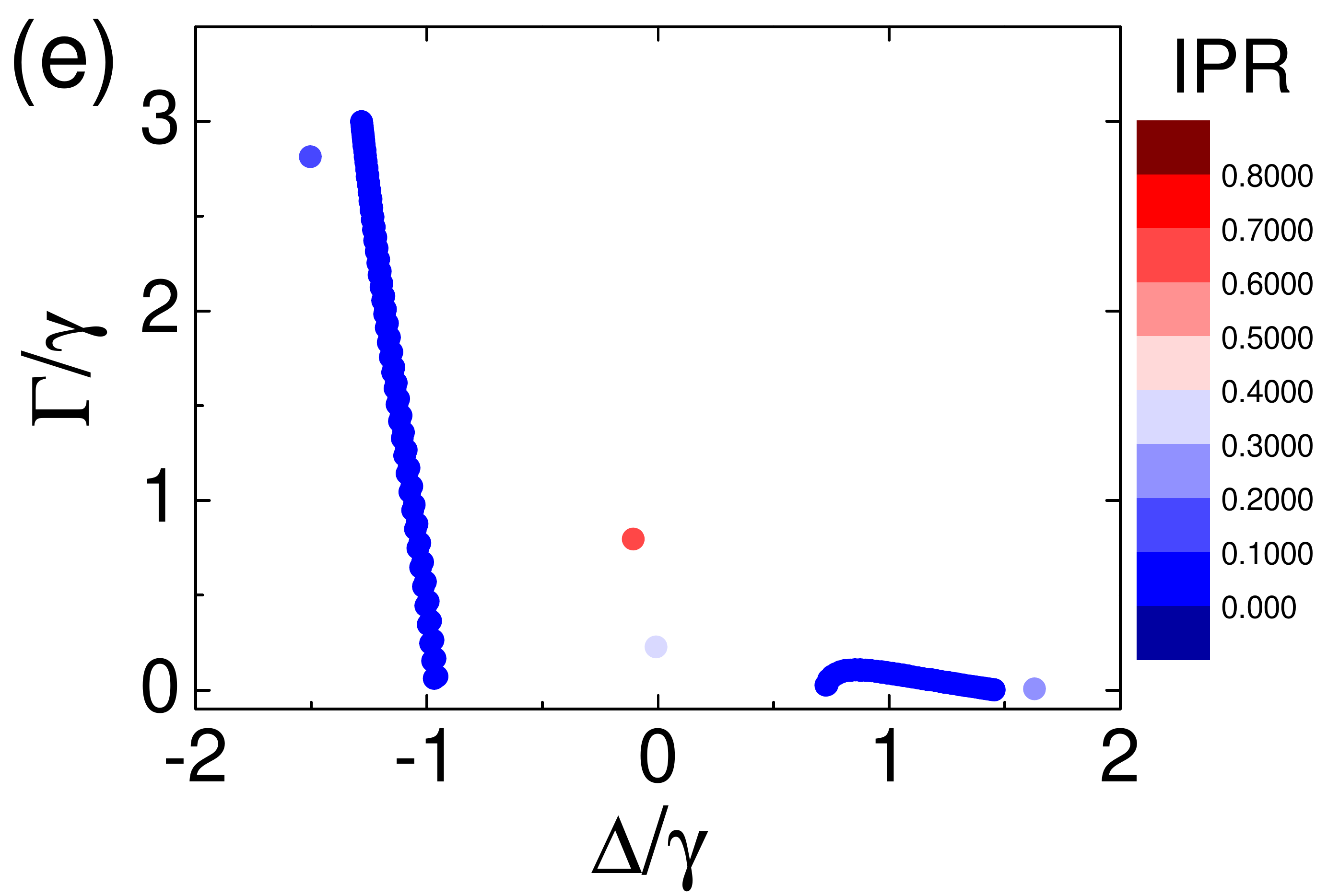}\label{interfacemodebeta06}
}
	\hspace{0.01in}
\subfloat{
	\includegraphics[width=0.46\linewidth]{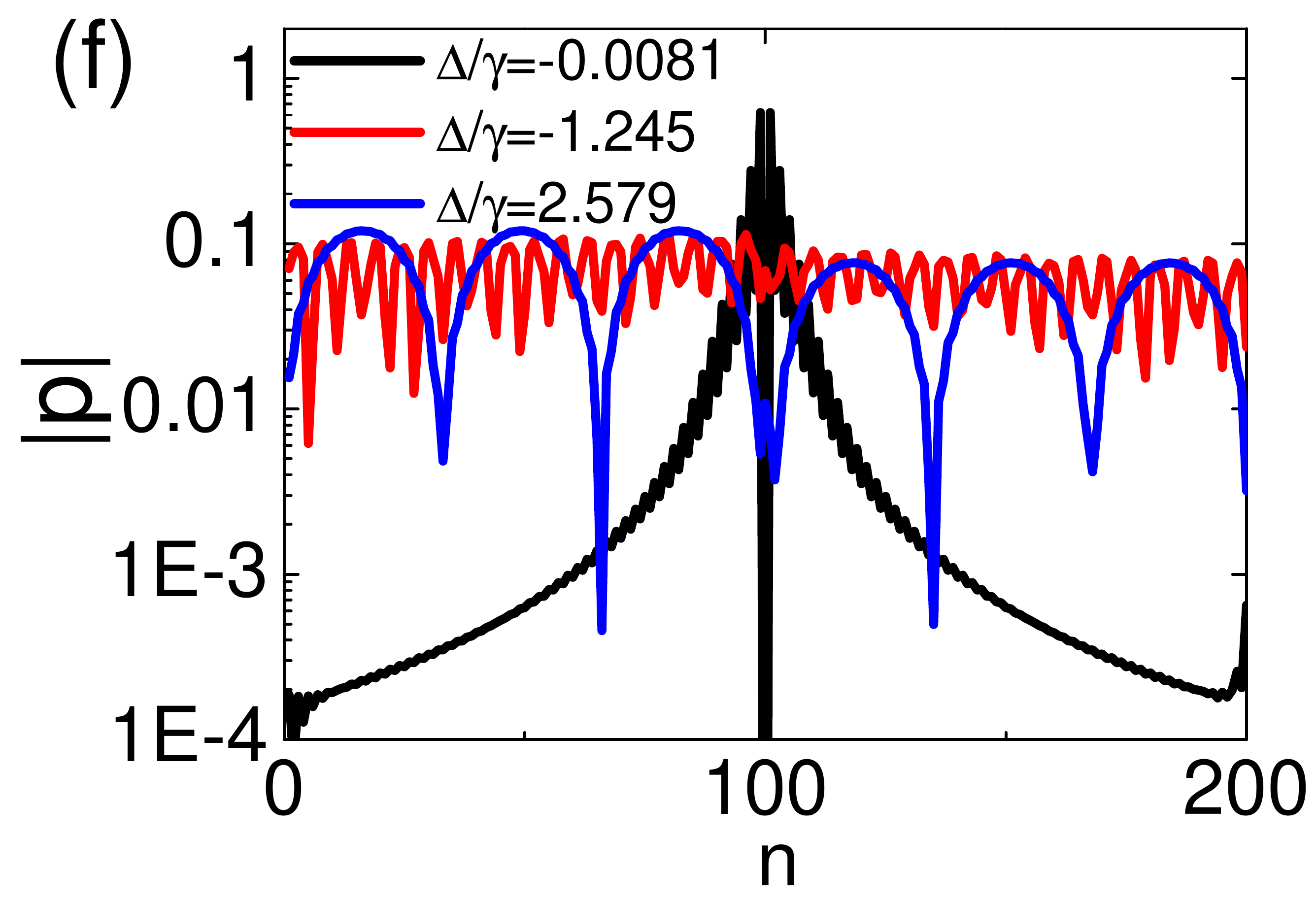}\label{interfacemodedipole}
}
	\caption{ (a) Bandstructure (eigenstate distribution) for a finite dimerized chain with $N=100$ atoms under $\beta=0.6$ and $d/\lambda_0=0.5$. There are two midgap states at ($\Delta/\gamma=-0.10548$, $\Gamma/\gamma=0.7960$), and ($\Delta/\gamma=-0.10558$, $\Gamma/\gamma=0.7963$). (b) Bandstructure (eigenstate distribution) of a topologically trivial dimerized chain with $N=100$ atoms under $\beta=0.4$ and $d/\lambda_0=0.5$. (c) Dipole moment distribution $|p_i|$ (logarithmic scale) of the two midgap edge states. (d) Schematic of the connected chain. (e) Eigenstate distribution for a connected chain.  (f) Dipole moment distribution of the interface state at ($\Delta/\gamma=-0.0081$, $\Gamma/\gamma=0.2258$) in (e) (logarithmic scale), compared with those of bulk states at ($\Delta/\gamma=-1.245$, $\Gamma/\gamma=2.579$) and ($\Delta/\gamma=1.453$, $\Gamma/\gamma=6.3740\times10^{-4}$).} 
	\label{edgemode}
\end{figure}

In Figs.\ref{beta06longband05} and \ref{beta04longband05}, we show the eigenstate distributions (bandstructures) for the cases of $\beta=0.6$ and $\beta=0.4$ with lattice period $d=0.5\lambda_0$, which are almost the same with those of infinitely long chains shown in Fig.\ref{figcda}. The lower-in-frequency band contains states with very high decay rates ($\Gamma/\gamma>1$) while the bulk states in the upper-in-frequency band are long-lived ($\Gamma/\gamma\ll1$). Therefore we can regard the lower band as the bright band while the upper band as the dark band. In the $\beta=0.6$ case, we find two midgap states at $\Delta/\gamma=-0.10548$ and $\Delta/\gamma=-0.10558$, which are found to be red-detuned from the single atom resonance. Note in the nearest-neighbor approximation, the two midgap states exactly have the same frequency at $\Delta=0$, which are so-called gapless zero-energy modes \cite{asboth2016short,zhangPRB2018}. The gapped edge states and the relevant frequency shift is the result of the shift of the bandgap, originating from the diagonal terms in the Hamiltonian. In Fig.\ref{beta06longband05} the midgap states both have a rather high IPR around 0.34. However, in Fig.\ref{beta04longband05} for $\beta=0.4$, there are not any midgap states. The dipole moment distributions of the two midgap edge states along the chain are shown in Fig.\ref{midgapmode} in the logarithmic scale, which are exponentially localized over the edge. It is thus concluded that the midgap states are indeed topologically protected edge states as guaranteed by the non-trivial complex Zak phase for $\beta=0.6$. Moreover, we also note that the midgap states are also subradiant with a decay rate $\Gamma/\gamma\sim0.796<1$, compared to the single atom resonance. For finite chains, the spectrum of $\beta=0.5$ is still ungapped (not shown here). This feature implies that $\beta=0.5$ is indeed the quantum phase transition point \cite{lieuPRB2018}.

Another feature of topological systems is that, for two systems with different topological invariants, there exist topologically protected interface states at the boundary between them \cite{ozawa2018topological}. In Fig.\ref{interfacemodebeta06} we calculate the eigenstate distribution for a connected chain, which comprises a topologically trivial chain with $\beta=0.4$ in the left and a topologically nontrivial right chain with $\beta=0.6$ in the right (schematically shown in Fig.\ref{interfaceschematic}). Here the inter-chain distance, i.e., the distance between the rightmost atom in the left chain and the leftmost atom in the right chain is set to be $d_2$. It is observed that a long-lived interface state emerges in the bandgap at $\Delta/\gamma=-0.0081$ with a decay rate of $\Gamma/\gamma=0.2258$. Another midgap state is the edge state at the right boundary of the right chain, with a frequency at $\Delta/\gamma=-0.10553$ and decay rate of $\Gamma/\gamma=0.79615$. Note the IPR of the right-edge state ($\mathrm{IPR}=0.6876$) is even higher than that of the interface state ($\mathrm{IPR}=0.3153$). In Fig.\ref{interfacemodedipole} we show the dipole moment distribution for the interface state, compared with those of two bulk states in the upper band ($\Delta/\gamma=1.453$) and lower band ($\Delta/\gamma=-1.245$) respectively. The bulk states periodically distributed in the left and right chains while the interface state is exponentially localized. Surprisingly, it is further seen that one midgap state further emerges beyond the upper band ($\Delta/\gamma=1.6303$, $\Gamma/\gamma=0.0059$) and another situates below the lower band ($\Delta/\gamma=-1.5034$, $\Gamma/\gamma=2.8137$). Actually, these two states are subradiant and superradiant edge states formed by the pair of atoms near the interface \cite{Schilder2016,guerin2017light}, namely, the rightmost atom in the left chain and the leftmost atom in the right chain. Such states arises from the strong near-field dipole-dipole interactions between the atom pair \cite{Schilder2016,guerin2017light}, which can be observed by considering the eigen-frequencies of the interaction matrix of a pair of atoms (not shown here) \cite{Skipetrov2014}. It is also found that the spectral position of these two states strongly depend on the inter-chain distance and when the inter-chain distance becomes larger than $d_1$ where near-field interactions are not strong enough, they disappear and become mixed in the bulk states. Therefore, these two states are not topologically protected states at all.

\begin{figure}[htbp]
	\centering
	\flushleft
	\subfloat{
		\includegraphics[width=0.46\linewidth]{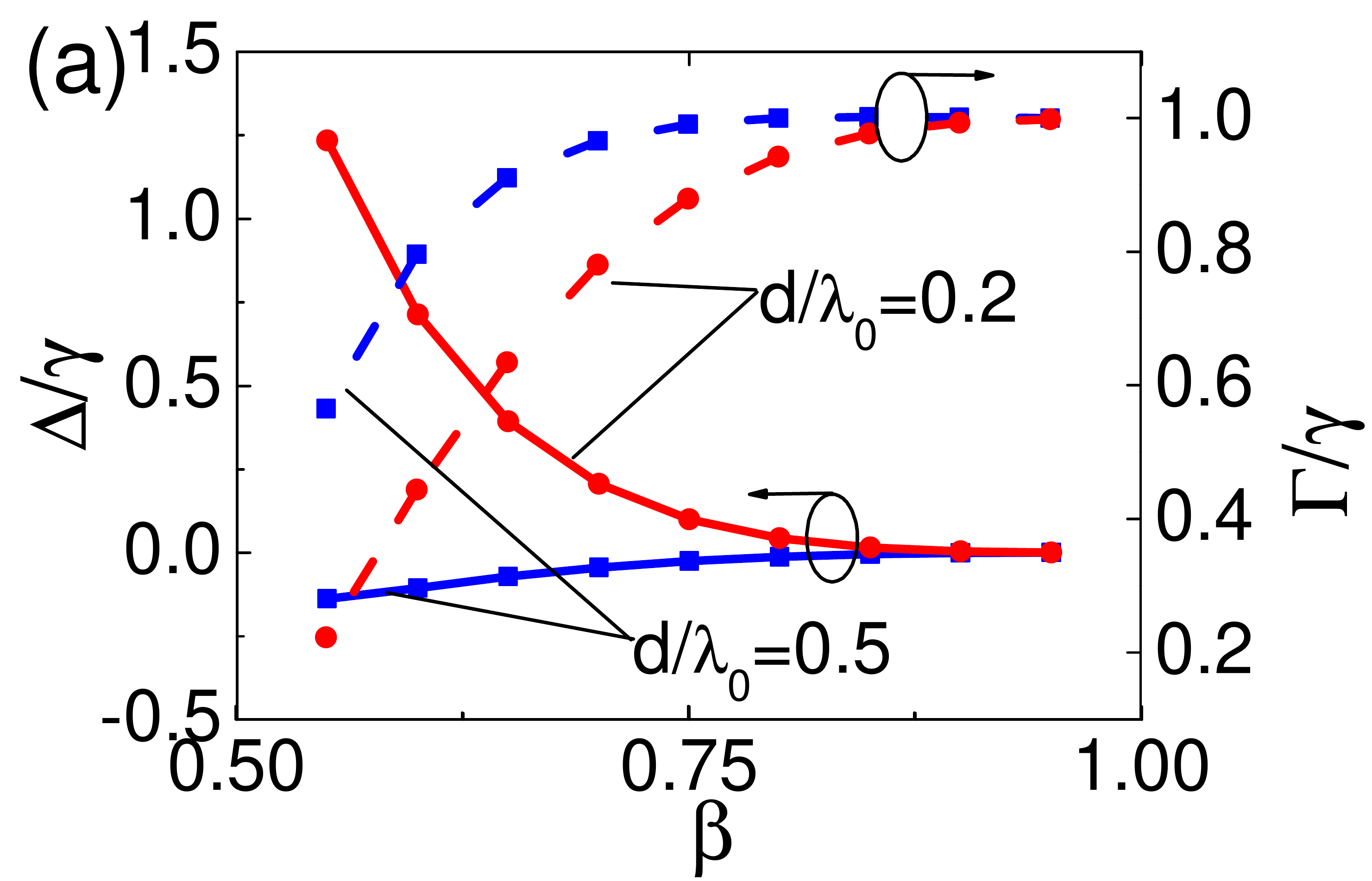}\label{deltagamma}
	}
	\hspace{0.01in}
	\subfloat{
		\includegraphics[width=0.41\linewidth]{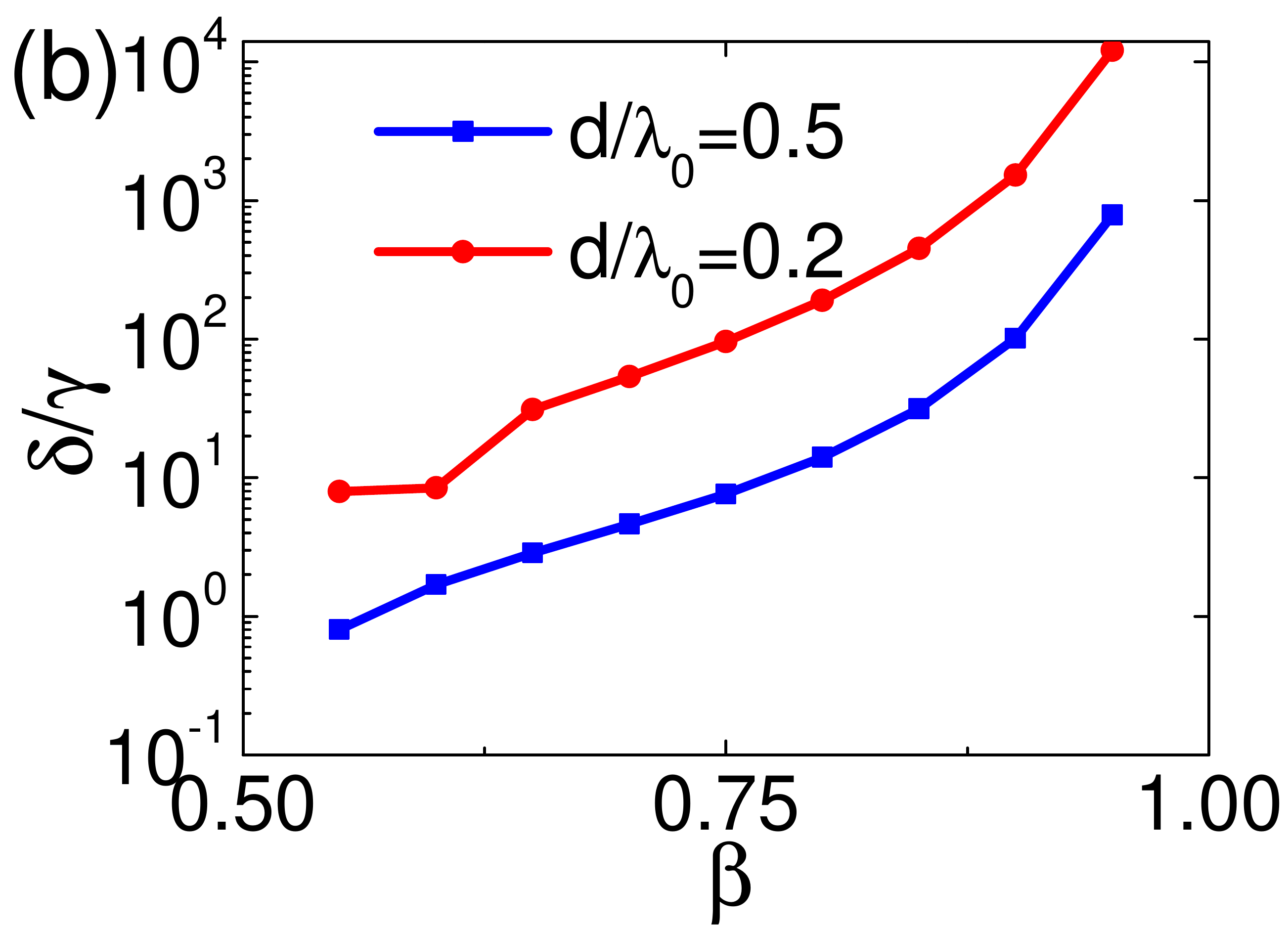}\label{gapsize}
	}
	\caption{(a) Edge state frequency (detuning) $\Delta/\gamma$ and decay rate $\Gamma/\gamma$ of topological nontrivial dimerized chain as a function of dimerization parameter $\beta$ under $d/\lambda_0=0.5$ and $d/\lambda_0=0.2$. (b) Band gap size $\delta/\gamma$ as a function of dimerization parameter $\beta$ under $d/\lambda_0=0.5$ and $d/\lambda_0=0.2$. }
	
	\label{figparameter}
\end{figure}

In order to reveal how the midgap edge states of a dimerized chain evolve with the dimerization parameter, in Fig.\ref{deltagamma} we show the detuning and decay rate as a function of $\beta$ for $d=0.5\lambda_0$ and $d=0.2\lambda_0$ respectively. It is observed that in both cases when the dimerization parameter increases from 0.5 to 1, the detunings of both edge states approach zero and their decay rates reach unity, leading to a behavior the same as the single-atom resonance. In fact, in dimerized plasmonic nanoparticle chains, it was usually found that the edge state frequency is the same as single-particle resonance frequency \cite{lingOE2015,downingPRB2017,pocockArxiv2017}. Thanks to the ultra-narrow linewidth of cold atoms, here we are able to observe the frequency shift of edge states from the single-atom resonance frequency, which is indeed substantial when $d$ is small and $\beta$ approaches 0.5. It is also noted that for $d=0.5\lambda_0$, the edge states are red-detuned from the single atom resonance frequency, while for $d=0.2\lambda_0$, they are in contrast blue-detuned.  This difference is definitely the result of interplay between the near-field and far-field dipole-dipole interactions. This phenomenon is consistent with some most recent researches on 1D plasmonic systems \cite{zhangPRB2018,downing2018topological}, where a blue-shift of the bandgap was found in Ref.\cite{zhangPRB2018} while a red-shift was found in Ref.\cite{downing2018topological}.  In Fig.\ref{gapsize}, we also plot the real bandgap width $\delta$, which drastically expands when the dimerization increases, resulting in a $10000\gamma$-wide gap for $\beta=0.95$ under $d=0.2\lambda_0$. Such a remarkable feature provides a very flexible platform for (real) bandgap engineering in nanophotonic systems, which is one of the main advantages of cold atomic systems.

%After calculating the EM responses of all scatterers based on above multiple scattering equations, we find the total scattering field of the random cluster of particles at an arbitrary position $\mathbf{r}\neq\mathbf{r}_j$ where $\mathbf{r}_j$ denotes the positions of scatterers, is computed as
%\begin{equation}\label{Es_eq}
%\mathbf{E}_s(\mathbf{r})=-k^2\sum_{i=1}^{N}\mathbf{G}_0(\mathbf{r},\mathbf{r}_i)\mathbf{d}_i
%\end{equation}
%Here $\mathbf{r}$ can be an arbitrary position in the space including the source position $\mathbf{r}_s$, except for the positions inside the scatterers. The subscript $c$ actually denotes a configurational average procedure.  Here the elements in scattering field tensor $\mathbf{S}(\mathbf{r},\mathbf{r}_s)$ can be calculated through the relation $\mathbf{E}_s(\mathbf{r})=\mathbf{S}(\mathbf{r},\mathbf{r}_s)\mathbf{p}_s$, by aligning the dipole moment with different axes.
 
\section{Discussion}
By this stage, we have already shown the non-trivial topological properties of the dimerized cold atomic chain, although it is described by a non-Hermitian Hamiltonian. In this section, we go further beyond the conclusions drawn above. First, we demonstrate the effect of disorder, which does not respect the chiral symmetry, to elucidate that such edges modes are indeed topologically protected. Then, we examine whether the topological states still persist when the lattice constant is increased to make far-field, long-range dipole-dipole interactions dominate. We further discuss the property of transverse eigenstates, and find that the existence of strong far-field dipole-dipole interactions (decaying slowly as $1/r$ where $r$ is the distance) in the transverse eigenstates still preserves the topological properties described by the quantized complex Zak phase, which is non-trivial for $\beta>0.5$. Finally, we propose possible experimental realizations for the dimerized atomic chain. 
\subsection{Effect of disorder}
\begin{figure}[htbp]
\centering
\flushleft
\subfloat{
		\includegraphics[width=0.46\linewidth]{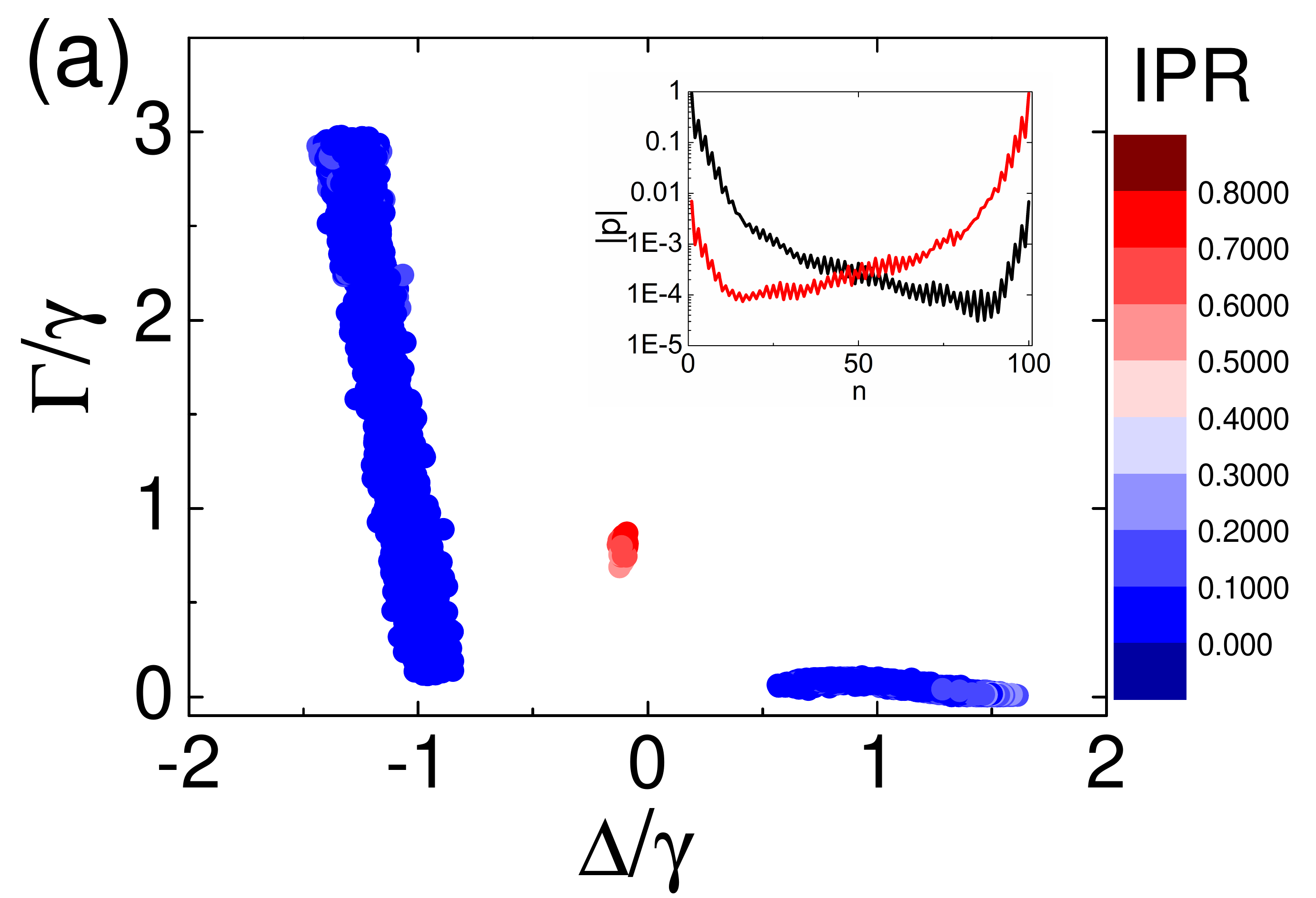}\label{edgemodebeta06disorder01}
}
\hspace{0.01in}
\subfloat{
		\includegraphics[width=0.46\linewidth]{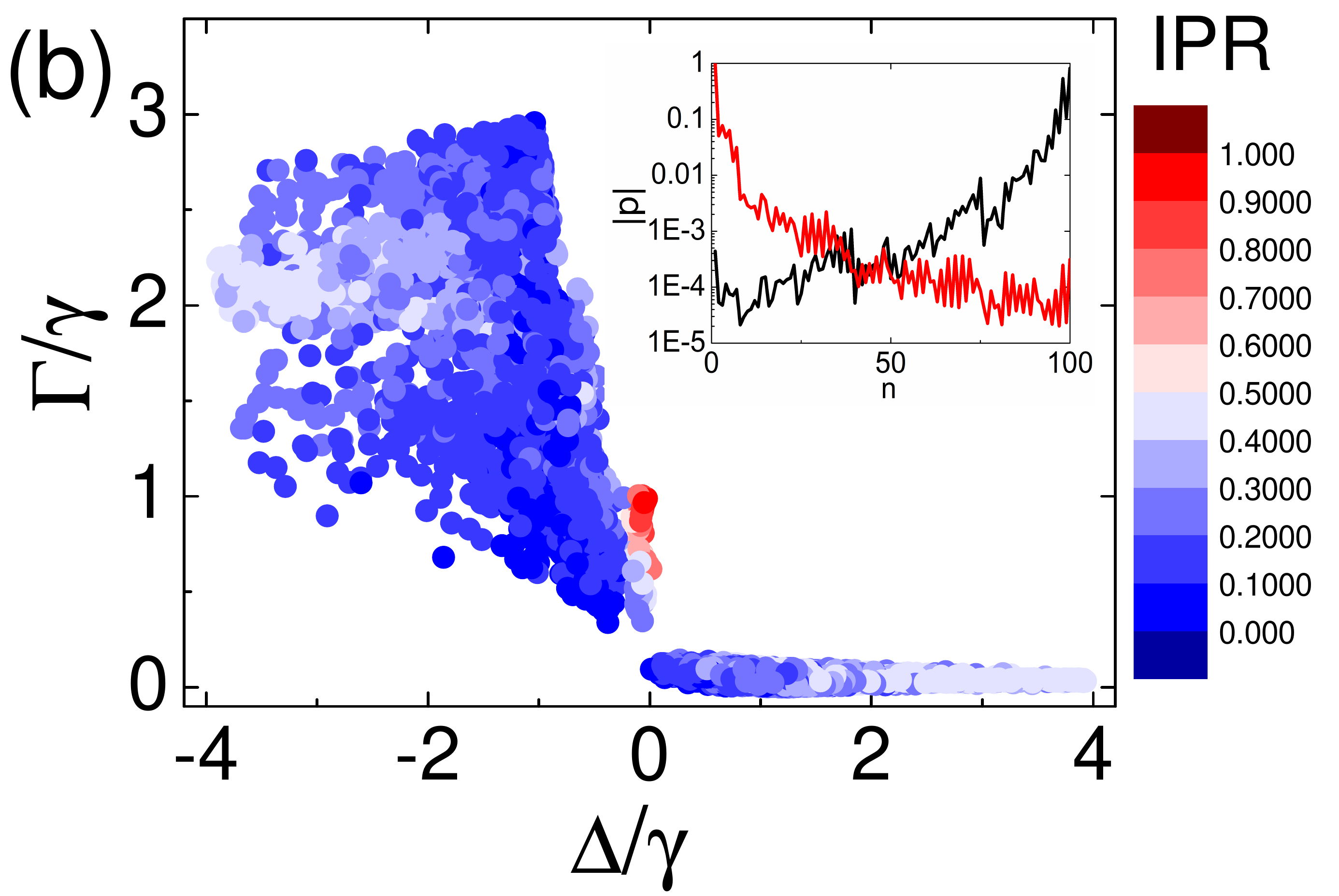}\label{edgemodebeta06disorder05}
}
\hspace{0.01in}
\subfloat{
	\includegraphics[width=0.46\linewidth]{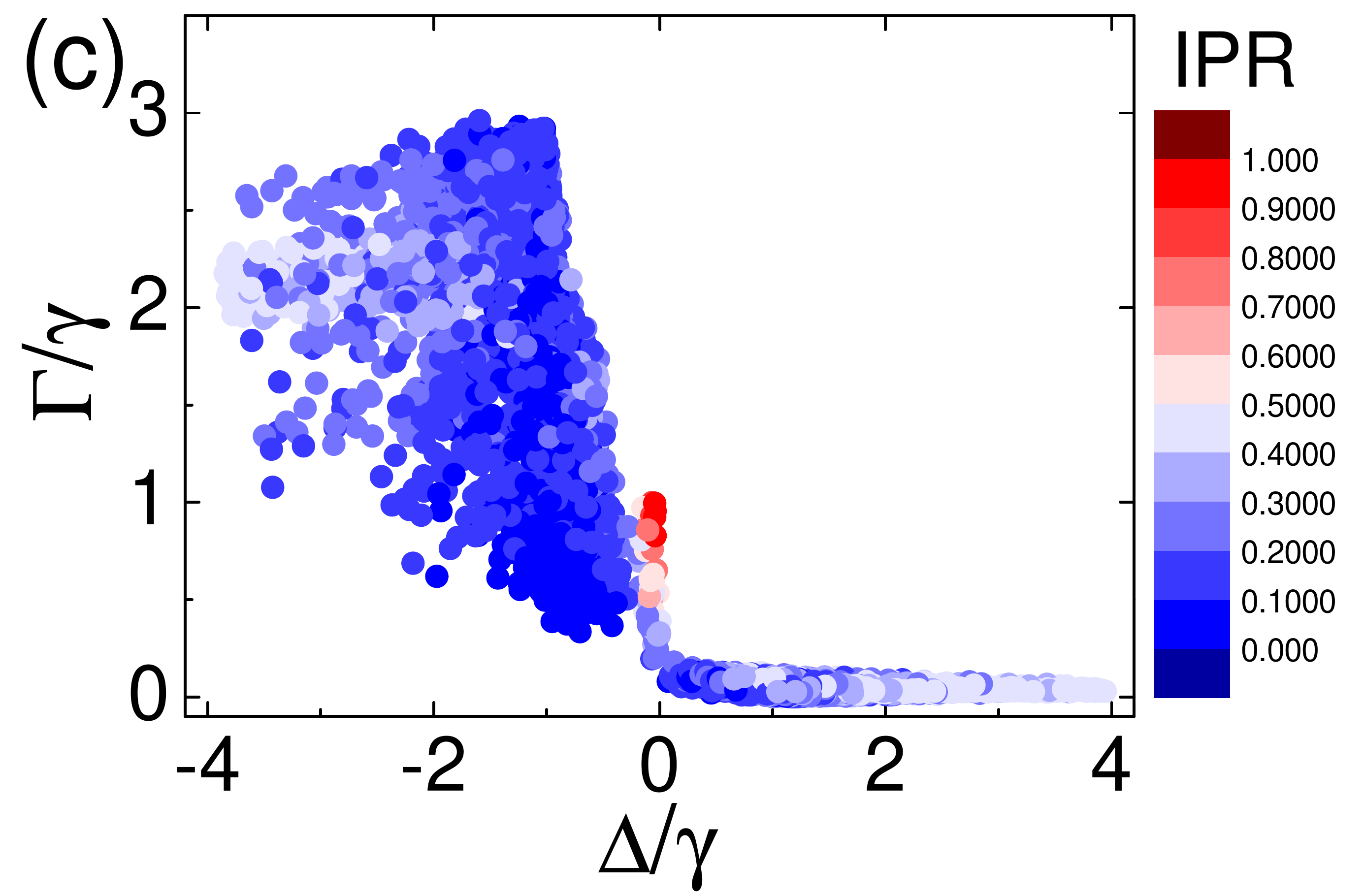}\label{interfacemodebeta06disorder05}
}
\subfloat{
	\includegraphics[width=0.46\linewidth]{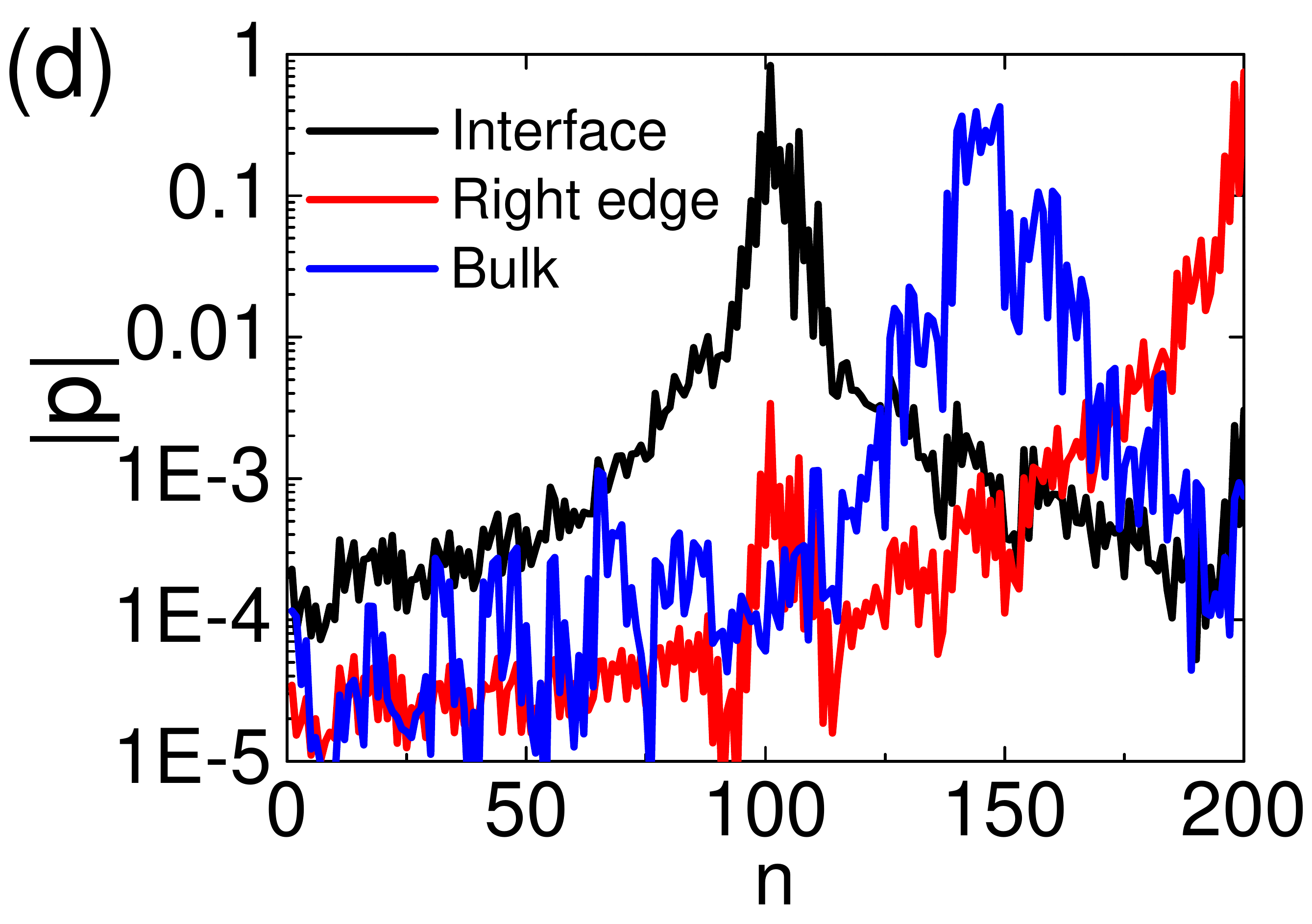}\label{interfacemodedisorder05dipolemoment}
}

\caption{ (a) Eigenstate distribution for weakly disordered ($\varepsilon=0.1$) dimerized chains with $N=100$ atoms under $\beta=0.6$ and $d=0.5\lambda_0$. Results of 100 random realizations are shown. Inset:  Dipole moment distributions (logarithmic scale) of the two edge states in a specific random configuration. (b) Eigenstate distribution for highly disordered ($\varepsilon=0.5$) dimerized chains with $N=100$ atoms under $\beta=0.6$ and $d=0.5\lambda_0$. Results of 100 random realizations are shown. Inset: Dipole moment distributions (logarithmic scale) of the two edge states in a specific random configuration. (c) Eigenstate distribution for connected dimerized chains comprising two disordered dimerized chains. Results of 100 random realizations are shown. (d) Dipole moment distributions of the interface state, as well as an edge state at the right boundary and an Anderson localized state.} 
	\label{disorder}
\end{figure}

To account for the circumstances in practice when the atoms are not perfectly trapped and fluctuate around the lattice sites, here we focus on the $\beta=0.6$, $d=0.5\lambda_0$ dimerized chain and discuss the effect of disorder on longitudinal eigenstates. Here disorder is introduced by shifting the positions of $A$ atoms randomly in the range $[-\varepsilon d_1/2,\varepsilon d_1/2]$ along the $x$-axis, while the positions of $B$ atoms are fixed. Note this kind of disorder can destroy the trivial chiral-symmetry-breaking condition, since it leads to a difference in the hoppings from $A$ to $A$ and $B$ to $B$, and then a nontrivial symmetry-breaking. If the midgap states in our system are robust to this kind of symmetry-breaking disorder, we expect them to be topologically stable \cite{stjeanNaturephoton2017}. In Figs.\ref{edgemodebeta06disorder01} and \ref{edgemodebeta06disorder05} we give the eigenstate distributions for $\varepsilon=0.1$ and $\varepsilon=0.5$ respectively, which are both obtained after 100 random realizations. It is found that disorder can broaden the bandstructure, leading to far-detuned states, especially for $\varepsilon=0.5$. This is because under strong disorder, many atoms get very close to each other, leading to significant near-field dipole-dipole interactions and thus large frequency shifts \cite{wangOL2018}. In both weak-disorder and strong-disorder cases, topologically protected edge states with high IPRs still persist. Moreover, their IPRs mostly distribute in the range of $[0.4,0.85]$, which are even larger than that of the ordered dimerized chain ($\mathrm{IPR}=0.34$), indicating a stronger spatial confinement. In the inset of Fig.\ref{edgemodebeta06disorder01}, the dipole moment distributions of the two edge states for a specific random realization are given for $\varepsilon=0.1$. It is noted that different from the ordered case, these states in a disorder dimerized chain not only split in frequency domain substantially (not shown here), but also separate with each other in their dipole moment spatial distributions. The two nearly identical two-sided edge states in the ordered case evolve into two different one-sided edge states in the disordered case. This is the direct consequence of the nontrivial chiral symmetry-breaking disorder. The same phenomenon is also observed for $\varepsilon=0.5$. When $\varepsilon>0.6$, we find that the edge states start to mix with bulk modes and the bandgap gets closed in both real and imaginary frequencies. However, in such a high disorder, it is still possible to observe highly localized edge states. 

In Fig.\ref{interfacemodebeta06disorder05} we calculate the eigenstate distribution of a connected chain which comprises a topologically nontrivial chain with $\beta=0.6$ and disorder $\varepsilon=0.5$ in the right, and a topologically trivial chain with $\beta=0.4$ and disorder $\varepsilon=0.5$ in the left, which is also obtained after 100 random realizations. In this circumstance, the topologically protected interface states with high IPRs still persist in the gap near $\Delta/\gamma\sim0$. In Fig.\ref{interfacemodedisorder05dipolemoment} the dipole moment distribution for a typical interface state is given. Remarkably, we find the dipole moment of this interface state is enhanced compared to that of the interface state in the ordered case. We may attribute this enhancement to the Anderson localization mechanism in strongly disordered systems \cite{Skipetrov2014}. In Fig.\ref{interfacemodedisorder05dipolemoment} we also show the edge state in the right chain, which also resides in the gap, and a typical Anderson localized bulk state, for a single random realization. It is found that the topologically protected edge and interface states are more spatially localized than the Anderson localized state, and result in stronger field enhancements.

\subsection{Effect of the increase of lattice constant}
\begin{figure}[htbp]
	\centering
	\subfloat{
	\includegraphics[width=0.46\linewidth]{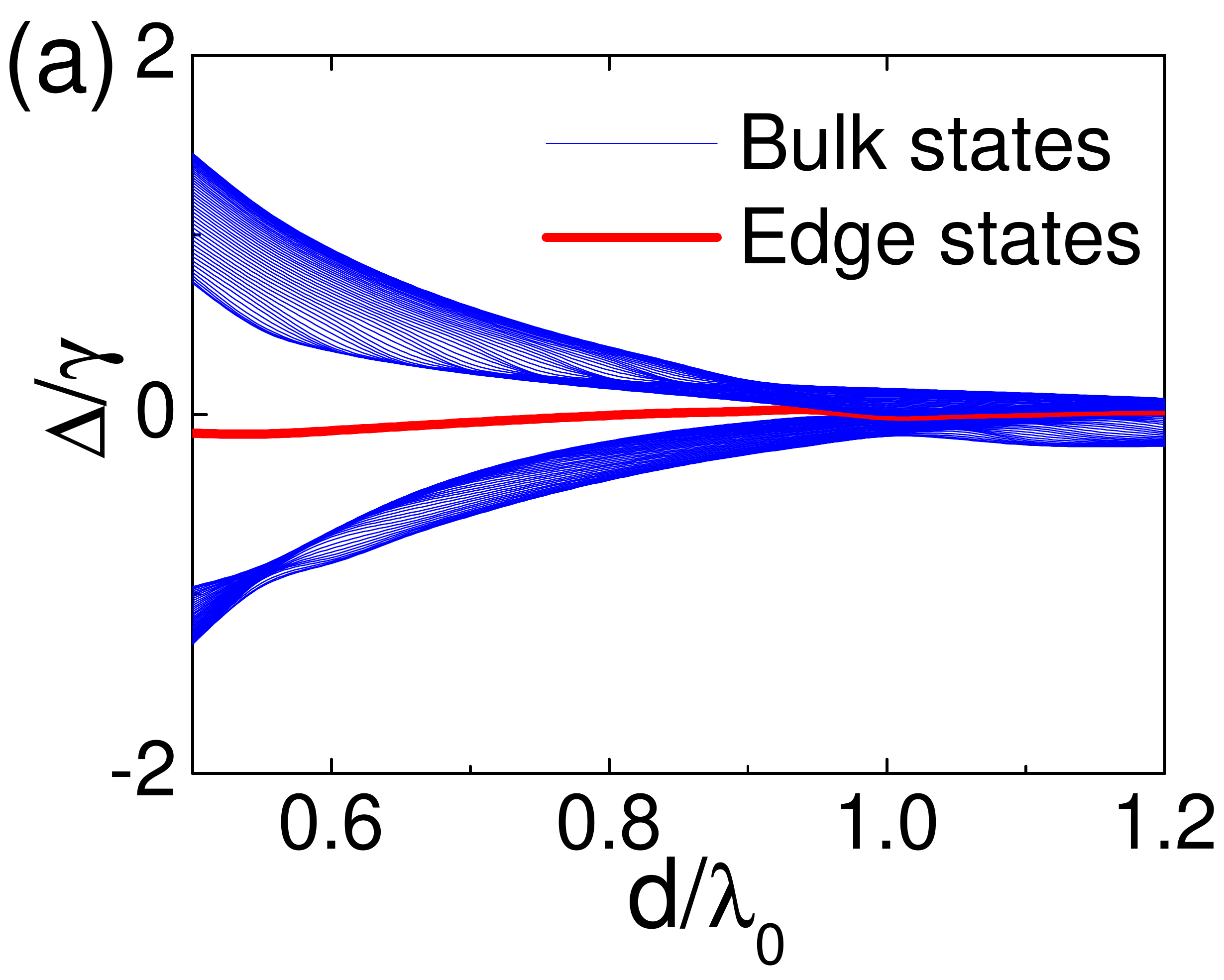}\label{longperioddependence06}
}
	\hspace{0.01in}
	\subfloat{
	\includegraphics[width=0.46\linewidth]{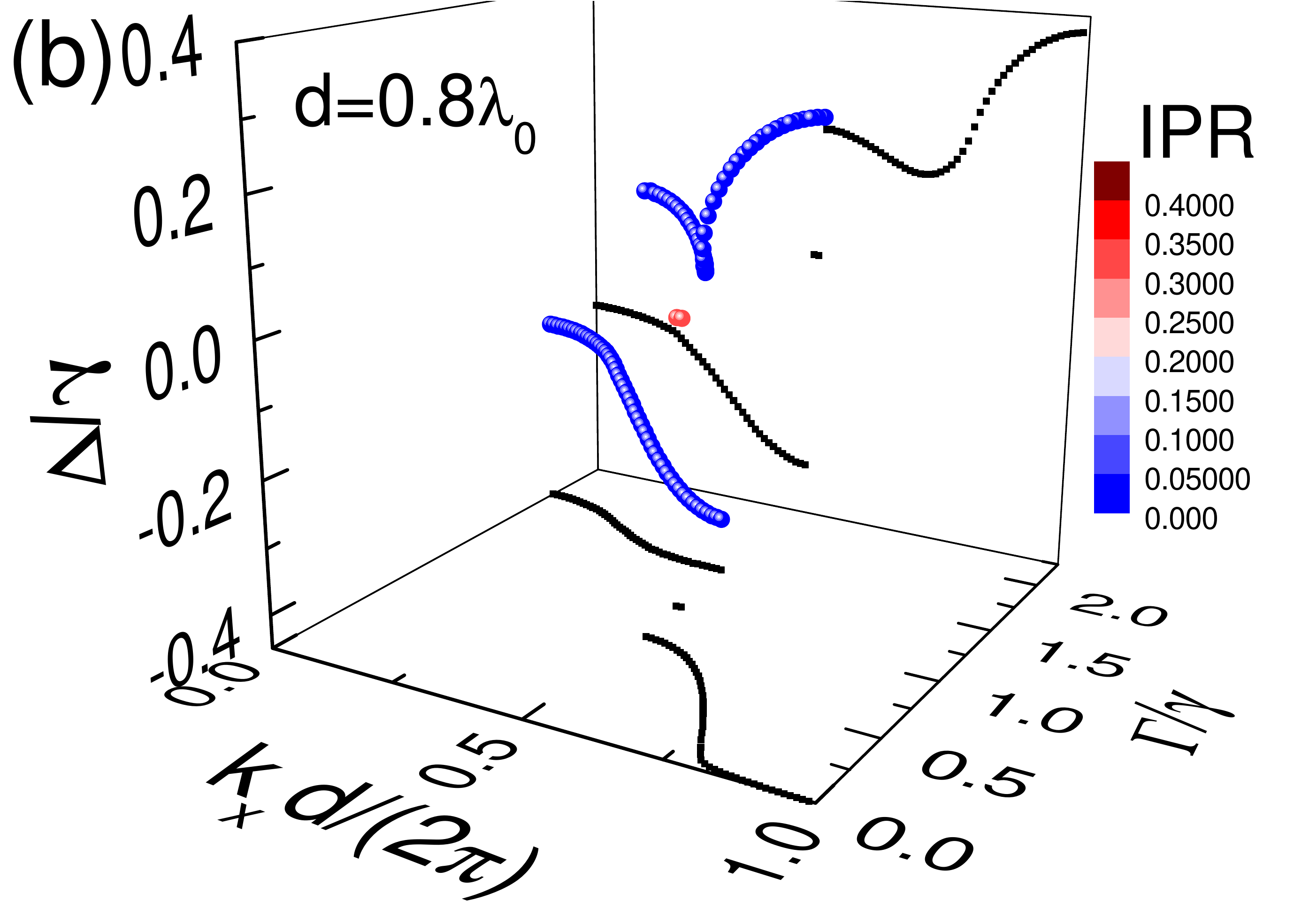}\label{beta06longband08}
}
	\hspace{0.01in}
\subfloat{
	\includegraphics[width=0.46\linewidth]{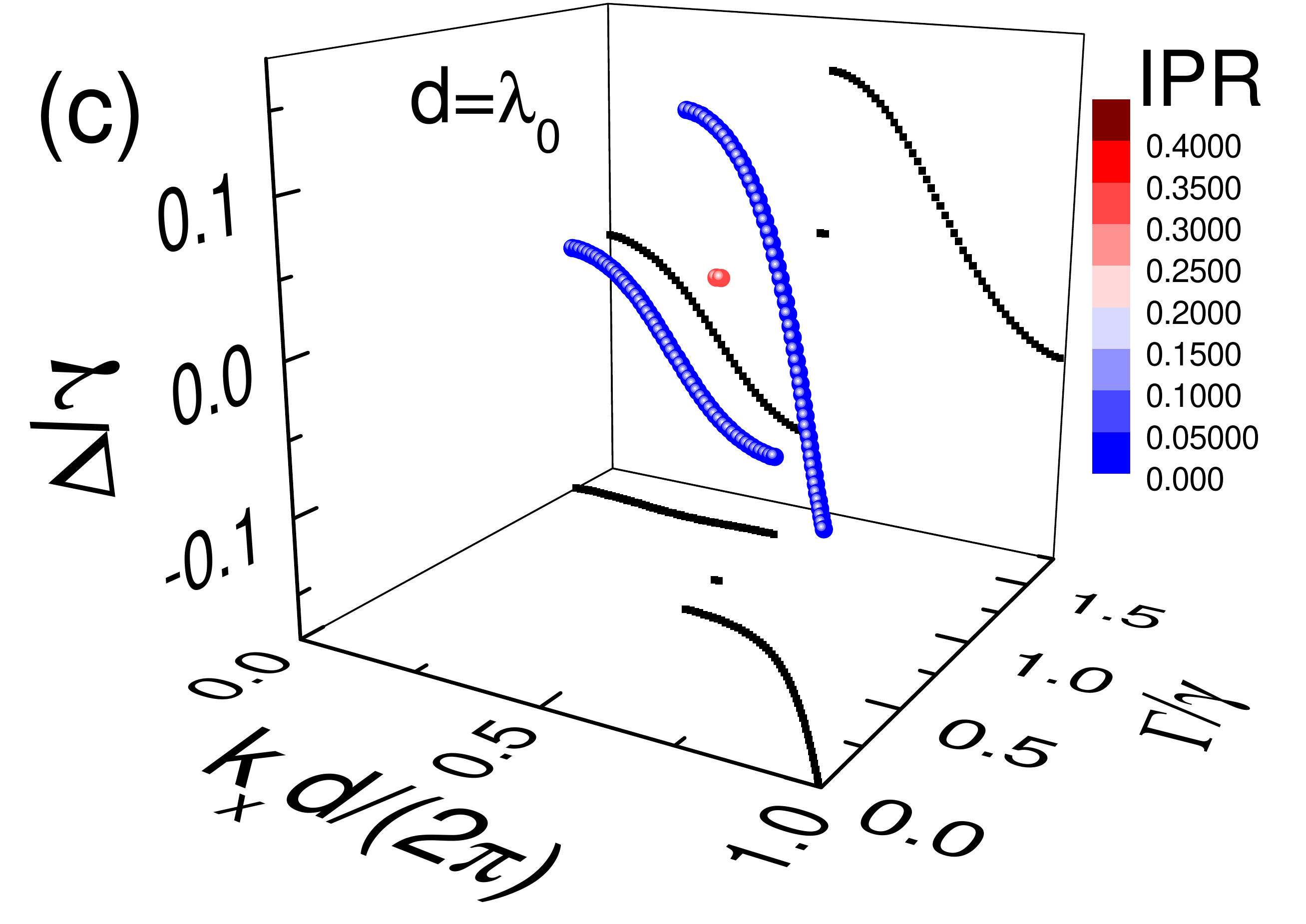}\label{beta06longband10}
}
	\hspace{0.01in}
\subfloat{
	\includegraphics[width=0.46\linewidth]{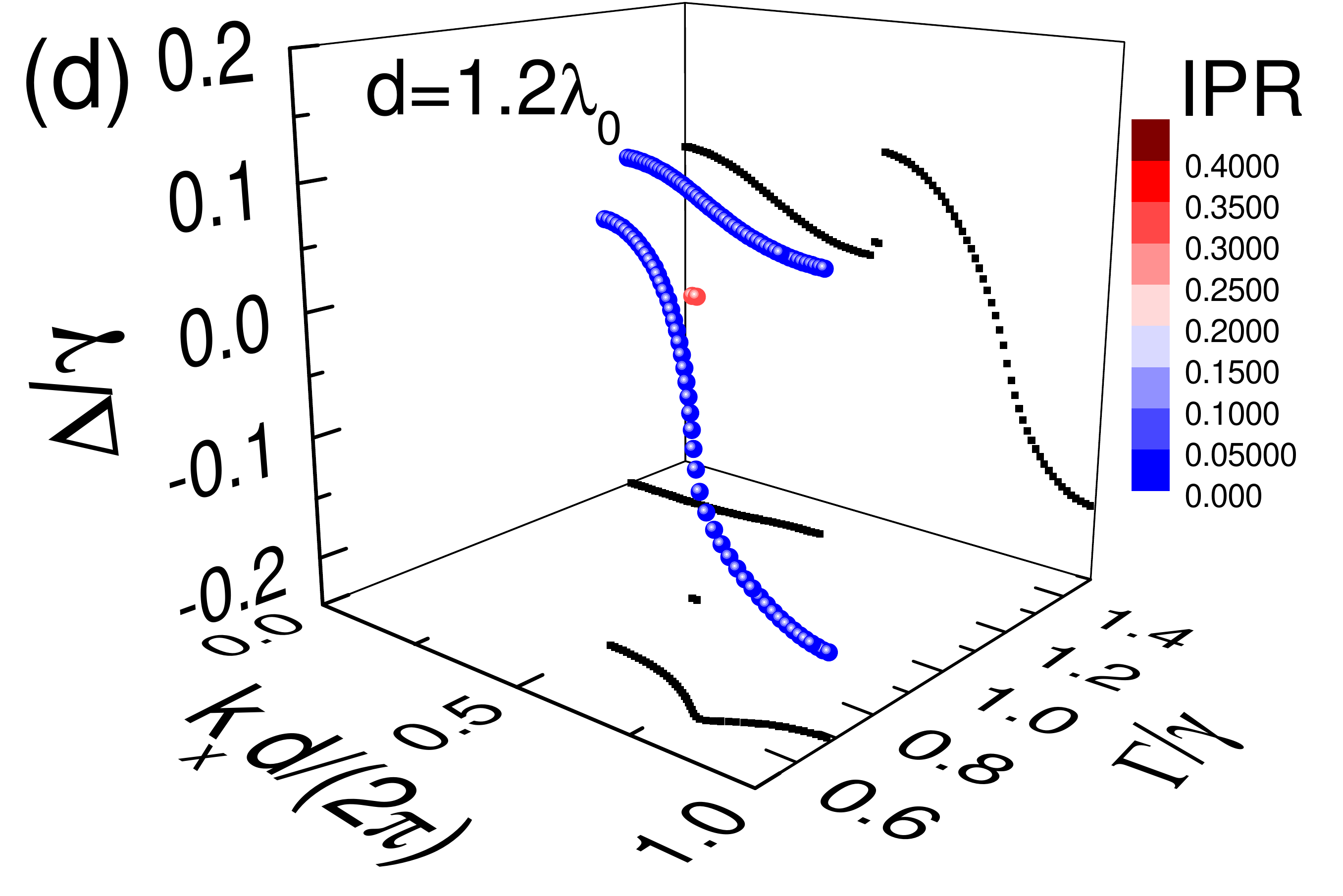}\label{beta06longband12}
}		
	\caption{ (a) Variation of eigenstate spectra of open systems with different lattice constants at $\beta=0.6$. Bandstructures calculated from finite chains with different lattice constants of (b) $d=0.8\lambda_0$, (c) $d=\lambda_0$ and (d) $d=1.2\lambda_0$.}\label{longband}
\end{figure}

In this subsection, we briefly examine whether the topological properties still persist when the lattice constant is increased to make far-field, long-range dipole-dipole interactions dominate, which give rise to a totally different picture from the conventional SSH model. In Fig.\ref{longperioddependence06} we calculate the real part of the spectrum of longitudinal eigenstates as a function of lattice constant $d$, where a finite chain with $N=100$ atoms is chosen and the dimerization parameter is $\beta$=0.6. The midgap edge states emerge, which are denoted by a red line indicating the high IPRs ($>0.3$) of these states. However, it is found that in the real spectrum, the bandgap actually closes when $d/\lambda_0\gtrsim1$, and the upper band overlaps with the lower band, indicating that the system might become ``metallic".  Since there is no topological phase transition, which requires a closure and re-opening of the bandgap \cite{krausPRL2012}, in a metallic system, it seems the emergence of these midgap edge states is, to some extent, counter-intuitive. 

\begin{figure}[htbp]
	\centering
	\subfloat{
		\includegraphics[width=0.3\linewidth]{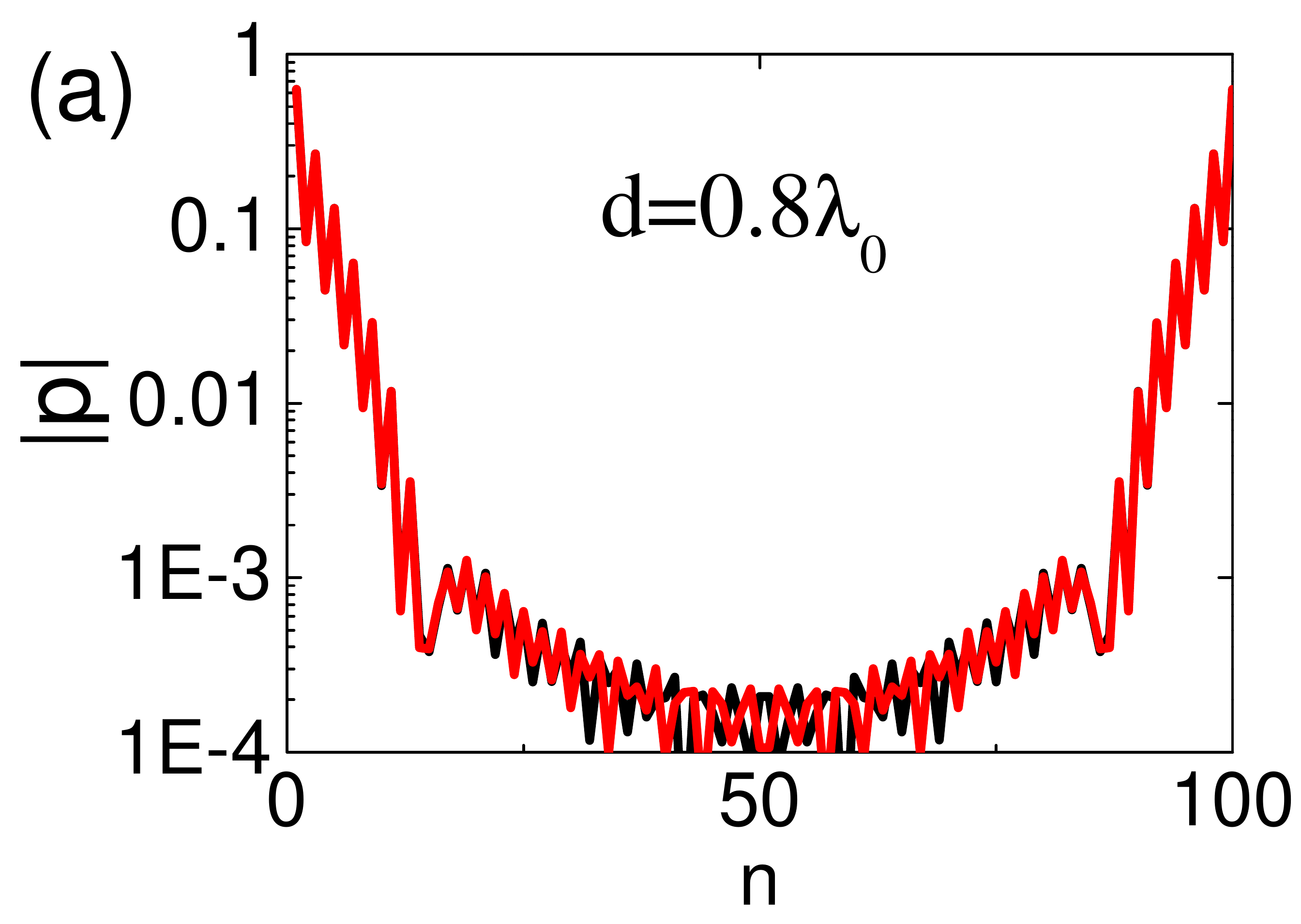}\label{longedgemode08}
	}
	\subfloat{
		\includegraphics[width=0.3\linewidth]{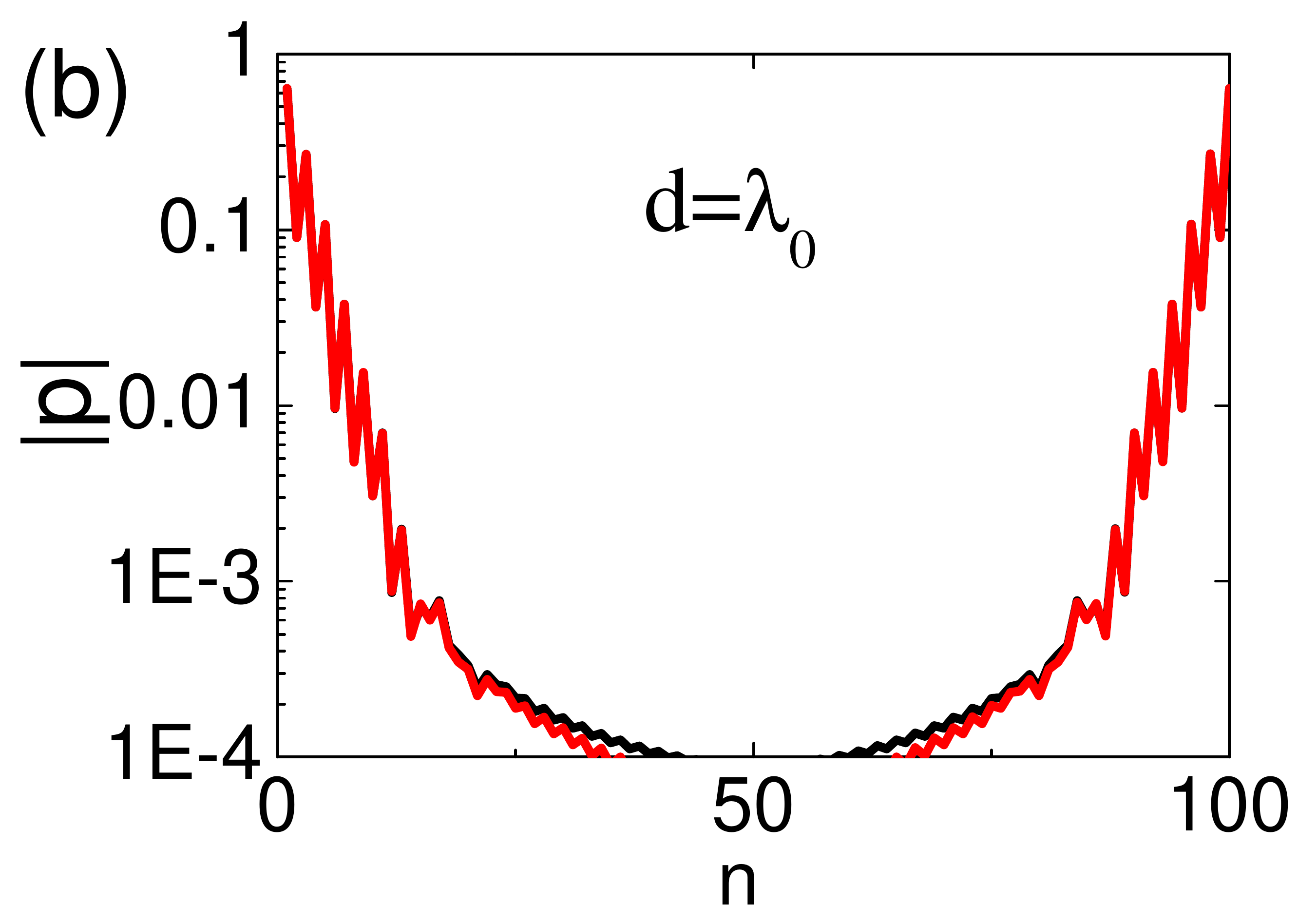}\label{longedgemode10}
	}
	\subfloat{
		\includegraphics[width=0.3\linewidth]{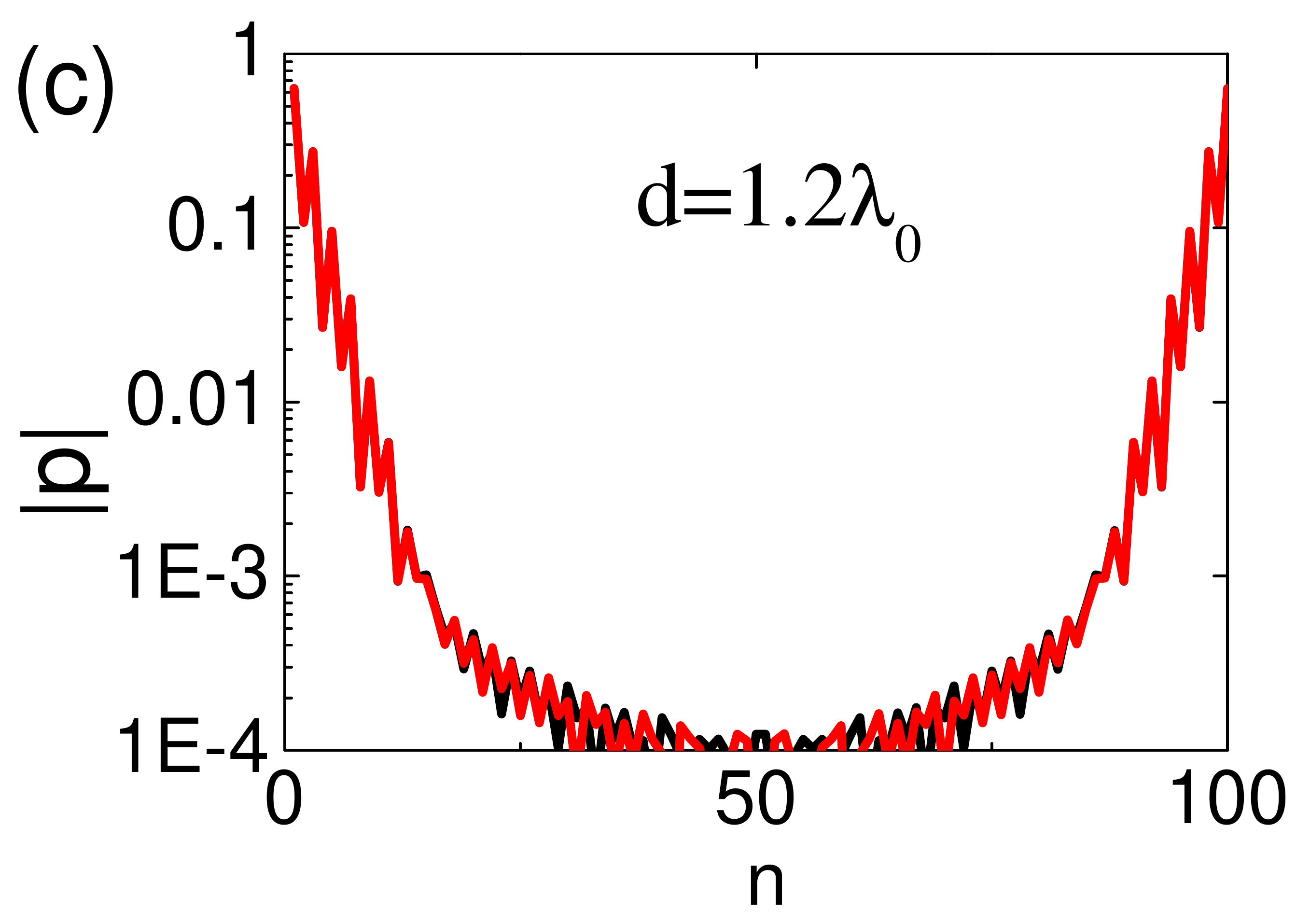}\label{longedgemode12}
	}
	\hspace{0.01in}
	\centering
\subfloat{
	\includegraphics[width=0.3\linewidth]{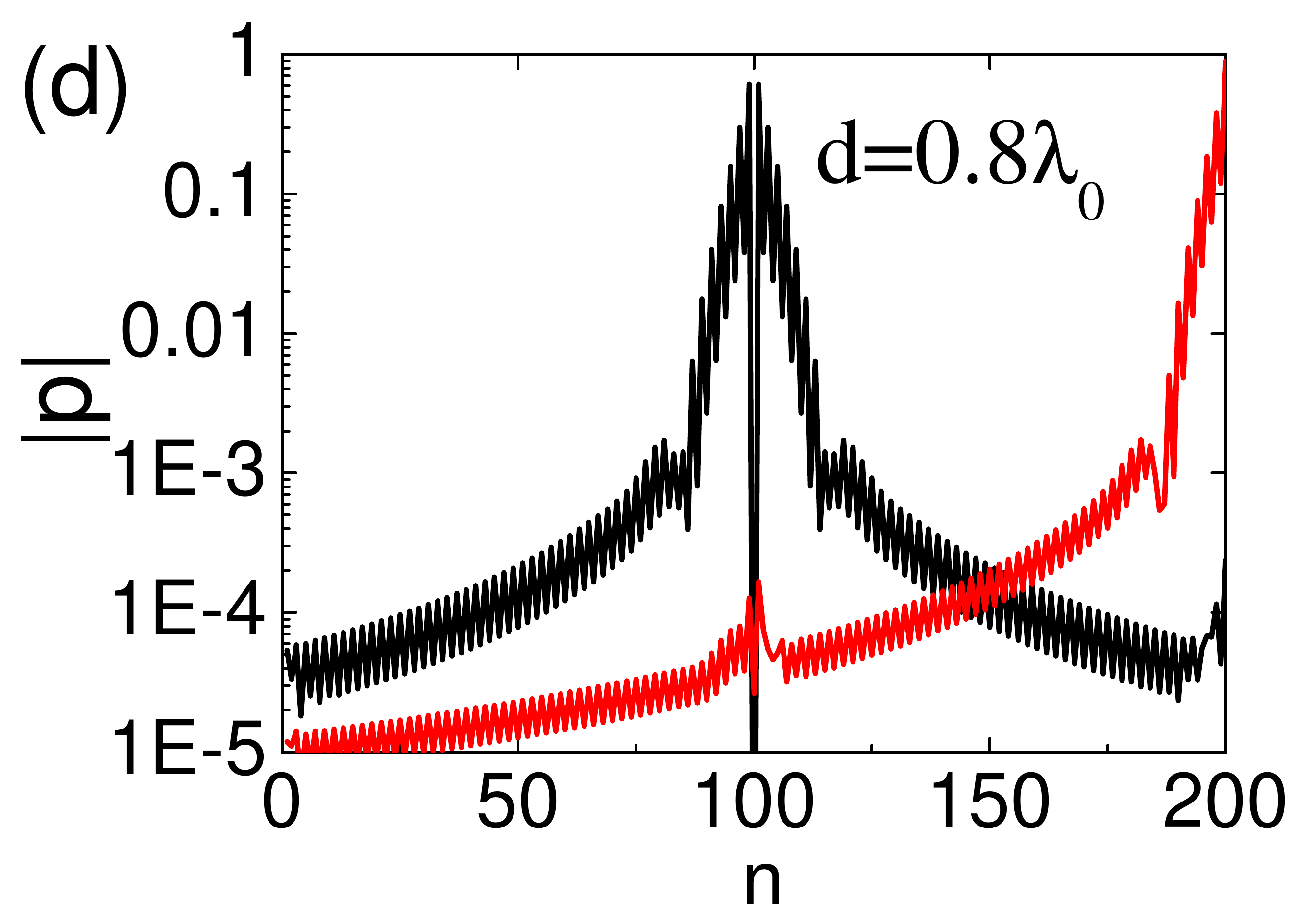}\label{longinterfacemode08}
}
\subfloat{
	\includegraphics[width=0.3\linewidth]{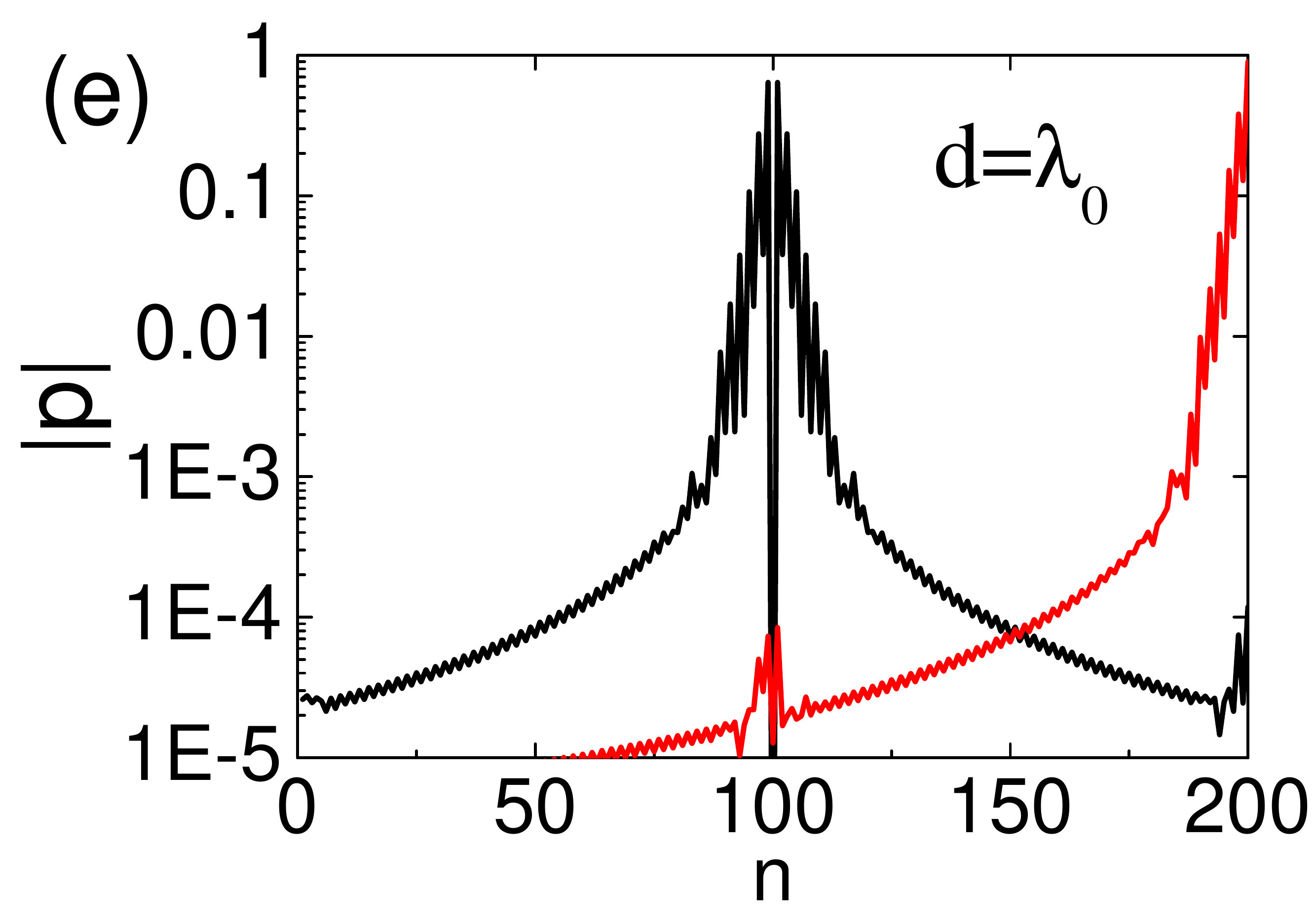}\label{longinterfacemode10}
}
\subfloat{
	\includegraphics[width=0.3\linewidth]{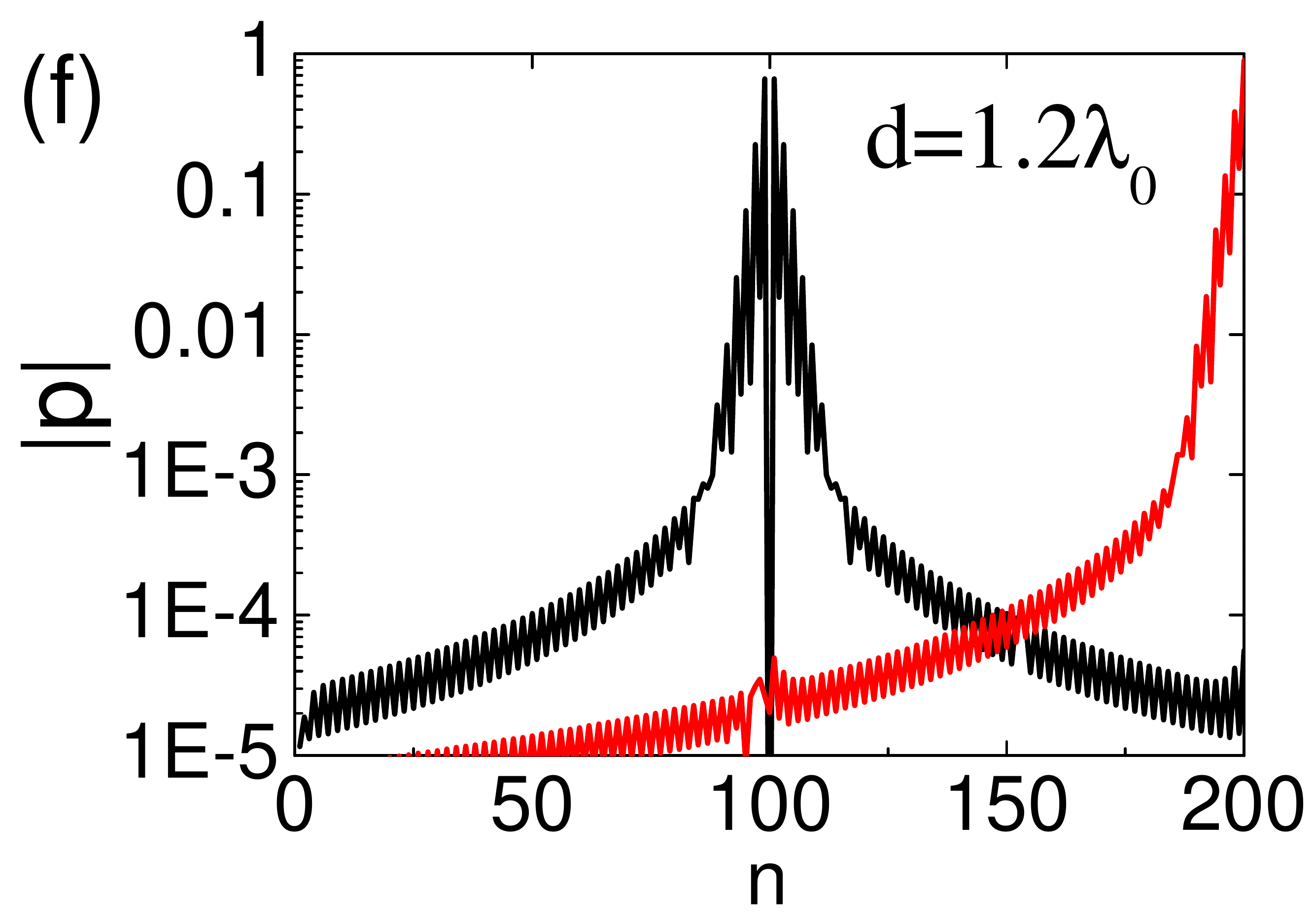}\label{longinterfacemode12}
}	
	\caption{(a-c) Dipole moment distributions of midgap edge states for finite chains with different lattice constants under the dimerization parameter of $\beta=0.6$. (d-f)Dipole moment distributions of the topologically protected interface and right-edge states of connected chains with different lattice constants comprising of a finite chain with $\beta=0.6$ and another with $\beta=0.4$.}\label{longedgmodes}
\end{figure}

To understand such a ``paradox'', we must again resort to the non-Hermitian definition of bandgap, which is already mentioned in Sec.\ref{topoinvariant}. In that sense, a complete bandgap is only possible when the bandgap is closed simultaneously in the real and imaginary spaces \cite{shenPRL2018}. Actually, if showing the calculated bandstructures of the eigenstates of these non-Hermitian systems in the complex plane in Figs.\ref{beta06longband08}-\ref{beta06longband12} for $d=0.8\lambda_0$, $\lambda_0$ and $1.2\lambda_0$ respectively, where $\beta=0.6$, we find that the complex bandgap is completely open in all the three cases, although the real spectra of the $d=\lambda_0$ and $1.2\lambda_0$ cases are ungapped. Notably, we indeed observe high-IPR discrete states reside in the complex bandgaps. In Figs.\ref{longedgemode08}-\ref{longedgemode12}, the dipole moment distributions of these midgap states are shown. We have also verified that the complex Zak phase is still quantized to be $\pi$ for $\beta>0.5$ and $0$ for $\beta<0.5$ (not shown here). Moreover, to further verify the topological nature of these systems, we also demonstrate the dipole moment distributions of the interface states emerging at the boundaries of topologically distinct chains, i.e., a $\beta=0.6$ chain in the right and $\beta=0.4$ in the left, shown in Figs.\ref{longinterfacemode08}-\ref{longinterfacemode12}.

%However, since the increase of lattice constant makes the real part of on-site term $a_{11}(k_x)$ be comparable with those of the hopping terms $a_{12}(k_x)$ and $a_{21}(k_x)$, the frequencies ($\Delta$) of edge states are lifted into the bulk spectrum due to the breaking of chiral symmetry. Nevertheless, this does not affect the topological nature of the band, because the eigenvectors are not affected by the on-site terms \cite{shenPRL2018,perezArxiv2018}. 

\subsection{Transverse eigenstates}
In this subsection, we investigate the topological properties of transverse eigenstates (polarized along the $y,z$ axes). Similar to the procedures in Eqs.(\ref{blocheqlong}-\ref{eigenvalue_long}), where using the transverse component of the Green's function ($G_{yy}$ or $G_{zz}$ for the present chain) 
\begin{equation}
G_{0}^{\bot}(x)=G_{0,yy}(x)=[\frac{i}{k|x|}-\frac{1}{(k|x|)^2}+1]\frac{\exp{(ik|x|)}}{4\pi |x|},
\end{equation}
we have
\begin{equation}\label{eigenvalue_trans}
\left(\begin{matrix}
a_{11}^{T}(k_x) & a_{12}^{T}(k_x)\\
a_{21}^{T}(k_x) & a_{22}^{T}(k_x)
\end{matrix}\right)\left(\begin{matrix}p_{A,k_x}\\p_{B,k_x}\end{matrix}\right)=\frac{E_{k_x}-\omega_0 + i \gamma/2}{-3\pi\gamma}\left(\begin{matrix}p_{A,k_x}\\p_{B,k_x}\end{matrix}\right),
\end{equation}
where the superscript $T$ stands for transverse. Note the transverse eigenstates are essentially different from the longitudinal states because the far-field, long range dipole-dipole interaction term which decays with the distance as $1/|x|$ plays a critical role. The diagonal elements of the transverse Bloch Hamiltonian $H^T(k_x)$ are calculated as
\begin{equation}
\begin{split}
&a_{11}^{T}(k_x)=a_{22}^{T}(k_x)=\frac{Li_1(z^+)+Li_1(z^-)}{4\pi kd}\\&+i\frac{Li_2(z^+)+Li_2(z^-)}{4\pi k^2d^2}-\frac{Li_3(z^+)+Li_3(z^-)}{4\pi k^3d^3}.
\end{split}
\end{equation}
The off-diagonal terms are given by
\begin{equation}
\begin{split}
&a_{12}^{T}(k_x)=\Big[i\frac{\Phi(z^+,2,\beta)}{4\pi k^2d^2}-\frac{\Phi(z^+,3,\beta)}{4\pi k^3d^3}+\frac{\Phi(z^+,1,\beta)}{4\pi kd}\Big]\\&\times\exp{(ik\beta d)}+\Big[i\frac{\Phi(z^-,2,1-\beta)}{4\pi k^2d^2}-\frac{\Phi(z^-,3,1-\beta)}{4\pi k^3d^3}\\&+\frac{\Phi(z^-,1,1-\beta)}{4\pi kd}\Big]z^-\exp{(-ik\beta d)},
\end{split}
\end{equation}
and
\begin{equation}
\begin{split}
&a_{21}^{T}(k_x)=\Big[i\frac{\Phi(z^+,2,1-\beta)}{4\pi k^2d^2}-\frac{\Phi(z^+,3,1-\beta)}{4\pi k^3d^3}\\&+\frac{\Phi(z^+,1,1-\beta)}{4\pi kd}\Big]z^+\exp{-(ik\beta d)}+\Big[i\frac{\Phi(z^-,2,\beta)}{4\pi k^2d^2}\\&-\frac{\Phi(z^-,3,\beta)}{4\pi k^3d^3}+\frac{\Phi(z^-,1,\beta)}{4\pi kd}\Big]\exp{(ik\beta d)}
\end{split}
\end{equation}

\begin{figure}[htbp]
	\centering
	\flushleft
%	\subfloat{
%	\includegraphics[width=0.46\linewidth]{transperioddependence06}\label{transperioddependence06}
%}
\hspace{0.01in}
	\subfloat{
		\includegraphics[width=0.46\linewidth]{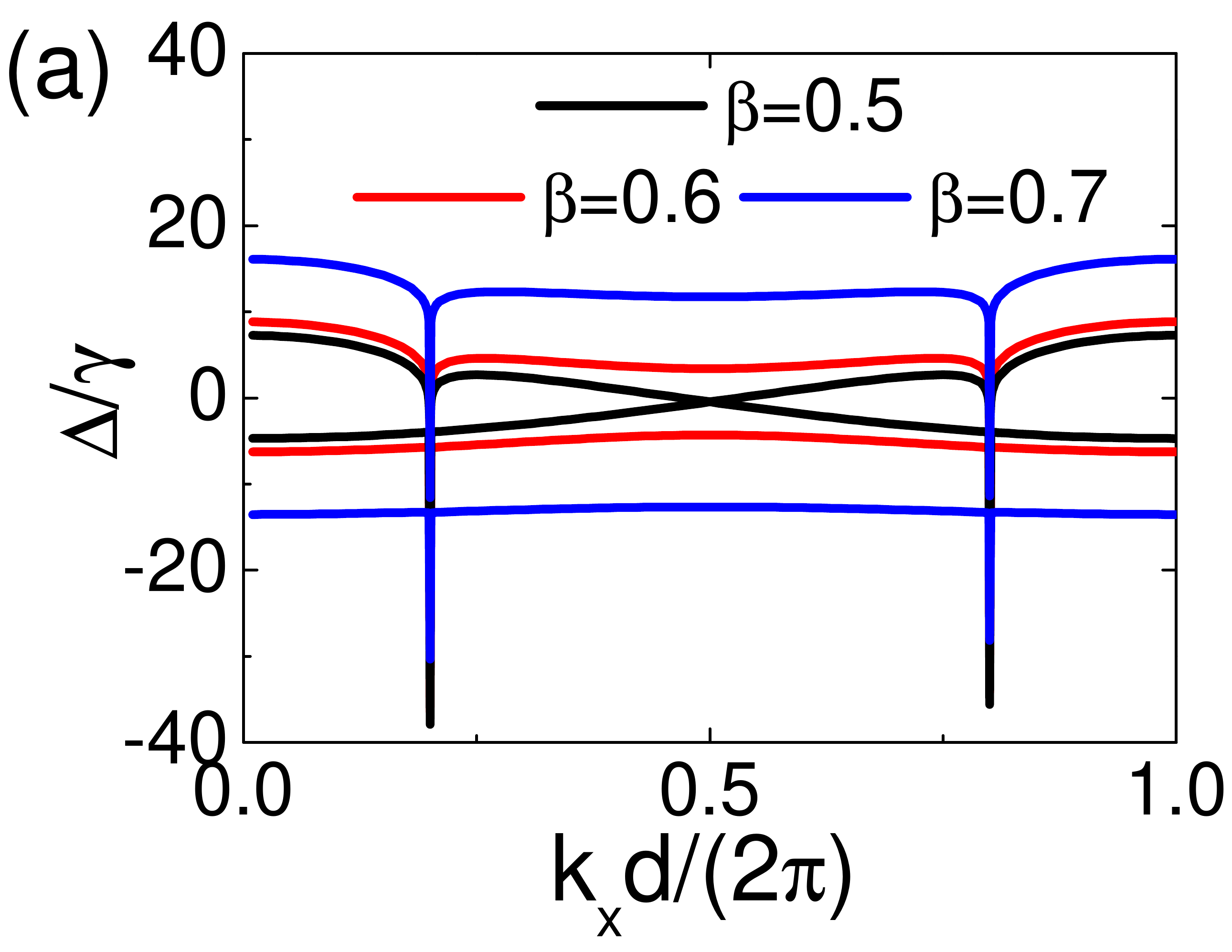}\label{transband02}
	}
	\hspace{0.01in}
	\subfloat{
		\includegraphics[width=0.46\linewidth]{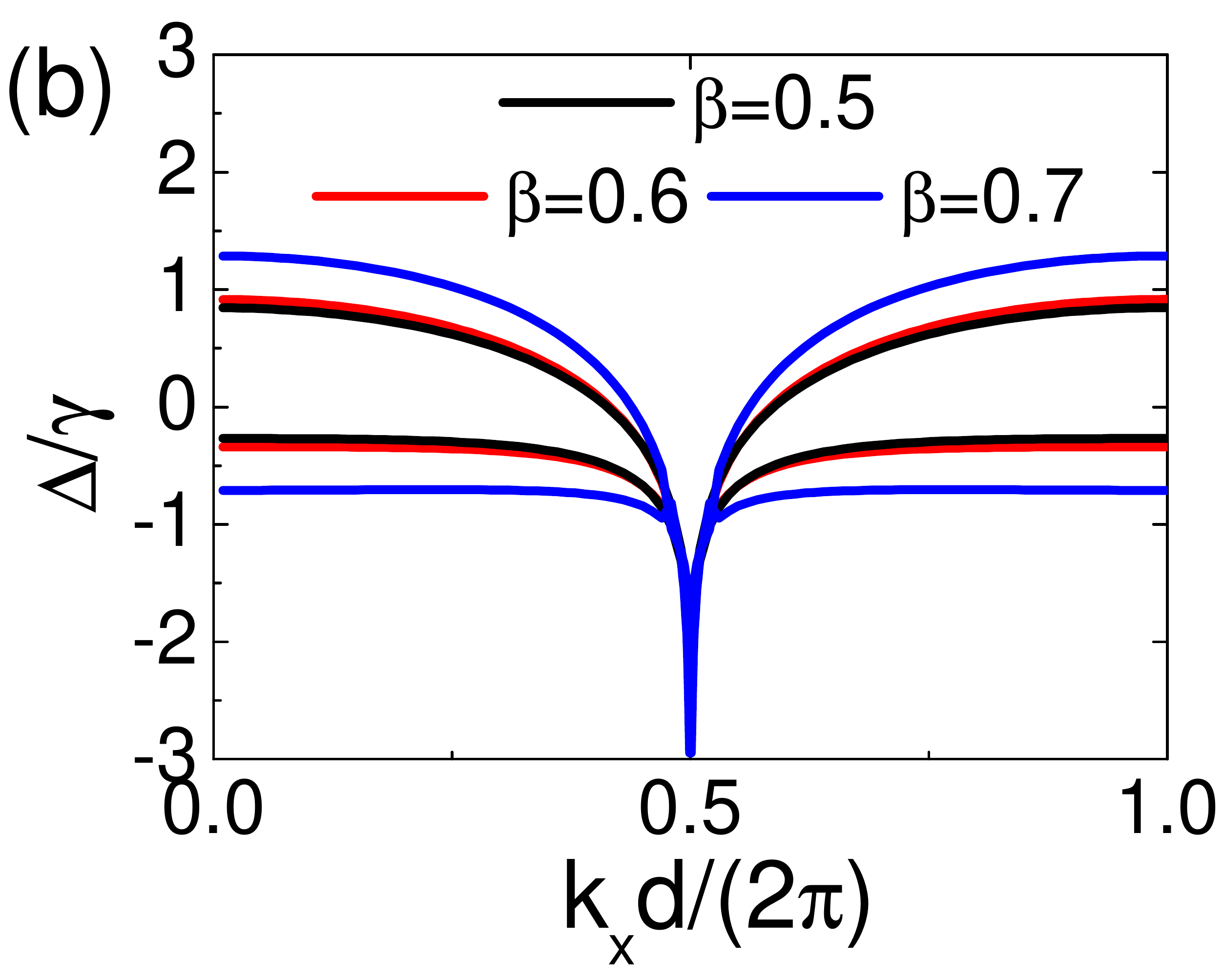}\label{transband05}
	}
	\hspace{0.01in}
%\subfloat{
%	\includegraphics[width=0.46\linewidth]{transband08}\label{transband08}
%}
	\subfloat{
	\includegraphics[width=0.46\linewidth]{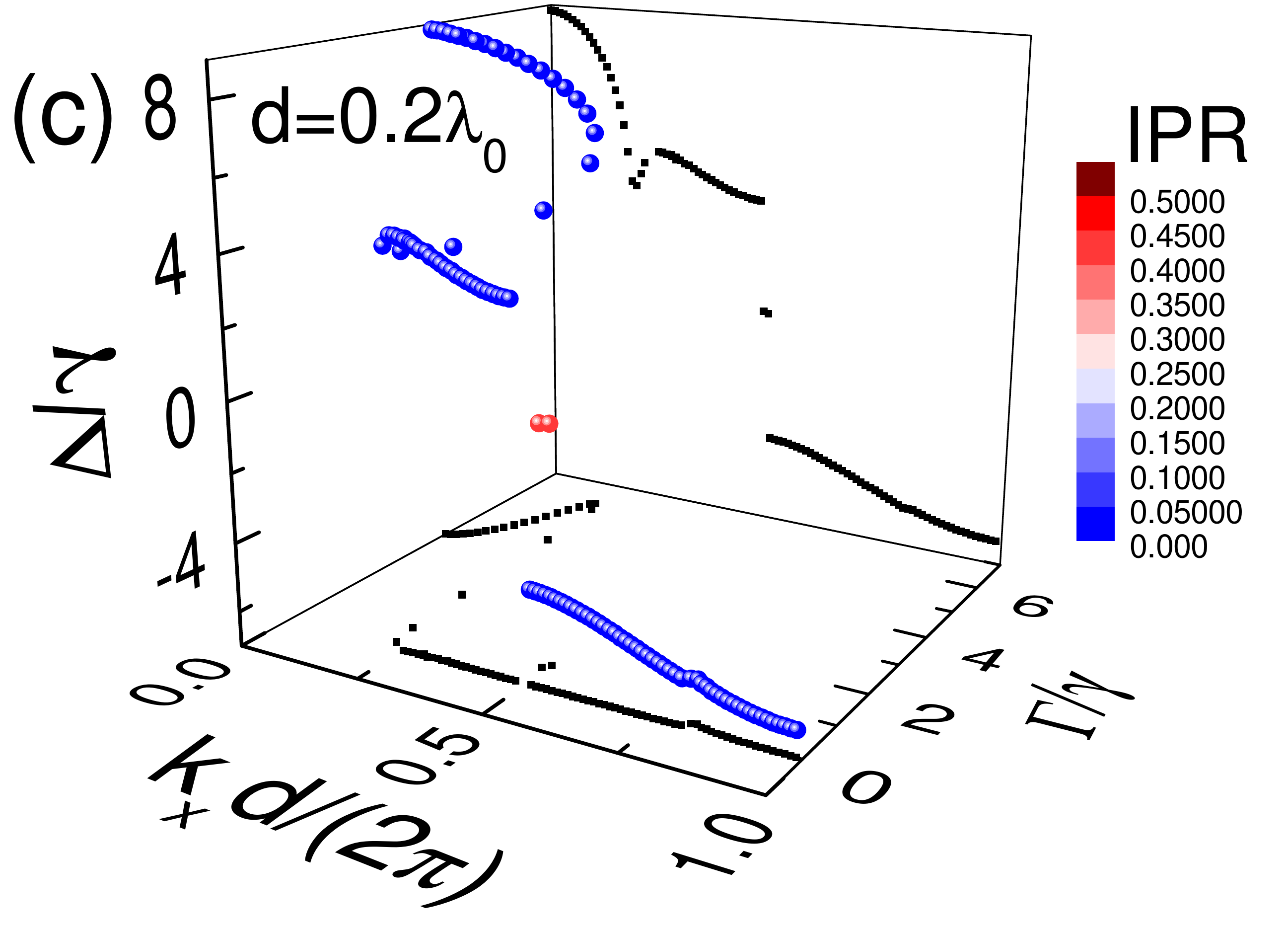}\label{beta06transband02}
}
\hspace{0.01in}
\subfloat{
	\includegraphics[width=0.46\linewidth]{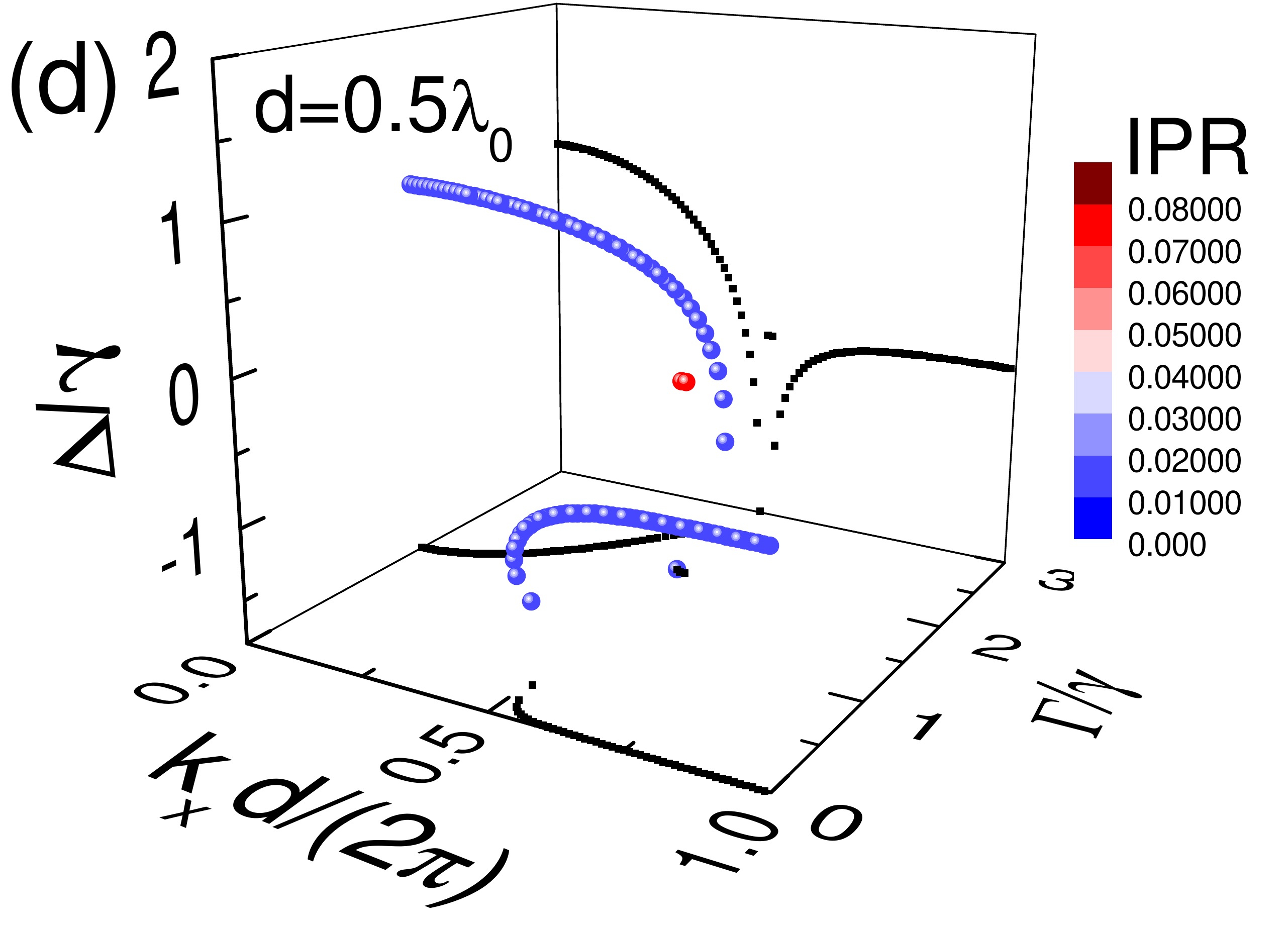}\label{beta06transband05}
}

	\caption{(a) Bandstructure for transverse eigenstates of a dimerized chain with $d=0.2\lambda_0$ for different dimerization parameters $\beta$. Here $\lambda_0=\omega_0/c$ is the wavelength at the single-atom resonance. (b) The same as that in (a) but with $d=0.5\lambda_0$. (c) Bandstructure for a finite chain with $N=100$ atoms where $\beta=0.6$ and $d=0.2\lambda_0$. (d) The same as that in (c) but with $d=0.5\lambda_0$. }
	
	\label{transband}
\end{figure}

In Figs.\ref{transband02} and \ref{transband05} we show the real parts of the transverse Bloch bandstructures for $d=0.2\lambda_0$ and $d=0.5\lambda_0$ with different dimerization parameters $\beta=0.5, 0.6, 0.7$. The bandstructures for $\beta=0.3$ and $\beta=0.4$ are the same as those for $\beta=0.7$ and $\beta=0.6$, respectively. It is noted that when the light line (at $k_x=2\pi/\lambda_0$) intersects the bandstructures, the transverse eigenstates are phase-matched to the free space radiation along the chain, and result in significant discontinuities \cite{weberPRB2004,pocockArxiv2017}. Apart from those discontinuities, the real bandstructures for $\beta\neq0.5$ are all gapped in the real space. We also calculate the corresponding bandstructures for finite chains with $N=100$ atoms in Figs.\ref{beta06transband02} and \ref{beta06transband05}. The high-IPR midgap states can be unambiguously identified at the edge of Brillouin zone ($k_x=\pi/d$) in the complex bandgaps. Notably, the IPR of the midgap states for the $d=0.2\lambda_0$ case is around 0.41, while that he $d=0.5\lambda_0$ case is much smaller, only around 0.07.

\begin{figure}[htbp]
	\centering
	\flushleft
\subfloat{
	\includegraphics[width=0.46\linewidth]{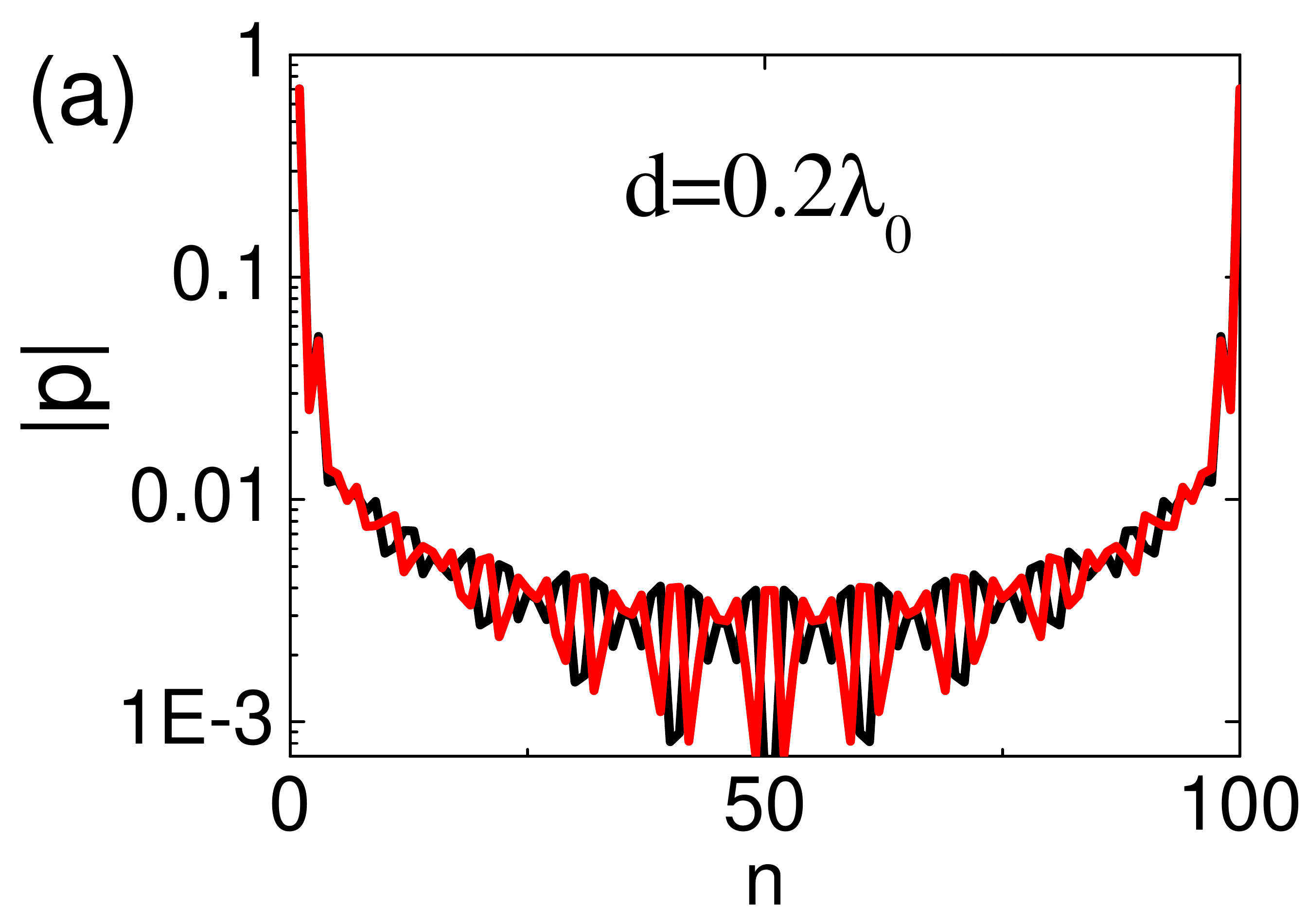}\label{transedgemode02}
}
\hspace{0.01in}
\subfloat{
	\includegraphics[width=0.46\linewidth]{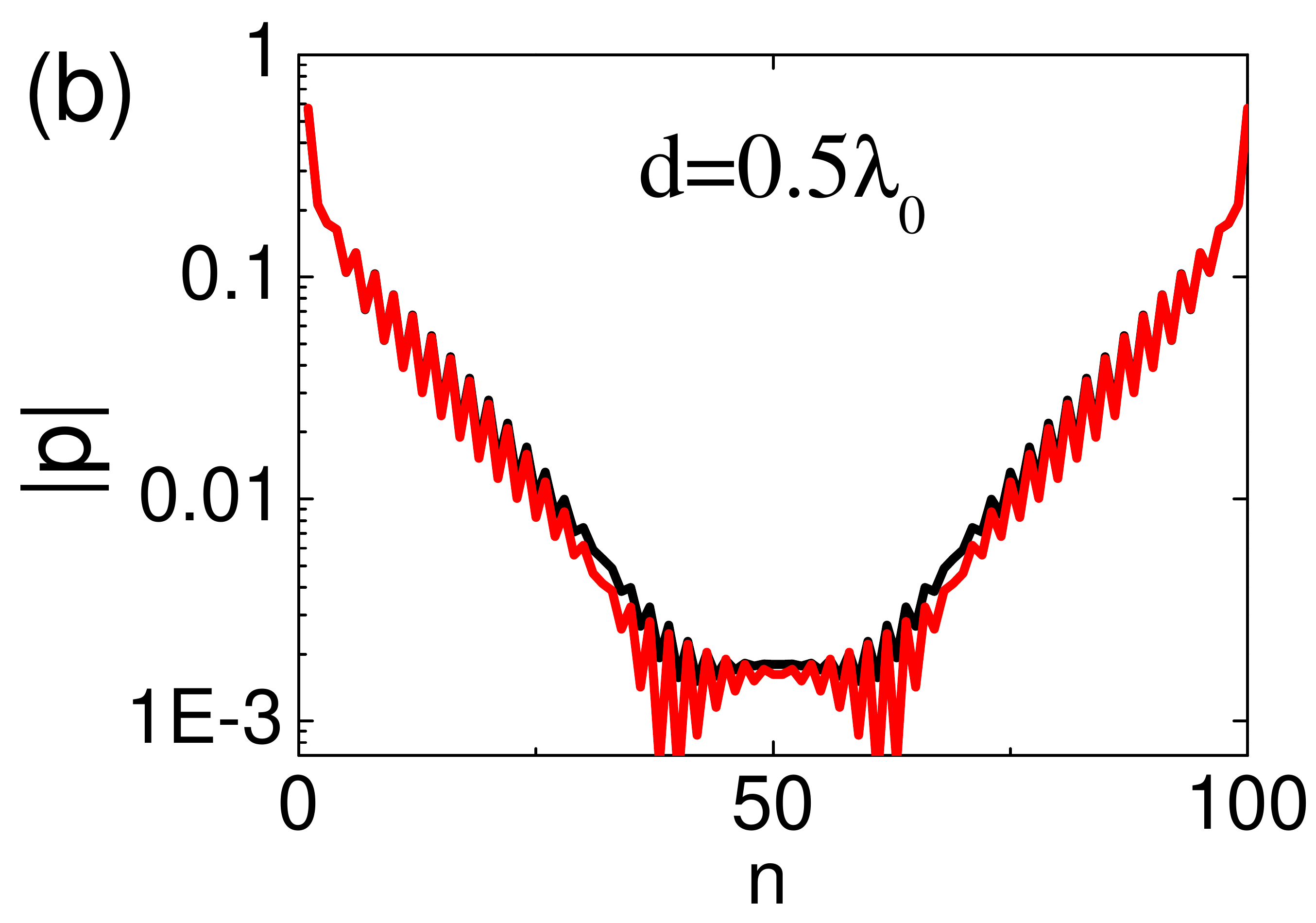}\label{transedgemode05}
}
\hspace{0.01in}
\subfloat{
	\includegraphics[width=0.46\linewidth]{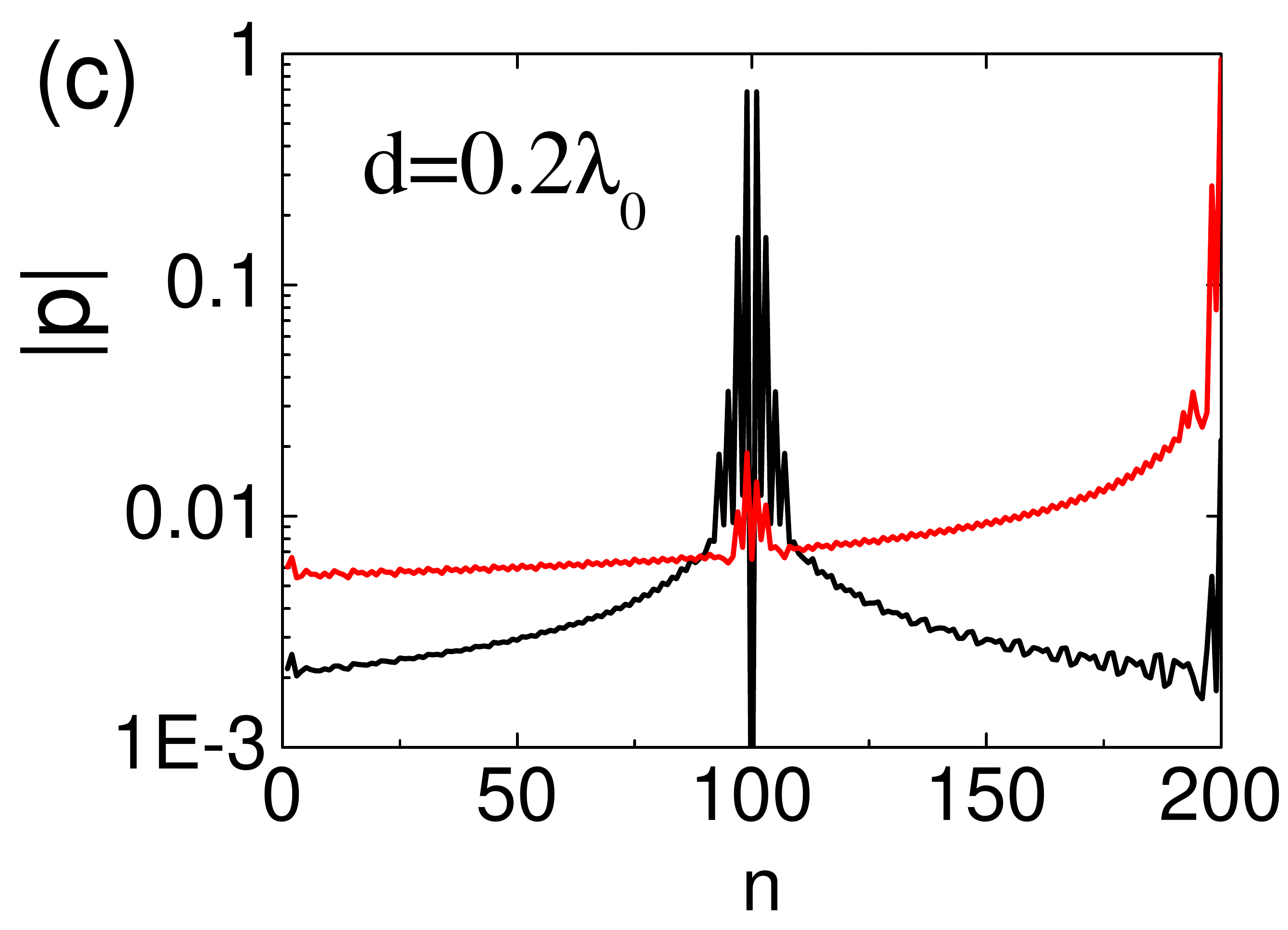}\label{beta06transinterfacemode02}
}
\hspace{0.01in}
\subfloat{
	\includegraphics[width=0.46\linewidth]{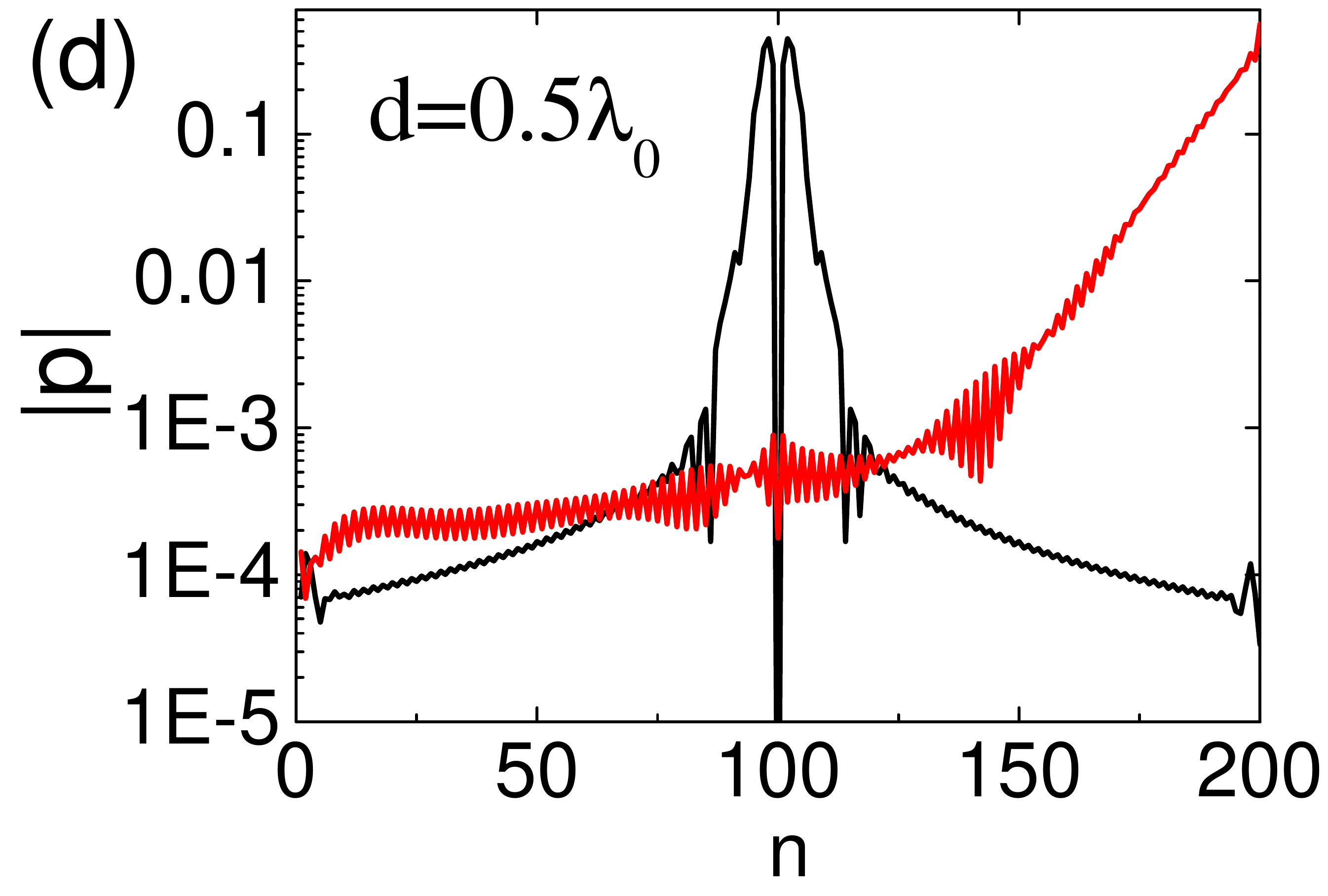}\label{beta06transinterfacemode05}
}
	\caption{(a) Dipole moment distributions of transverse midgap edge states for a finite chain with a lattice constant of $d=0.2\lambda_0$ and dimerization parameter of $\beta=0.6$. (b) The same as (a) but with a lattice constant $d=0.5\lambda_0$. (c) Dipole moment distributions of topologically protected interface and right-edge states of connected chains with a lattice constant of $d=0.2\lambda_0$ comprising of a finite chain with a dimerization parameter of $\beta=0.6$ and another with $\beta=0.4$. (d) The same as (c) but with a lattice constant $d=0.5\lambda_0$.}
	
	\label{transfiniteband}
\end{figure}

The dipole moment distributions for these midgap states are shown in Figs.\ref{transedgemode02} and \ref{transedgemode05} respectively, in the logarithmic scale. It is found that in both cases, the midgap states are exponentially localized over the edges. However, the localization length of the midgap states in the $d=0.5\lambda_0$ case is much longer than that in the $d=0.2\lambda_0$ case. This is consistent with the recent study of P\'erez-Gonz\'alez et al \cite{perezArxiv2018}, which take long-range hoppings in the Hermitian SSH model into account. In that study, it is found that when the amplitude of even hoppings (second neighbor, fourth neighbor, etc.), which enter into the diagonal terms in the Hamiltonian, is comparable to that of odd hoppings (nearest neighbor, third neighbor, etc.), which enter into the off-diagonal terms, the localization length of the edge states is strongly increased. This is the case in our system, where a longer lattice period makes $a_{11}(k_x)$ comparable with $a_{12}(k_x)$ and $a_{21}(k_x)$, because the far-field, long-range dipole-dipole interactions in transverse eigenstates are strong.
 
We also calculate the complex Zak phase for the transverse bandstructures (not shown here). In this situation, we still find that for $\beta>0.5$, the complex Zak phase is $\pi$ while for $\beta<0.5$ it is $0$. We thus conclude that in the transverse case, the bulk-boundary correspondence is still applicable although dipole-dipole interactions are considerable, because the bandstructures for infinite atoms and finite atoms are almost the same, and the complex Zak phase is able to indicate the emergence of midgap edge states. In Figs.\ref{beta06transinterfacemode02} and \ref{beta06transinterfacemode05}, when we combine two such systems with different complex Zak phases, we still observe highly localized interface states. As expected, the localization lengths of the interface and right edge states in the $d=0.5\lambda_0$ case are much longer.

\begin{figure}[htbp]
	\centering
	\includegraphics[width=0.6\linewidth]{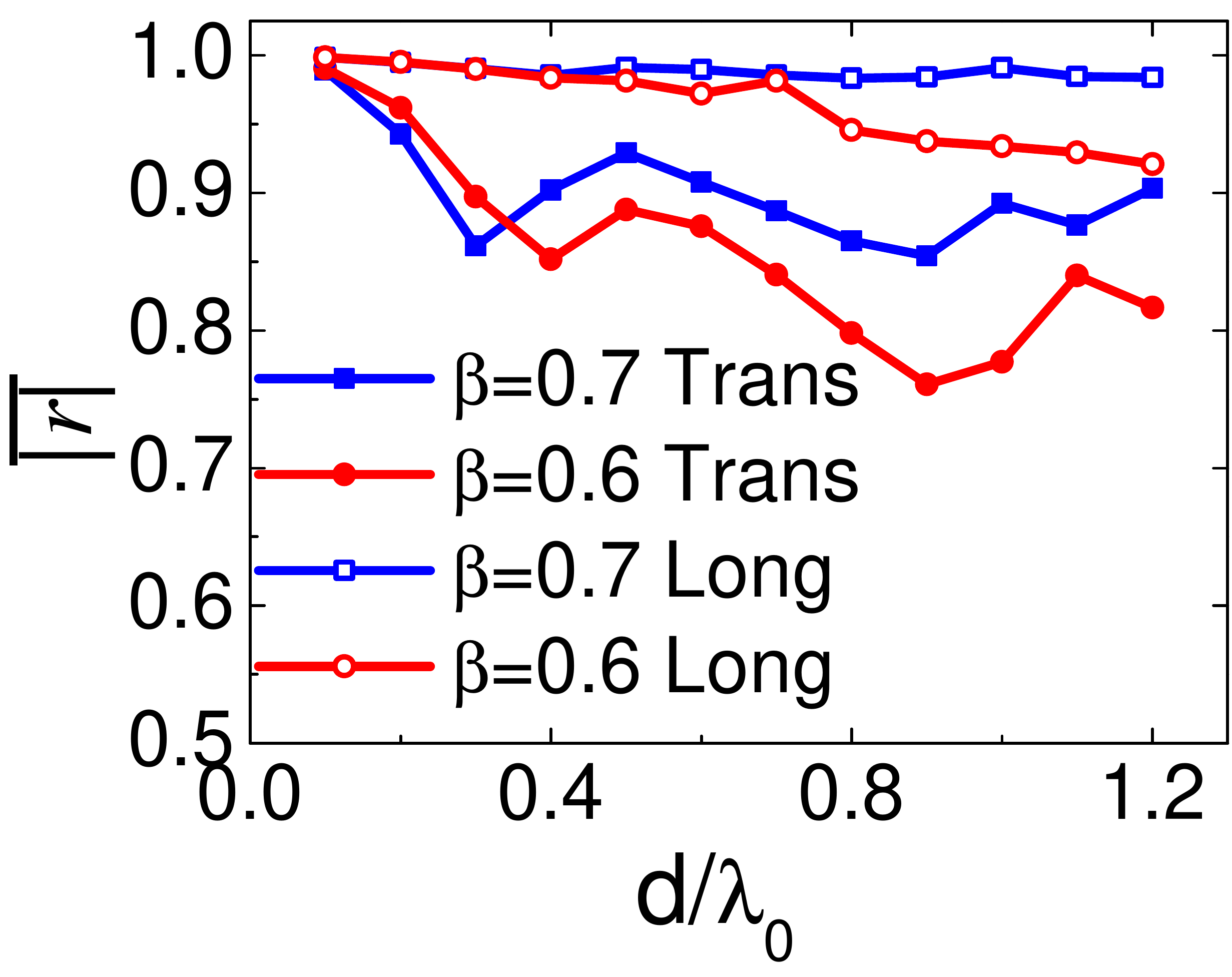}

	\caption{The averaged phase rigidity of eigenstates, which acts as a measure of Hermiticity of the non-Hermitian system, as a function of the lattice constant under different dimerization parameters for both longitudinal and transverse eigenstates of a finite chain with 100 atoms.}\label{phaserigidity}
	
	\label{nonhermiticity}
\end{figure}

Finally, to understand, at least partially, why in the present non-Hermitian systems the complex Zak phase and the principle of bulk-boundary correspondence are still valid, we calculate the measure of biorthogonality of an eigenstate $\phi$ of a finite chain, namely, the phase rigidity, defined as the ratio between biorthogonality and orthogonality \cite{alvarezPRB2018,eleuchPRA2016}
\begin{equation}
r_\phi=\frac{\langle \mathbf{p}_\phi^L|\mathbf{p}_\phi^R\rangle}{\langle \mathbf{p}_\phi^R|\mathbf{p}_\phi^R\rangle},
\end{equation}
where $\langle\mathbf{p}_\phi^L|$ is the eigenvector of the left eigenstate of the Hamiltonian of a finite chain of atoms. For Hermitian Hamiltonians, this quantity is exactly 1 for all eigenstates, while for a non-Hermitian Hamiltonian where $|\mathbf{p}_\phi^L\rangle\neq|\mathbf{p}_\phi^R\rangle$, $|r_\phi|$ is generally less than 1, and at and near an EP it takes its minimum value $r_\phi\rightarrow0$. In Fig.\ref{phaserigidity} we show the averaged phase rigidity of the eigenstates $\bar{|r|}=\Sigma_{\phi}|r_\phi|/N$ for transverse and longitudinal eigenstates under $\beta=0.6$ and $\beta=0.7$ as a function of the lattice constant $d$. It can be regarded as a measure of Hermiticity, and a smaller $|r|$ indicates stronger non-Hermiticity of the Hamiltonian \cite{alvarezPRB2018,eleuchPRA2016}. In all the cases investigated in the present study, we find that the although phase rigidity shows a decreasing trend with the increase of lattice constant, it is still considerably high. Therefore, in the present system, the non-Hermiticity is weak, and the methods and principles developed from Hermitian systems can be generalized and then applied here.

\subsection{On the experimental realization}
Our proposal of 1D dimerized atomic arrays is experimentally available in the state-of-art quantum simulation techniques.  Actually, recent advances in optical lattices, which are made by optical standing waves, can create high-filling-factor (exceeding 90\%) atomic arrays with a single atom per lattice site, by utilizing the superfluid-Mott insulator transition at high trap potential depths \cite{shersonNature2010,bakrNature2011,bakrScience2010,weitenbergNature2011}. The lattice spacing can be controlled to reach deep-subwavelength scale \cite{syzranovNaturecomms2016,olmosPRL2013,wangPRL2018}, compared to the dipole transition wavelength of the two-level atom. For instance, for the bosonic strontium atom $\mathrm{^{84}Sr}$, the $^3P_0$-$^3D_1$ dipole transition wavelength is 2.6$\mathrm{\mu m}$, and using the optical lattice with the ``magic wavelength" (here a blue-detuned 412.8nm laser) can achieve a deep-subwavelength lattice spacing of $a=206.4\mathrm{nm}$ \cite{olmosPRL2013}. More specifically, such a dimerized atomic array can be appropriately realized by the 1D double-well potential (superlattice potential) composed of two superimposed optical lattices with different wavelengths (usually the wavelength of one optical lattice is twice of the other's) \cite{follingNaturephys2007,wirthNaturephys2011,atalaNaturephys2013}, where the dimerization parameter can be well-controlled by tuning the phases and amplitudes of two optical lattices.

Moreover, cutting-edge developments in the manipulation of single atoms make the fabrication of atomic lattices with \textit{arbitrary} ordered geometries feasible. It was recently shown that 1D $\mathrm{^{87}Rb}$ atomic arrays with desired arrangements in a large scale (more than 50 atoms) can be assembled in an atom by atom way, through a fast, real-time control of an array of tightly focused optical tweezers \cite{endresScience2016,bernienNature2017}. Two-dimensional $\mathrm{^{87}Rb}$ arrays with user-defined geometries can be also fabricated in a similar way \cite{lahayeScience2016}. This technique offers an alternative and ideal platform for the research of atom-made photonic structures. Other modern quantum simulation techniques, for instance, the dynamic modulation for spin-dependent optical lattice \cite{nascimbenePRL2015} and nanophotonic atom lattices using dielectric photonic crystals \cite{gonzalezNaturephoton2015} or plasmonic nanoparticle arrays \cite{gullansPRL2012}, are also possible ways to create ultracold atom-based topological photonic systems with desired periodic patterns.

\section{Conclusion}
In conclusion, we propose one-dimensional dimerized ultracold atomic chains as a non-Hermitian analogy of the conventional SSH model. We show that the complex Zak phase is still quantized and becomes nontrivial when the dimerization parameter $\beta>0.5$ despite the non-Hermitian Hamiltonian and the breaking of chiral symmetry. Topologically protected optical states with high IPRs are found at the edges of dimerized chains with nontrivial complex Zak phases ($\theta_B=\pi$), as well as at the interface between chains with different complex Zak phases, based on an analysis of the eigenstate distribution for finite chains. These phenomena confirm the validity of the bulk-boundary correspondence in the present non-Hermitian system. We further show that such topological edge states are robust under disorders which do not respect the chiral symmetry, elucidating that such edges modes are indeed topologically protected. For transverse eigenstates, we discover that the existence of strong far-field dipole-dipole interactions still preserves the topological properties described by a quantized complex Zak phase, which is nontrivial for $\beta>0.5$, while the localization length of topological edge states is increased with the increase of lattice constant.
	
On the other hand, the ultra-strong scattering cross section and ultra-narrow linewidth of a single cold atom allows us to observe in more detail about topological states than conventional systems, such as the frequency shift with respect to the single-atom resonance and the largely tunable bandgap. We also expect that topological photonic states in such dimerized ultracold atomic chains can provide an efficient interface for studying the topological states of light and matter \cite{zollerNaturephys2016}. Moreover, if nonlinear effects in the atoms are included at high-intensity excitation, such dimerized chains can be applied to study the many-body physics of interacting photons \cite{yelinPRL2017,yelinPRL20172}. We also envisage some general features of non-Hermitian topological photonic systems composed of highly resonant point dipoles, including quantum dots, plasmonic nanoparticles, etc., can be obtained from our study. 

\begin{acknowledgments}
We thank the financial support from the National Natural Science Foundation of China (51636004, 51476097), Shanghai Key Fundamental Research Grant (18JC1413300, 16JC1403200), and the Foundation for Innovative Research Groups of the National Natural Science Foundation of China (51521004).
\end{acknowledgments}

\bibliography{ssh}

%merlin.mbs apsrev4-1.bst 2010-07-25 4.21a (PWD, AO, DPC) hacked
%Control: key (0)
%Control: author (72) initials jnrlst
%Control: editor formatted (1) identically to author
%Control: production of article title (-1) disabled
%Control: page (0) single
%Control: year (1) truncated
%Control: production of eprint (0) enabled
\begin{thebibliography}{90}%
\makeatletter
\providecommand \@ifxundefined [1]{%
 \@ifx{#1\undefined}
}%
\providecommand \@ifnum [1]{%
 \ifnum #1\expandafter \@firstoftwo
 \else \expandafter \@secondoftwo
 \fi
}%
\providecommand \@ifx [1]{%
 \ifx #1\expandafter \@firstoftwo
 \else \expandafter \@secondoftwo
 \fi
}%
\providecommand \natexlab [1]{#1}%
\providecommand \enquote  [1]{``#1''}%
\providecommand \bibnamefont  [1]{#1}%
\providecommand \bibfnamefont [1]{#1}%
\providecommand \citenamefont [1]{#1}%
\providecommand \href@noop [0]{\@secondoftwo}%
\providecommand \href [0]{\begingroup \@sanitize@url \@href}%
\providecommand \@href[1]{\@@startlink{#1}\@@href}%
\providecommand \@@href[1]{\endgroup#1\@@endlink}%
\providecommand \@sanitize@url [0]{\catcode `\\12\catcode `\$12\catcode
  `\&12\catcode `\#12\catcode `\^12\catcode `\_12\catcode `\%12\relax}%
\providecommand \@@startlink[1]{}%
\providecommand \@@endlink[0]{}%
\providecommand \url  [0]{\begingroup\@sanitize@url \@url }%
\providecommand \@url [1]{\endgroup\@href {#1}{\urlprefix }}%
\providecommand \urlprefix  [0]{URL }%
\providecommand \Eprint [0]{\href }%
\providecommand \doibase [0]{http://dx.doi.org/}%
\providecommand \selectlanguage [0]{\@gobble}%
\providecommand \bibinfo  [0]{\@secondoftwo}%
\providecommand \bibfield  [0]{\@secondoftwo}%
\providecommand \translation [1]{[#1]}%
\providecommand \BibitemOpen [0]{}%
\providecommand \bibitemStop [0]{}%
\providecommand \bibitemNoStop [0]{.\EOS\space}%
\providecommand \EOS [0]{\spacefactor3000\relax}%
\providecommand \BibitemShut  [1]{\csname bibitem#1\endcsname}%
\let\auto@bib@innerbib\@empty
%</preamble>
\bibitem [{\citenamefont {Lu}\ \emph {et~al.}(2014)\citenamefont {Lu},
  \citenamefont {Joannopoulos},\ and\ \citenamefont
  {Solja{\v{c}}i{\'c}}}]{luNPhoton2014}%
  \BibitemOpen
  \bibfield  {author} {\bibinfo {author} {\bibfnamefont {L.}~\bibnamefont
  {Lu}}, \bibinfo {author} {\bibfnamefont {J.~D.}\ \bibnamefont
  {Joannopoulos}}, \ and\ \bibinfo {author} {\bibfnamefont {M.}~\bibnamefont
  {Solja{\v{c}}i{\'c}}},\ }\href@noop {} {\bibfield  {journal} {\bibinfo
  {journal} {Nature Photonics}\ }\textbf {\bibinfo {volume} {8}},\ \bibinfo
  {pages} {821} (\bibinfo {year} {2014})}\BibitemShut {NoStop}%
\bibitem [{\citenamefont {Khanikaev}\ and\ \citenamefont
  {Shvets}(2017)}]{khanikaevNPhoton2017}%
  \BibitemOpen
  \bibfield  {author} {\bibinfo {author} {\bibfnamefont {A.~B.}\ \bibnamefont
  {Khanikaev}}\ and\ \bibinfo {author} {\bibfnamefont {G.}~\bibnamefont
  {Shvets}},\ }\href@noop {} {\bibfield  {journal} {\bibinfo  {journal} {Nature
  Photonics}\ }\textbf {\bibinfo {volume} {11}},\ \bibinfo {pages} {763}
  (\bibinfo {year} {2017})}\BibitemShut {NoStop}%
\bibitem [{\citenamefont {Ozawa}\ \emph {et~al.}(2018)\citenamefont {Ozawa},
  \citenamefont {Price}, \citenamefont {Amo}, \citenamefont {Goldman},
  \citenamefont {Hafezi}, \citenamefont {Lu}, \citenamefont {Rechtsman},
  \citenamefont {Schuster}, \citenamefont {Simon}, \citenamefont {Zilberberg}
  \emph {et~al.}}]{ozawa2018topological}%
  \BibitemOpen
  \bibfield  {author} {\bibinfo {author} {\bibfnamefont {T.}~\bibnamefont
  {Ozawa}}, \bibinfo {author} {\bibfnamefont {H.~M.}\ \bibnamefont {Price}},
  \bibinfo {author} {\bibfnamefont {A.}~\bibnamefont {Amo}}, \bibinfo {author}
  {\bibfnamefont {N.}~\bibnamefont {Goldman}}, \bibinfo {author} {\bibfnamefont
  {M.}~\bibnamefont {Hafezi}}, \bibinfo {author} {\bibfnamefont
  {L.}~\bibnamefont {Lu}}, \bibinfo {author} {\bibfnamefont {M.}~\bibnamefont
  {Rechtsman}}, \bibinfo {author} {\bibfnamefont {D.}~\bibnamefont {Schuster}},
  \bibinfo {author} {\bibfnamefont {J.}~\bibnamefont {Simon}}, \bibinfo
  {author} {\bibfnamefont {O.}~\bibnamefont {Zilberberg}},  \emph {et~al.},\
  }\href@noop {} {\bibfield  {journal} {\bibinfo  {journal} {arXiv preprint
  arXiv:1802.04173}\ } (\bibinfo {year} {2018})}\BibitemShut {NoStop}%
\bibitem [{\citenamefont {Poli}\ \emph {et~al.}(2015)\citenamefont {Poli},
  \citenamefont {Bellec}, \citenamefont {Kuhl}, \citenamefont {Mortessagne},\
  and\ \citenamefont {Schomerus}}]{poliNComms2015}%
  \BibitemOpen
  \bibfield  {author} {\bibinfo {author} {\bibfnamefont {C.}~\bibnamefont
  {Poli}}, \bibinfo {author} {\bibfnamefont {M.}~\bibnamefont {Bellec}},
  \bibinfo {author} {\bibfnamefont {U.}~\bibnamefont {Kuhl}}, \bibinfo {author}
  {\bibfnamefont {F.}~\bibnamefont {Mortessagne}}, \ and\ \bibinfo {author}
  {\bibfnamefont {H.}~\bibnamefont {Schomerus}},\ }\href@noop {} {\bibfield
  {journal} {\bibinfo  {journal} {Nature communications}\ }\textbf {\bibinfo
  {volume} {6}},\ \bibinfo {pages} {6710} (\bibinfo {year} {2015})}\BibitemShut
  {NoStop}%
\bibitem [{\citenamefont {El-Ganainy}\ and\ \citenamefont
  {Levy}(2015)}]{el-GanainyOL2015}%
  \BibitemOpen
  \bibfield  {author} {\bibinfo {author} {\bibfnamefont {R.}~\bibnamefont
  {El-Ganainy}}\ and\ \bibinfo {author} {\bibfnamefont {M.}~\bibnamefont
  {Levy}},\ }\href {\doibase 10.1364/OL.40.005275} {\bibfield  {journal}
  {\bibinfo  {journal} {Opt. Lett.}\ }\textbf {\bibinfo {volume} {40}},\
  \bibinfo {pages} {5275} (\bibinfo {year} {2015})}\BibitemShut {NoStop}%
\bibitem [{\citenamefont {Parto}\ \emph {et~al.}(2018)\citenamefont {Parto},
  \citenamefont {Wittek}, \citenamefont {Hodaei}, \citenamefont {Harari},
  \citenamefont {Bandres}, \citenamefont {Ren}, \citenamefont {Rechtsman},
  \citenamefont {Segev}, \citenamefont {Christodoulides},\ and\ \citenamefont
  {Khajavikhan}}]{partoPRL2018}%
  \BibitemOpen
  \bibfield  {author} {\bibinfo {author} {\bibfnamefont {M.}~\bibnamefont
  {Parto}}, \bibinfo {author} {\bibfnamefont {S.}~\bibnamefont {Wittek}},
  \bibinfo {author} {\bibfnamefont {H.}~\bibnamefont {Hodaei}}, \bibinfo
  {author} {\bibfnamefont {G.}~\bibnamefont {Harari}}, \bibinfo {author}
  {\bibfnamefont {M.~A.}\ \bibnamefont {Bandres}}, \bibinfo {author}
  {\bibfnamefont {J.}~\bibnamefont {Ren}}, \bibinfo {author} {\bibfnamefont
  {M.~C.}\ \bibnamefont {Rechtsman}}, \bibinfo {author} {\bibfnamefont
  {M.}~\bibnamefont {Segev}}, \bibinfo {author} {\bibfnamefont {D.~N.}\
  \bibnamefont {Christodoulides}}, \ and\ \bibinfo {author} {\bibfnamefont
  {M.}~\bibnamefont {Khajavikhan}},\ }\href {\doibase
  10.1103/PhysRevLett.120.113901} {\bibfield  {journal} {\bibinfo  {journal}
  {Phys. Rev. Lett.}\ }\textbf {\bibinfo {volume} {120}},\ \bibinfo {pages}
  {113901} (\bibinfo {year} {2018})}\BibitemShut {NoStop}%
\bibitem [{\citenamefont {Ling}\ \emph {et~al.}(2015)\citenamefont {Ling},
  \citenamefont {Xiao}, \citenamefont {Chan}, \citenamefont {Yu},\ and\
  \citenamefont {Fung}}]{lingOE2015}%
  \BibitemOpen
  \bibfield  {author} {\bibinfo {author} {\bibfnamefont {C.~W.}\ \bibnamefont
  {Ling}}, \bibinfo {author} {\bibfnamefont {M.}~\bibnamefont {Xiao}}, \bibinfo
  {author} {\bibfnamefont {C.~T.}\ \bibnamefont {Chan}}, \bibinfo {author}
  {\bibfnamefont {S.~F.}\ \bibnamefont {Yu}}, \ and\ \bibinfo {author}
  {\bibfnamefont {K.~H.}\ \bibnamefont {Fung}},\ }\href {\doibase
  10.1364/OE.23.002021} {\bibfield  {journal} {\bibinfo  {journal} {Opt.
  Express}\ }\textbf {\bibinfo {volume} {23}},\ \bibinfo {pages} {2021}
  (\bibinfo {year} {2015})}\BibitemShut {NoStop}%
\bibitem [{\citenamefont {Downing}\ and\ \citenamefont
  {Weick}(2017)}]{downingPRB2017}%
  \BibitemOpen
  \bibfield  {author} {\bibinfo {author} {\bibfnamefont {C.~A.}\ \bibnamefont
  {Downing}}\ and\ \bibinfo {author} {\bibfnamefont {G.}~\bibnamefont
  {Weick}},\ }\href {\doibase 10.1103/PhysRevB.95.125426} {\bibfield  {journal}
  {\bibinfo  {journal} {Phys. Rev. B}\ }\textbf {\bibinfo {volume} {95}},\
  \bibinfo {pages} {125426} (\bibinfo {year} {2017})}\BibitemShut {NoStop}%
\bibitem [{\citenamefont {Longhi}(2013)}]{longhiOL2013}%
  \BibitemOpen
  \bibfield  {author} {\bibinfo {author} {\bibfnamefont {S.}~\bibnamefont
  {Longhi}},\ }\href {\doibase 10.1364/OL.38.003716} {\bibfield  {journal}
  {\bibinfo  {journal} {Opt. Lett.}\ }\textbf {\bibinfo {volume} {38}},\
  \bibinfo {pages} {3716} (\bibinfo {year} {2013})}\BibitemShut {NoStop}%
\bibitem [{\citenamefont {Blanco-Redondo}\ \emph {et~al.}(2016)\citenamefont
  {Blanco-Redondo}, \citenamefont {Andonegui}, \citenamefont {Collins},
  \citenamefont {Harari}, \citenamefont {Lumer}, \citenamefont {Rechtsman},
  \citenamefont {Eggleton},\ and\ \citenamefont
  {Segev}}]{blanco-RedondoPRL2016}%
  \BibitemOpen
  \bibfield  {author} {\bibinfo {author} {\bibfnamefont {A.}~\bibnamefont
  {Blanco-Redondo}}, \bibinfo {author} {\bibfnamefont {I.}~\bibnamefont
  {Andonegui}}, \bibinfo {author} {\bibfnamefont {M.~J.}\ \bibnamefont
  {Collins}}, \bibinfo {author} {\bibfnamefont {G.}~\bibnamefont {Harari}},
  \bibinfo {author} {\bibfnamefont {Y.}~\bibnamefont {Lumer}}, \bibinfo
  {author} {\bibfnamefont {M.~C.}\ \bibnamefont {Rechtsman}}, \bibinfo {author}
  {\bibfnamefont {B.~J.}\ \bibnamefont {Eggleton}}, \ and\ \bibinfo {author}
  {\bibfnamefont {M.}~\bibnamefont {Segev}},\ }\href {\doibase
  10.1103/PhysRevLett.116.163901} {\bibfield  {journal} {\bibinfo  {journal}
  {Phys. Rev. Lett.}\ }\textbf {\bibinfo {volume} {116}},\ \bibinfo {pages}
  {163901} (\bibinfo {year} {2016})}\BibitemShut {NoStop}%
\bibitem [{\citenamefont {Asb{\'o}th}\ \emph {et~al.}(2016)\citenamefont
  {Asb{\'o}th}, \citenamefont {Oroszl{\'a}ny},\ and\ \citenamefont
  {P{\'a}lyi}}]{asboth2016short}%
  \BibitemOpen
  \bibfield  {author} {\bibinfo {author} {\bibfnamefont {J.~K.}\ \bibnamefont
  {Asb{\'o}th}}, \bibinfo {author} {\bibfnamefont {L.}~\bibnamefont
  {Oroszl{\'a}ny}}, \ and\ \bibinfo {author} {\bibfnamefont {A.~P.}\
  \bibnamefont {P{\'a}lyi}},\ }\href {\doibase 10.1007/978-3-319-25607-8}
  {\emph {\bibinfo {title} {A Short Course on Topological Insulators: Band
  Structure and Edge States in One and Two Dimensions}}}\ (\bibinfo
  {publisher} {Springer},\ \bibinfo {year} {2016})\BibitemShut {NoStop}%
\bibitem [{\citenamefont {Deng}\ \emph {et~al.}(2016)\citenamefont {Deng},
  \citenamefont {Chen}, \citenamefont {Panoiu},\ and\ \citenamefont
  {Ye}}]{dengOL2016}%
  \BibitemOpen
  \bibfield  {author} {\bibinfo {author} {\bibfnamefont {H.}~\bibnamefont
  {Deng}}, \bibinfo {author} {\bibfnamefont {X.}~\bibnamefont {Chen}}, \bibinfo
  {author} {\bibfnamefont {N.~C.}\ \bibnamefont {Panoiu}}, \ and\ \bibinfo
  {author} {\bibfnamefont {F.}~\bibnamefont {Ye}},\ }\href {\doibase
  10.1364/OL.41.004281} {\bibfield  {journal} {\bibinfo  {journal} {Opt.
  Lett.}\ }\textbf {\bibinfo {volume} {41}},\ \bibinfo {pages} {4281} (\bibinfo
  {year} {2016})}\BibitemShut {NoStop}%
\bibitem [{\citenamefont {Meier}\ \emph {et~al.}(2016)\citenamefont {Meier},
  \citenamefont {An},\ and\ \citenamefont {Gadway}}]{meierNC2016}%
  \BibitemOpen
  \bibfield  {author} {\bibinfo {author} {\bibfnamefont {E.~J.}\ \bibnamefont
  {Meier}}, \bibinfo {author} {\bibfnamefont {F.~A.}\ \bibnamefont {An}}, \
  and\ \bibinfo {author} {\bibfnamefont {B.}~\bibnamefont {Gadway}},\
  }\href@noop {} {\bibfield  {journal} {\bibinfo  {journal} {Nature
  communications}\ }\textbf {\bibinfo {volume} {7}},\ \bibinfo {pages} {13986}
  (\bibinfo {year} {2016})}\BibitemShut {NoStop}%
\bibitem [{\citenamefont {Perczel}\ \emph
  {et~al.}(2017{\natexlab{a}})\citenamefont {Perczel}, \citenamefont
  {Borregaard}, \citenamefont {Chang}, \citenamefont {Pichler}, \citenamefont
  {Yelin}, \citenamefont {Zoller},\ and\ \citenamefont
  {Lukin}}]{yelinPRL20172}%
  \BibitemOpen
  \bibfield  {author} {\bibinfo {author} {\bibfnamefont {J.}~\bibnamefont
  {Perczel}}, \bibinfo {author} {\bibfnamefont {J.}~\bibnamefont {Borregaard}},
  \bibinfo {author} {\bibfnamefont {D.~E.}\ \bibnamefont {Chang}}, \bibinfo
  {author} {\bibfnamefont {H.}~\bibnamefont {Pichler}}, \bibinfo {author}
  {\bibfnamefont {S.~F.}\ \bibnamefont {Yelin}}, \bibinfo {author}
  {\bibfnamefont {P.}~\bibnamefont {Zoller}}, \ and\ \bibinfo {author}
  {\bibfnamefont {M.~D.}\ \bibnamefont {Lukin}},\ }\href {\doibase
  10.1103/PhysRevLett.119.023603} {\bibfield  {journal} {\bibinfo  {journal}
  {Phys. Rev. Lett.}\ }\textbf {\bibinfo {volume} {119}},\ \bibinfo {pages}
  {023603} (\bibinfo {year} {2017}{\natexlab{a}})}\BibitemShut {NoStop}%
\bibitem [{\citenamefont {Perczel}\ \emph
  {et~al.}(2017{\natexlab{b}})\citenamefont {Perczel}, \citenamefont
  {Borregaard}, \citenamefont {Chang}, \citenamefont {Pichler}, \citenamefont
  {Yelin}, \citenamefont {Zoller},\ and\ \citenamefont
  {Lukin}}]{perczelPRA2017}%
  \BibitemOpen
  \bibfield  {author} {\bibinfo {author} {\bibfnamefont {J.}~\bibnamefont
  {Perczel}}, \bibinfo {author} {\bibfnamefont {J.}~\bibnamefont {Borregaard}},
  \bibinfo {author} {\bibfnamefont {D.~E.}\ \bibnamefont {Chang}}, \bibinfo
  {author} {\bibfnamefont {H.}~\bibnamefont {Pichler}}, \bibinfo {author}
  {\bibfnamefont {S.~F.}\ \bibnamefont {Yelin}}, \bibinfo {author}
  {\bibfnamefont {P.}~\bibnamefont {Zoller}}, \ and\ \bibinfo {author}
  {\bibfnamefont {M.~D.}\ \bibnamefont {Lukin}},\ }\href {\doibase
  10.1103/PhysRevA.96.063801} {\bibfield  {journal} {\bibinfo  {journal} {Phys.
  Rev. A}\ }\textbf {\bibinfo {volume} {96}},\ \bibinfo {pages} {063801}
  (\bibinfo {year} {2017}{\natexlab{b}})}\BibitemShut {NoStop}%
\bibitem [{\citenamefont {Bloch}(2005)}]{blochNaturephys2005}%
  \BibitemOpen
  \bibfield  {author} {\bibinfo {author} {\bibfnamefont {I.}~\bibnamefont
  {Bloch}},\ }\href@noop {} {\bibfield  {journal} {\bibinfo  {journal} {Nature
  Physics}\ }\textbf {\bibinfo {volume} {1}},\ \bibinfo {pages} {23} (\bibinfo
  {year} {2005})}\BibitemShut {NoStop}%
\bibitem [{\citenamefont {Lester}\ \emph {et~al.}(2015)\citenamefont {Lester},
  \citenamefont {Luick}, \citenamefont {Kaufman}, \citenamefont {Reynolds},\
  and\ \citenamefont {Regal}}]{lesterPRL2015}%
  \BibitemOpen
  \bibfield  {author} {\bibinfo {author} {\bibfnamefont {B.~J.}\ \bibnamefont
  {Lester}}, \bibinfo {author} {\bibfnamefont {N.}~\bibnamefont {Luick}},
  \bibinfo {author} {\bibfnamefont {A.~M.}\ \bibnamefont {Kaufman}}, \bibinfo
  {author} {\bibfnamefont {C.~M.}\ \bibnamefont {Reynolds}}, \ and\ \bibinfo
  {author} {\bibfnamefont {C.~A.}\ \bibnamefont {Regal}},\ }\href@noop {}
  {\bibfield  {journal} {\bibinfo  {journal} {Physical review letters}\
  }\textbf {\bibinfo {volume} {115}},\ \bibinfo {pages} {073003} (\bibinfo
  {year} {2015})}\BibitemShut {NoStop}%
\bibitem [{\citenamefont {Goldman}\ \emph {et~al.}(2016)\citenamefont
  {Goldman}, \citenamefont {Budich},\ and\ \citenamefont
  {Zoller}}]{zollerNaturephys2016}%
  \BibitemOpen
  \bibfield  {author} {\bibinfo {author} {\bibfnamefont {N.}~\bibnamefont
  {Goldman}}, \bibinfo {author} {\bibfnamefont {J.}~\bibnamefont {Budich}}, \
  and\ \bibinfo {author} {\bibfnamefont {P.}~\bibnamefont {Zoller}},\
  }\href@noop {} {\bibfield  {journal} {\bibinfo  {journal} {Nature Physics}\
  }\textbf {\bibinfo {volume} {12}},\ \bibinfo {pages} {639} (\bibinfo {year}
  {2016})}\BibitemShut {NoStop}%
\bibitem [{\citenamefont {Greiner}\ \emph {et~al.}(2002)\citenamefont
  {Greiner}, \citenamefont {Mandel}, \citenamefont {Esslinger}, \citenamefont
  {H{\"a}nsch},\ and\ \citenamefont {Bloch}}]{greinerNature2002}%
  \BibitemOpen
  \bibfield  {author} {\bibinfo {author} {\bibfnamefont {M.}~\bibnamefont
  {Greiner}}, \bibinfo {author} {\bibfnamefont {O.}~\bibnamefont {Mandel}},
  \bibinfo {author} {\bibfnamefont {T.}~\bibnamefont {Esslinger}}, \bibinfo
  {author} {\bibfnamefont {T.~W.}\ \bibnamefont {H{\"a}nsch}}, \ and\ \bibinfo
  {author} {\bibfnamefont {I.}~\bibnamefont {Bloch}},\ }\href@noop {}
  {\bibfield  {journal} {\bibinfo  {journal} {nature}\ }\textbf {\bibinfo
  {volume} {415}},\ \bibinfo {pages} {39} (\bibinfo {year} {2002})}\BibitemShut
  {NoStop}%
\bibitem [{\citenamefont {Sherson}\ \emph {et~al.}(2010)\citenamefont
  {Sherson}, \citenamefont {Weitenberg}, \citenamefont {Endres}, \citenamefont
  {Cheneau}, \citenamefont {Bloch},\ and\ \citenamefont
  {Kuhr}}]{shersonNature2010}%
  \BibitemOpen
  \bibfield  {author} {\bibinfo {author} {\bibfnamefont {J.~F.}\ \bibnamefont
  {Sherson}}, \bibinfo {author} {\bibfnamefont {C.}~\bibnamefont {Weitenberg}},
  \bibinfo {author} {\bibfnamefont {M.}~\bibnamefont {Endres}}, \bibinfo
  {author} {\bibfnamefont {M.}~\bibnamefont {Cheneau}}, \bibinfo {author}
  {\bibfnamefont {I.}~\bibnamefont {Bloch}}, \ and\ \bibinfo {author}
  {\bibfnamefont {S.}~\bibnamefont {Kuhr}},\ }\href@noop {} {\bibfield
  {journal} {\bibinfo  {journal} {Nature}\ }\textbf {\bibinfo {volume} {467}},\
  \bibinfo {pages} {68} (\bibinfo {year} {2010})}\BibitemShut {NoStop}%
\bibitem [{\citenamefont {Bakr}\ \emph {et~al.}(2011)\citenamefont {Bakr},
  \citenamefont {Preiss}, \citenamefont {Tai}, \citenamefont {Ma},
  \citenamefont {Simon},\ and\ \citenamefont {Greiner}}]{bakrNature2011}%
  \BibitemOpen
  \bibfield  {author} {\bibinfo {author} {\bibfnamefont {W.~S.}\ \bibnamefont
  {Bakr}}, \bibinfo {author} {\bibfnamefont {P.~M.}\ \bibnamefont {Preiss}},
  \bibinfo {author} {\bibfnamefont {M.~E.}\ \bibnamefont {Tai}}, \bibinfo
  {author} {\bibfnamefont {R.}~\bibnamefont {Ma}}, \bibinfo {author}
  {\bibfnamefont {J.}~\bibnamefont {Simon}}, \ and\ \bibinfo {author}
  {\bibfnamefont {M.}~\bibnamefont {Greiner}},\ }\href@noop {} {\bibfield
  {journal} {\bibinfo  {journal} {Nature}\ }\textbf {\bibinfo {volume} {480}},\
  \bibinfo {pages} {500} (\bibinfo {year} {2011})}\BibitemShut {NoStop}%
\bibitem [{\citenamefont {Bakr}\ \emph {et~al.}(2010)\citenamefont {Bakr},
  \citenamefont {Peng}, \citenamefont {Tai}, \citenamefont {Ma}, \citenamefont
  {Simon}, \citenamefont {Gillen}, \citenamefont {Foelling}, \citenamefont
  {Pollet},\ and\ \citenamefont {Greiner}}]{bakrScience2010}%
  \BibitemOpen
  \bibfield  {author} {\bibinfo {author} {\bibfnamefont {W.~S.}\ \bibnamefont
  {Bakr}}, \bibinfo {author} {\bibfnamefont {A.}~\bibnamefont {Peng}}, \bibinfo
  {author} {\bibfnamefont {M.~E.}\ \bibnamefont {Tai}}, \bibinfo {author}
  {\bibfnamefont {R.}~\bibnamefont {Ma}}, \bibinfo {author} {\bibfnamefont
  {J.}~\bibnamefont {Simon}}, \bibinfo {author} {\bibfnamefont {J.~I.}\
  \bibnamefont {Gillen}}, \bibinfo {author} {\bibfnamefont {S.}~\bibnamefont
  {Foelling}}, \bibinfo {author} {\bibfnamefont {L.}~\bibnamefont {Pollet}}, \
  and\ \bibinfo {author} {\bibfnamefont {M.}~\bibnamefont {Greiner}},\
  }\href@noop {} {\bibfield  {journal} {\bibinfo  {journal} {Science}\ }\textbf
  {\bibinfo {volume} {329}},\ \bibinfo {pages} {547} (\bibinfo {year}
  {2010})}\BibitemShut {NoStop}%
\bibitem [{\citenamefont {Weitenberg}\ \emph {et~al.}(2011)\citenamefont
  {Weitenberg}, \citenamefont {Endres}, \citenamefont {Sherson}, \citenamefont
  {Cheneau}, \citenamefont {Schau{\ss}}, \citenamefont {Fukuhara},
  \citenamefont {Bloch},\ and\ \citenamefont {Kuhr}}]{weitenbergNature2011}%
  \BibitemOpen
  \bibfield  {author} {\bibinfo {author} {\bibfnamefont {C.}~\bibnamefont
  {Weitenberg}}, \bibinfo {author} {\bibfnamefont {M.}~\bibnamefont {Endres}},
  \bibinfo {author} {\bibfnamefont {J.~F.}\ \bibnamefont {Sherson}}, \bibinfo
  {author} {\bibfnamefont {M.}~\bibnamefont {Cheneau}}, \bibinfo {author}
  {\bibfnamefont {P.}~\bibnamefont {Schau{\ss}}}, \bibinfo {author}
  {\bibfnamefont {T.}~\bibnamefont {Fukuhara}}, \bibinfo {author}
  {\bibfnamefont {I.}~\bibnamefont {Bloch}}, \ and\ \bibinfo {author}
  {\bibfnamefont {S.}~\bibnamefont {Kuhr}},\ }\href@noop {} {\bibfield
  {journal} {\bibinfo  {journal} {Nature}\ }\textbf {\bibinfo {volume} {471}},\
  \bibinfo {pages} {319} (\bibinfo {year} {2011})}\BibitemShut {NoStop}%
\bibitem [{\citenamefont {Endres}\ \emph {et~al.}(2016)\citenamefont {Endres},
  \citenamefont {Bernien}, \citenamefont {Keesling}, \citenamefont {Levine},
  \citenamefont {Anschuetz}, \citenamefont {Krajenbrink}, \citenamefont
  {Senko}, \citenamefont {Vuletic}, \citenamefont {Greiner},\ and\
  \citenamefont {Lukin}}]{endresScience2016}%
  \BibitemOpen
  \bibfield  {author} {\bibinfo {author} {\bibfnamefont {M.}~\bibnamefont
  {Endres}}, \bibinfo {author} {\bibfnamefont {H.}~\bibnamefont {Bernien}},
  \bibinfo {author} {\bibfnamefont {A.}~\bibnamefont {Keesling}}, \bibinfo
  {author} {\bibfnamefont {H.}~\bibnamefont {Levine}}, \bibinfo {author}
  {\bibfnamefont {E.~R.}\ \bibnamefont {Anschuetz}}, \bibinfo {author}
  {\bibfnamefont {A.}~\bibnamefont {Krajenbrink}}, \bibinfo {author}
  {\bibfnamefont {C.}~\bibnamefont {Senko}}, \bibinfo {author} {\bibfnamefont
  {V.}~\bibnamefont {Vuletic}}, \bibinfo {author} {\bibfnamefont
  {M.}~\bibnamefont {Greiner}}, \ and\ \bibinfo {author} {\bibfnamefont
  {M.~D.}\ \bibnamefont {Lukin}},\ }\href@noop {} {\bibfield  {journal}
  {\bibinfo  {journal} {Science}\ ,\ \bibinfo {pages} {aah3752}} (\bibinfo
  {year} {2016})}\BibitemShut {NoStop}%
\bibitem [{\citenamefont {Barredo}\ \emph {et~al.}(2016)\citenamefont
  {Barredo}, \citenamefont {De~L{\'e}s{\'e}leuc}, \citenamefont {Lienhard},
  \citenamefont {Lahaye},\ and\ \citenamefont {Browaeys}}]{lahayeScience2016}%
  \BibitemOpen
  \bibfield  {author} {\bibinfo {author} {\bibfnamefont {D.}~\bibnamefont
  {Barredo}}, \bibinfo {author} {\bibfnamefont {S.}~\bibnamefont
  {De~L{\'e}s{\'e}leuc}}, \bibinfo {author} {\bibfnamefont {V.}~\bibnamefont
  {Lienhard}}, \bibinfo {author} {\bibfnamefont {T.}~\bibnamefont {Lahaye}}, \
  and\ \bibinfo {author} {\bibfnamefont {A.}~\bibnamefont {Browaeys}},\
  }\href@noop {} {\bibfield  {journal} {\bibinfo  {journal} {Science}\ }\textbf
  {\bibinfo {volume} {354}},\ \bibinfo {pages} {1021} (\bibinfo {year}
  {2016})}\BibitemShut {NoStop}%
\bibitem [{\citenamefont {Guerin}\ \emph {et~al.}(2017)\citenamefont {Guerin},
  \citenamefont {Rouabah},\ and\ \citenamefont {Kaiser}}]{guerin2017light}%
  \BibitemOpen
  \bibfield  {author} {\bibinfo {author} {\bibfnamefont {W.}~\bibnamefont
  {Guerin}}, \bibinfo {author} {\bibfnamefont {M.}~\bibnamefont {Rouabah}}, \
  and\ \bibinfo {author} {\bibfnamefont {R.}~\bibnamefont {Kaiser}},\
  }\href@noop {} {\bibfield  {journal} {\bibinfo  {journal} {Journal of Modern
  Optics}\ }\textbf {\bibinfo {volume} {64}},\ \bibinfo {pages} {895} (\bibinfo
  {year} {2017})}\BibitemShut {NoStop}%
\bibitem [{\citenamefont {Jenkins}\ and\ \citenamefont
  {Ruostekoski}(2012)}]{jenkinsPRA2012}%
  \BibitemOpen
  \bibfield  {author} {\bibinfo {author} {\bibfnamefont {S.~D.}\ \bibnamefont
  {Jenkins}}\ and\ \bibinfo {author} {\bibfnamefont {J.}~\bibnamefont
  {Ruostekoski}},\ }\href {\doibase 10.1103/PhysRevA.86.031602} {\bibfield
  {journal} {\bibinfo  {journal} {Phys. Rev. A}\ }\textbf {\bibinfo {volume}
  {86}},\ \bibinfo {pages} {031602} (\bibinfo {year} {2012})}\BibitemShut
  {NoStop}%
\bibitem [{\citenamefont {Bettles}\ \emph {et~al.}(2016)\citenamefont
  {Bettles}, \citenamefont {Gardiner},\ and\ \citenamefont
  {Adams}}]{bettlesPRL2016}%
  \BibitemOpen
  \bibfield  {author} {\bibinfo {author} {\bibfnamefont {R.~J.}\ \bibnamefont
  {Bettles}}, \bibinfo {author} {\bibfnamefont {S.~A.}\ \bibnamefont
  {Gardiner}}, \ and\ \bibinfo {author} {\bibfnamefont {C.~S.}\ \bibnamefont
  {Adams}},\ }\href {\doibase 10.1103/PhysRevLett.116.103602} {\bibfield
  {journal} {\bibinfo  {journal} {Phys. Rev. Lett.}\ }\textbf {\bibinfo
  {volume} {116}},\ \bibinfo {pages} {103602} (\bibinfo {year}
  {2016})}\BibitemShut {NoStop}%
\bibitem [{\citenamefont {Shahmoon}\ \emph {et~al.}(2017)\citenamefont
  {Shahmoon}, \citenamefont {Wild}, \citenamefont {Lukin},\ and\ \citenamefont
  {Yelin}}]{yelinPRL2017}%
  \BibitemOpen
  \bibfield  {author} {\bibinfo {author} {\bibfnamefont {E.}~\bibnamefont
  {Shahmoon}}, \bibinfo {author} {\bibfnamefont {D.~S.}\ \bibnamefont {Wild}},
  \bibinfo {author} {\bibfnamefont {M.~D.}\ \bibnamefont {Lukin}}, \ and\
  \bibinfo {author} {\bibfnamefont {S.~F.}\ \bibnamefont {Yelin}},\ }\href
  {\doibase 10.1103/PhysRevLett.118.113601} {\bibfield  {journal} {\bibinfo
  {journal} {Phys. Rev. Lett.}\ }\textbf {\bibinfo {volume} {118}},\ \bibinfo
  {pages} {113601} (\bibinfo {year} {2017})}\BibitemShut {NoStop}%
\bibitem [{\citenamefont {Wang}\ \emph {et~al.}(2017)\citenamefont {Wang},
  \citenamefont {Zhao}, \citenamefont {Kan},\ and\ \citenamefont
  {Huang}}]{wangOE2017}%
  \BibitemOpen
  \bibfield  {author} {\bibinfo {author} {\bibfnamefont {B.~X.}\ \bibnamefont
  {Wang}}, \bibinfo {author} {\bibfnamefont {C.~Y.}\ \bibnamefont {Zhao}},
  \bibinfo {author} {\bibfnamefont {Y.~H.}\ \bibnamefont {Kan}}, \ and\
  \bibinfo {author} {\bibfnamefont {T.~C.}\ \bibnamefont {Huang}},\ }\href
  {\doibase 10.1364/OE.25.018760} {\bibfield  {journal} {\bibinfo  {journal}
  {Opt. Express}\ }\textbf {\bibinfo {volume} {25}},\ \bibinfo {pages} {18760}
  (\bibinfo {year} {2017})}\BibitemShut {NoStop}%
\bibitem [{\citenamefont {Schilder}\ \emph {et~al.}(2016)\citenamefont
  {Schilder}, \citenamefont {Sauvan}, \citenamefont {Hugonin}, \citenamefont
  {Jennewein}, \citenamefont {Sortais}, \citenamefont {Browaeys},\ and\
  \citenamefont {Greffet}}]{Schilder2016}%
  \BibitemOpen
  \bibfield  {author} {\bibinfo {author} {\bibfnamefont {N.~J.}\ \bibnamefont
  {Schilder}}, \bibinfo {author} {\bibfnamefont {C.}~\bibnamefont {Sauvan}},
  \bibinfo {author} {\bibfnamefont {J.-P.}\ \bibnamefont {Hugonin}}, \bibinfo
  {author} {\bibfnamefont {S.}~\bibnamefont {Jennewein}}, \bibinfo {author}
  {\bibfnamefont {Y.~R.~P.}\ \bibnamefont {Sortais}}, \bibinfo {author}
  {\bibfnamefont {A.}~\bibnamefont {Browaeys}}, \ and\ \bibinfo {author}
  {\bibfnamefont {J.-J.}\ \bibnamefont {Greffet}},\ }\href {\doibase
  10.1103/PhysRevA.93.063835} {\bibfield  {journal} {\bibinfo  {journal} {Phys.
  Rev. A}\ }\textbf {\bibinfo {volume} {93}},\ \bibinfo {pages} {063835}
  (\bibinfo {year} {2016})}\BibitemShut {NoStop}%
\bibitem [{\citenamefont {Chang}\ \emph {et~al.}(2007)\citenamefont {Chang},
  \citenamefont {S{\o}rensen}, \citenamefont {Demler},\ and\ \citenamefont
  {Lukin}}]{changNaturephys2007}%
  \BibitemOpen
  \bibfield  {author} {\bibinfo {author} {\bibfnamefont {D.~E.}\ \bibnamefont
  {Chang}}, \bibinfo {author} {\bibfnamefont {A.~S.}\ \bibnamefont
  {S{\o}rensen}}, \bibinfo {author} {\bibfnamefont {E.~A.}\ \bibnamefont
  {Demler}}, \ and\ \bibinfo {author} {\bibfnamefont {M.~D.}\ \bibnamefont
  {Lukin}},\ }\href@noop {} {\bibfield  {journal} {\bibinfo  {journal} {Nature
  Physics}\ }\textbf {\bibinfo {volume} {3}},\ \bibinfo {pages} {807} (\bibinfo
  {year} {2007})}\BibitemShut {NoStop}%
\bibitem [{\citenamefont {Downing}\ and\ \citenamefont
  {Weick}(2018)}]{downing2018topological}%
  \BibitemOpen
  \bibfield  {author} {\bibinfo {author} {\bibfnamefont {C.~A.}\ \bibnamefont
  {Downing}}\ and\ \bibinfo {author} {\bibfnamefont {G.}~\bibnamefont
  {Weick}},\ }\href@noop {} {\bibfield  {journal} {\bibinfo  {journal} {arXiv
  preprint arXiv:1803.08872}\ } (\bibinfo {year} {2018})}\BibitemShut {NoStop}%
\bibitem [{\citenamefont {Pocock}\ \emph {et~al.}(0)\citenamefont {Pocock},
  \citenamefont {XIAO}, \citenamefont {Huidobro},\ and\ \citenamefont
  {Giannini}}]{pocockArxiv2017}%
  \BibitemOpen
  \bibfield  {author} {\bibinfo {author} {\bibfnamefont {S.}~\bibnamefont
  {Pocock}}, \bibinfo {author} {\bibfnamefont {X.}~\bibnamefont {XIAO}},
  \bibinfo {author} {\bibfnamefont {P.~A.}\ \bibnamefont {Huidobro}}, \ and\
  \bibinfo {author} {\bibfnamefont {V.}~\bibnamefont {Giannini}},\ }\href
  {\doibase 10.1021/acsphotonics.8b00117} {\bibfield  {journal} {\bibinfo
  {journal} {ACS Photonics}\ }\textbf {\bibinfo {volume} {0}},\ \bibinfo
  {pages} {null} (\bibinfo {year} {0})},\ \Eprint
  {http://arxiv.org/abs/https://doi.org/10.1021/acsphotonics.8b00117}
  {https://doi.org/10.1021/acsphotonics.8b00117} \BibitemShut {NoStop}%
\bibitem [{\citenamefont {Garrison}\ and\ \citenamefont
  {Wright}(1988)}]{garrisonPLA1988}%
  \BibitemOpen
  \bibfield  {author} {\bibinfo {author} {\bibfnamefont {J.}~\bibnamefont
  {Garrison}}\ and\ \bibinfo {author} {\bibfnamefont {E.}~\bibnamefont
  {Wright}},\ }\href {\doibase https://doi.org/10.1016/0375-9601(88)90905-X}
  {\bibfield  {journal} {\bibinfo  {journal} {Physics Letters A}\ }\textbf
  {\bibinfo {volume} {128}},\ \bibinfo {pages} {177 } (\bibinfo {year}
  {1988})}\BibitemShut {NoStop}%
\bibitem [{\citenamefont {Rudner}\ and\ \citenamefont
  {Levitov}(2009)}]{rudnerPRL2009}%
  \BibitemOpen
  \bibfield  {author} {\bibinfo {author} {\bibfnamefont {M.~S.}\ \bibnamefont
  {Rudner}}\ and\ \bibinfo {author} {\bibfnamefont {L.~S.}\ \bibnamefont
  {Levitov}},\ }\href {\doibase 10.1103/PhysRevLett.102.065703} {\bibfield
  {journal} {\bibinfo  {journal} {Phys. Rev. Lett.}\ }\textbf {\bibinfo
  {volume} {102}},\ \bibinfo {pages} {065703} (\bibinfo {year}
  {2009})}\BibitemShut {NoStop}%
\bibitem [{\citenamefont {Hu}\ and\ \citenamefont {Hughes}(2011)}]{huPRB2011}%
  \BibitemOpen
  \bibfield  {author} {\bibinfo {author} {\bibfnamefont {Y.~C.}\ \bibnamefont
  {Hu}}\ and\ \bibinfo {author} {\bibfnamefont {T.~L.}\ \bibnamefont
  {Hughes}},\ }\href {\doibase 10.1103/PhysRevB.84.153101} {\bibfield
  {journal} {\bibinfo  {journal} {Phys. Rev. B}\ }\textbf {\bibinfo {volume}
  {84}},\ \bibinfo {pages} {153101} (\bibinfo {year} {2011})}\BibitemShut
  {NoStop}%
\bibitem [{\citenamefont {Esaki}\ \emph {et~al.}(2011)\citenamefont {Esaki},
  \citenamefont {Sato}, \citenamefont {Hasebe},\ and\ \citenamefont
  {Kohmoto}}]{esakiPRB2011}%
  \BibitemOpen
  \bibfield  {author} {\bibinfo {author} {\bibfnamefont {K.}~\bibnamefont
  {Esaki}}, \bibinfo {author} {\bibfnamefont {M.}~\bibnamefont {Sato}},
  \bibinfo {author} {\bibfnamefont {K.}~\bibnamefont {Hasebe}}, \ and\ \bibinfo
  {author} {\bibfnamefont {M.}~\bibnamefont {Kohmoto}},\ }\href {\doibase
  10.1103/PhysRevB.84.205128} {\bibfield  {journal} {\bibinfo  {journal} {Phys.
  Rev. B}\ }\textbf {\bibinfo {volume} {84}},\ \bibinfo {pages} {205128}
  (\bibinfo {year} {2011})}\BibitemShut {NoStop}%
\bibitem [{\citenamefont {Liang}\ and\ \citenamefont
  {Huang}(2013)}]{liangPRA2013}%
  \BibitemOpen
  \bibfield  {author} {\bibinfo {author} {\bibfnamefont {S.-D.}\ \bibnamefont
  {Liang}}\ and\ \bibinfo {author} {\bibfnamefont {G.-Y.}\ \bibnamefont
  {Huang}},\ }\href {\doibase 10.1103/PhysRevA.87.012118} {\bibfield  {journal}
  {\bibinfo  {journal} {Phys. Rev. A}\ }\textbf {\bibinfo {volume} {87}},\
  \bibinfo {pages} {012118} (\bibinfo {year} {2013})}\BibitemShut {NoStop}%
\bibitem [{\citenamefont {Schomerus}(2013)}]{schomerusOL2013}%
  \BibitemOpen
  \bibfield  {author} {\bibinfo {author} {\bibfnamefont {H.}~\bibnamefont
  {Schomerus}},\ }\href {\doibase 10.1364/OL.38.001912} {\bibfield  {journal}
  {\bibinfo  {journal} {Opt. Lett.}\ }\textbf {\bibinfo {volume} {38}},\
  \bibinfo {pages} {1912} (\bibinfo {year} {2013})}\BibitemShut {NoStop}%
\bibitem [{\citenamefont {Lee}(2016)}]{leePRL2016}%
  \BibitemOpen
  \bibfield  {author} {\bibinfo {author} {\bibfnamefont {T.~E.}\ \bibnamefont
  {Lee}},\ }\href {\doibase 10.1103/PhysRevLett.116.133903} {\bibfield
  {journal} {\bibinfo  {journal} {Phys. Rev. Lett.}\ }\textbf {\bibinfo
  {volume} {116}},\ \bibinfo {pages} {133903} (\bibinfo {year}
  {2016})}\BibitemShut {NoStop}%
\bibitem [{\citenamefont {Ling}\ \emph {et~al.}(2016)\citenamefont {Ling},
  \citenamefont {Choi}, \citenamefont {Mok}, \citenamefont {Zhang},\ and\
  \citenamefont {Fung}}]{lingSR2016}%
  \BibitemOpen
  \bibfield  {author} {\bibinfo {author} {\bibfnamefont {C.}~\bibnamefont
  {Ling}}, \bibinfo {author} {\bibfnamefont {K.~H.}\ \bibnamefont {Choi}},
  \bibinfo {author} {\bibfnamefont {T.}~\bibnamefont {Mok}}, \bibinfo {author}
  {\bibfnamefont {Z.-Q.}\ \bibnamefont {Zhang}}, \ and\ \bibinfo {author}
  {\bibfnamefont {K.~H.}\ \bibnamefont {Fung}},\ }\href@noop {} {\bibfield
  {journal} {\bibinfo  {journal} {Scientific reports}\ }\textbf {\bibinfo
  {volume} {6}},\ \bibinfo {pages} {38049} (\bibinfo {year}
  {2016})}\BibitemShut {NoStop}%
\bibitem [{\citenamefont {Leykam}\ \emph {et~al.}(2017)\citenamefont {Leykam},
  \citenamefont {Bliokh}, \citenamefont {Huang}, \citenamefont {Chong},\ and\
  \citenamefont {Nori}}]{leykamPRL2017}%
  \BibitemOpen
  \bibfield  {author} {\bibinfo {author} {\bibfnamefont {D.}~\bibnamefont
  {Leykam}}, \bibinfo {author} {\bibfnamefont {K.~Y.}\ \bibnamefont {Bliokh}},
  \bibinfo {author} {\bibfnamefont {C.}~\bibnamefont {Huang}}, \bibinfo
  {author} {\bibfnamefont {Y.~D.}\ \bibnamefont {Chong}}, \ and\ \bibinfo
  {author} {\bibfnamefont {F.}~\bibnamefont {Nori}},\ }\href {\doibase
  10.1103/PhysRevLett.118.040401} {\bibfield  {journal} {\bibinfo  {journal}
  {Phys. Rev. Lett.}\ }\textbf {\bibinfo {volume} {118}},\ \bibinfo {pages}
  {040401} (\bibinfo {year} {2017})}\BibitemShut {NoStop}%
\bibitem [{\citenamefont {Jin}(2017)}]{jinPRA2017}%
  \BibitemOpen
  \bibfield  {author} {\bibinfo {author} {\bibfnamefont {L.}~\bibnamefont
  {Jin}},\ }\href {\doibase 10.1103/PhysRevA.96.032103} {\bibfield  {journal}
  {\bibinfo  {journal} {Phys. Rev. A}\ }\textbf {\bibinfo {volume} {96}},\
  \bibinfo {pages} {032103} (\bibinfo {year} {2017})}\BibitemShut {NoStop}%
\bibitem [{\citenamefont {Weimann}\ \emph {et~al.}(2017)\citenamefont
  {Weimann}, \citenamefont {Kremer}, \citenamefont {Plotnik}, \citenamefont
  {Lumer}, \citenamefont {Nolte}, \citenamefont {Makris}, \citenamefont
  {Segev}, \citenamefont {Rechtsman},\ and\ \citenamefont
  {Szameit}}]{weimannNaturemat2017}%
  \BibitemOpen
  \bibfield  {author} {\bibinfo {author} {\bibfnamefont {S.}~\bibnamefont
  {Weimann}}, \bibinfo {author} {\bibfnamefont {M.}~\bibnamefont {Kremer}},
  \bibinfo {author} {\bibfnamefont {Y.}~\bibnamefont {Plotnik}}, \bibinfo
  {author} {\bibfnamefont {Y.}~\bibnamefont {Lumer}}, \bibinfo {author}
  {\bibfnamefont {S.}~\bibnamefont {Nolte}}, \bibinfo {author} {\bibfnamefont
  {K.}~\bibnamefont {Makris}}, \bibinfo {author} {\bibfnamefont
  {M.}~\bibnamefont {Segev}}, \bibinfo {author} {\bibfnamefont
  {M.}~\bibnamefont {Rechtsman}}, \ and\ \bibinfo {author} {\bibfnamefont
  {A.}~\bibnamefont {Szameit}},\ }\href@noop {} {\bibfield  {journal} {\bibinfo
   {journal} {Nature materials}\ }\textbf {\bibinfo {volume} {16}},\ \bibinfo
  {pages} {433} (\bibinfo {year} {2017})}\BibitemShut {NoStop}%
\bibitem [{\citenamefont {Lieu}(2018)}]{lieuPRB2018}%
  \BibitemOpen
  \bibfield  {author} {\bibinfo {author} {\bibfnamefont {S.}~\bibnamefont
  {Lieu}},\ }\href {\doibase 10.1103/PhysRevB.97.045106} {\bibfield  {journal}
  {\bibinfo  {journal} {Phys. Rev. B}\ }\textbf {\bibinfo {volume} {97}},\
  \bibinfo {pages} {045106} (\bibinfo {year} {2018})}\BibitemShut {NoStop}%
\bibitem [{\citenamefont {Yuce}(2018)}]{yucePRA2018}%
  \BibitemOpen
  \bibfield  {author} {\bibinfo {author} {\bibfnamefont {C.}~\bibnamefont
  {Yuce}},\ }\href {\doibase 10.1103/PhysRevA.97.042118} {\bibfield  {journal}
  {\bibinfo  {journal} {Phys. Rev. A}\ }\textbf {\bibinfo {volume} {97}},\
  \bibinfo {pages} {042118} (\bibinfo {year} {2018})}\BibitemShut {NoStop}%
\bibitem [{\citenamefont {Xiong}(2018)}]{xiongJPC2018}%
  \BibitemOpen
  \bibfield  {author} {\bibinfo {author} {\bibfnamefont {Y.}~\bibnamefont
  {Xiong}},\ }\href {http://stacks.iop.org/2399-6528/2/i=3/a=035043} {\bibfield
   {journal} {\bibinfo  {journal} {Journal of Physics Communications}\ }\textbf
  {\bibinfo {volume} {2}},\ \bibinfo {pages} {035043} (\bibinfo {year}
  {2018})}\BibitemShut {NoStop}%
\bibitem [{\citenamefont {Shen}\ \emph {et~al.}(2018)\citenamefont {Shen},
  \citenamefont {Zhen},\ and\ \citenamefont {Fu}}]{shenPRL2018}%
  \BibitemOpen
  \bibfield  {author} {\bibinfo {author} {\bibfnamefont {H.}~\bibnamefont
  {Shen}}, \bibinfo {author} {\bibfnamefont {B.}~\bibnamefont {Zhen}}, \ and\
  \bibinfo {author} {\bibfnamefont {L.}~\bibnamefont {Fu}},\ }\href {\doibase
  10.1103/PhysRevLett.120.146402} {\bibfield  {journal} {\bibinfo  {journal}
  {Phys. Rev. Lett.}\ }\textbf {\bibinfo {volume} {120}},\ \bibinfo {pages}
  {146402} (\bibinfo {year} {2018})}\BibitemShut {NoStop}%
\bibitem [{\citenamefont {Yao}\ and\ \citenamefont {Wang}(2018)}]{yao2018edge}%
  \BibitemOpen
  \bibfield  {author} {\bibinfo {author} {\bibfnamefont {S.}~\bibnamefont
  {Yao}}\ and\ \bibinfo {author} {\bibfnamefont {Z.}~\bibnamefont {Wang}},\
  }\href@noop {} {\bibfield  {journal} {\bibinfo  {journal} {arXiv preprint
  arXiv:1803.01876}\ } (\bibinfo {year} {2018})}\BibitemShut {NoStop}%
\bibitem [{\citenamefont {Yin}\ \emph {et~al.}(2018)\citenamefont {Yin},
  \citenamefont {Jiang}, \citenamefont {Li}, \citenamefont {L\"u},\ and\
  \citenamefont {Chen}}]{yinPRA2018}%
  \BibitemOpen
  \bibfield  {author} {\bibinfo {author} {\bibfnamefont {C.}~\bibnamefont
  {Yin}}, \bibinfo {author} {\bibfnamefont {H.}~\bibnamefont {Jiang}}, \bibinfo
  {author} {\bibfnamefont {L.}~\bibnamefont {Li}}, \bibinfo {author}
  {\bibfnamefont {R.}~\bibnamefont {L\"u}}, \ and\ \bibinfo {author}
  {\bibfnamefont {S.}~\bibnamefont {Chen}},\ }\href {\doibase
  10.1103/PhysRevA.97.052115} {\bibfield  {journal} {\bibinfo  {journal} {Phys.
  Rev. A}\ }\textbf {\bibinfo {volume} {97}},\ \bibinfo {pages} {052115}
  (\bibinfo {year} {2018})}\BibitemShut {NoStop}%
\bibitem [{\citenamefont {Alvarez}\ \emph {et~al.}(2018)\citenamefont
  {Alvarez}, \citenamefont {Vargas}, \citenamefont {Berdakin},\ and\
  \citenamefont {Torres}}]{alvarez2018topologicalreview}%
  \BibitemOpen
  \bibfield  {author} {\bibinfo {author} {\bibfnamefont {V.}~\bibnamefont
  {Alvarez}}, \bibinfo {author} {\bibfnamefont {J.}~\bibnamefont {Vargas}},
  \bibinfo {author} {\bibfnamefont {M.}~\bibnamefont {Berdakin}}, \ and\
  \bibinfo {author} {\bibfnamefont {L.}~\bibnamefont {Torres}},\ }\href@noop {}
  {\bibfield  {journal} {\bibinfo  {journal} {arXiv preprint arXiv:1805.08200}\
  } (\bibinfo {year} {2018})}\BibitemShut {NoStop}%
\bibitem [{\citenamefont {Dangel}\ \emph {et~al.}(2018)\citenamefont {Dangel},
  \citenamefont {Wagner}, \citenamefont {Cartarius}, \citenamefont {Main},\
  and\ \citenamefont {Wunner}}]{dangel2018topological}%
  \BibitemOpen
  \bibfield  {author} {\bibinfo {author} {\bibfnamefont {F.}~\bibnamefont
  {Dangel}}, \bibinfo {author} {\bibfnamefont {M.}~\bibnamefont {Wagner}},
  \bibinfo {author} {\bibfnamefont {H.}~\bibnamefont {Cartarius}}, \bibinfo
  {author} {\bibfnamefont {J.}~\bibnamefont {Main}}, \ and\ \bibinfo {author}
  {\bibfnamefont {G.}~\bibnamefont {Wunner}},\ }\href@noop {} {\bibfield
  {journal} {\bibinfo  {journal} {arXiv preprint arXiv:1803.02636}\ } (\bibinfo
  {year} {2018})}\BibitemShut {NoStop}%
\bibitem [{\citenamefont {Kunst}\ \emph {et~al.}(2018)\citenamefont {Kunst},
  \citenamefont {Edvardsson}, \citenamefont {Budich},\ and\ \citenamefont
  {Bergholtz}}]{kunst2018biorthogonal}%
  \BibitemOpen
  \bibfield  {author} {\bibinfo {author} {\bibfnamefont {F.~K.}\ \bibnamefont
  {Kunst}}, \bibinfo {author} {\bibfnamefont {E.}~\bibnamefont {Edvardsson}},
  \bibinfo {author} {\bibfnamefont {J.~C.}\ \bibnamefont {Budich}}, \ and\
  \bibinfo {author} {\bibfnamefont {E.~J.}\ \bibnamefont {Bergholtz}},\
  }\href@noop {} {\bibfield  {journal} {\bibinfo  {journal} {arXiv preprint
  arXiv:1805.06492}\ } (\bibinfo {year} {2018})}\BibitemShut {NoStop}%
\bibitem [{\citenamefont {Gong}\ \emph {et~al.}(2018)\citenamefont {Gong},
  \citenamefont {Ashida}, \citenamefont {Kawabata}, \citenamefont {Takasan},
  \citenamefont {Higashikawa},\ and\ \citenamefont
  {Ueda}}]{gong2018topological}%
  \BibitemOpen
  \bibfield  {author} {\bibinfo {author} {\bibfnamefont {Z.}~\bibnamefont
  {Gong}}, \bibinfo {author} {\bibfnamefont {Y.}~\bibnamefont {Ashida}},
  \bibinfo {author} {\bibfnamefont {K.}~\bibnamefont {Kawabata}}, \bibinfo
  {author} {\bibfnamefont {K.}~\bibnamefont {Takasan}}, \bibinfo {author}
  {\bibfnamefont {S.}~\bibnamefont {Higashikawa}}, \ and\ \bibinfo {author}
  {\bibfnamefont {M.}~\bibnamefont {Ueda}},\ }\href@noop {} {\bibfield
  {journal} {\bibinfo  {journal} {arXiv preprint arXiv:1802.07964}\ } (\bibinfo
  {year} {2018})}\BibitemShut {NoStop}%
\bibitem [{\citenamefont {Kawabata}\ \emph {et~al.}(2018)\citenamefont
  {Kawabata}, \citenamefont {Shiozaki},\ and\ \citenamefont
  {Ueda}}]{kawabata2018nonhermitian}%
  \BibitemOpen
  \bibfield  {author} {\bibinfo {author} {\bibfnamefont {K.}~\bibnamefont
  {Kawabata}}, \bibinfo {author} {\bibfnamefont {K.}~\bibnamefont {Shiozaki}},
  \ and\ \bibinfo {author} {\bibfnamefont {M.}~\bibnamefont {Ueda}},\
  }\href@noop {} {\bibfield  {journal} {\bibinfo  {journal} {arXiv preprint
  arXiv:1805.09632}\ } (\bibinfo {year} {2018})}\BibitemShut {NoStop}%
\bibitem [{\citenamefont {Ryu}\ \emph {et~al.}(2010)\citenamefont {Ryu},
  \citenamefont {Schnyder}, \citenamefont {Furusaki},\ and\ \citenamefont
  {Ludwig}}]{ryuNJP2010}%
  \BibitemOpen
  \bibfield  {author} {\bibinfo {author} {\bibfnamefont {S.}~\bibnamefont
  {Ryu}}, \bibinfo {author} {\bibfnamefont {A.~P.}\ \bibnamefont {Schnyder}},
  \bibinfo {author} {\bibfnamefont {A.}~\bibnamefont {Furusaki}}, \ and\
  \bibinfo {author} {\bibfnamefont {A.~W.~W.}\ \bibnamefont {Ludwig}},\ }\href
  {http://stacks.iop.org/1367-2630/12/i=6/a=065010} {\bibfield  {journal}
  {\bibinfo  {journal} {New Journal of Physics}\ }\textbf {\bibinfo {volume}
  {12}},\ \bibinfo {pages} {065010} (\bibinfo {year} {2010})}\BibitemShut
  {NoStop}%
\bibitem [{\citenamefont {Antezza}\ and\ \citenamefont
  {Castin}(2009{\natexlab{a}})}]{antezzaPRA2009}%
  \BibitemOpen
  \bibfield  {author} {\bibinfo {author} {\bibfnamefont {M.}~\bibnamefont
  {Antezza}}\ and\ \bibinfo {author} {\bibfnamefont {Y.}~\bibnamefont
  {Castin}},\ }\href {\doibase 10.1103/PhysRevA.80.013816} {\bibfield
  {journal} {\bibinfo  {journal} {Phys. Rev. A}\ }\textbf {\bibinfo {volume}
  {80}},\ \bibinfo {pages} {013816} (\bibinfo {year}
  {2009}{\natexlab{a}})}\BibitemShut {NoStop}%
\bibitem [{\citenamefont {Bienaimé}\ \emph {et~al.}(2011)\citenamefont
  {Bienaimé}, \citenamefont {Petruzzo}, \citenamefont {Bigerni}, \citenamefont
  {Piovella},\ and\ \citenamefont {Kaiser}}]{kaiserJMO2011}%
  \BibitemOpen
  \bibfield  {author} {\bibinfo {author} {\bibfnamefont {T.}~\bibnamefont
  {Bienaimé}}, \bibinfo {author} {\bibfnamefont {M.}~\bibnamefont {Petruzzo}},
  \bibinfo {author} {\bibfnamefont {D.}~\bibnamefont {Bigerni}}, \bibinfo
  {author} {\bibfnamefont {N.}~\bibnamefont {Piovella}}, \ and\ \bibinfo
  {author} {\bibfnamefont {R.}~\bibnamefont {Kaiser}},\ }\href {\doibase
  10.1080/09500340.2011.594911} {\bibfield  {journal} {\bibinfo  {journal}
  {Journal of Modern Optics}\ }\textbf {\bibinfo {volume} {58}},\ \bibinfo
  {pages} {1942} (\bibinfo {year} {2011})}\BibitemShut {NoStop}%
\bibitem [{\citenamefont {Bienaimé}\ \emph {et~al.}(2012)\citenamefont
  {Bienaimé}, \citenamefont {Bachelard}, \citenamefont {Piovella},\ and\
  \citenamefont {Kaiser}}]{kaiserFP2012}%
  \BibitemOpen
  \bibfield  {author} {\bibinfo {author} {\bibfnamefont {T.}~\bibnamefont
  {Bienaimé}}, \bibinfo {author} {\bibfnamefont {R.}~\bibnamefont
  {Bachelard}}, \bibinfo {author} {\bibfnamefont {N.}~\bibnamefont {Piovella}},
  \ and\ \bibinfo {author} {\bibfnamefont {R.}~\bibnamefont {Kaiser}},\ }\href
  {\doibase 10.1002/prop.201200089} {\bibfield  {journal} {\bibinfo  {journal}
  {Fortschritte der Physik}\ }\textbf {\bibinfo {volume} {61}},\ \bibinfo
  {pages} {377} (\bibinfo {year} {2012})}\BibitemShut {NoStop}%
\bibitem [{\citenamefont {Guerin}\ \emph {et~al.}(2016)\citenamefont {Guerin},
  \citenamefont {Ara\'ujo},\ and\ \citenamefont {Kaiser}}]{guerinPRL2016}%
  \BibitemOpen
  \bibfield  {author} {\bibinfo {author} {\bibfnamefont {W.}~\bibnamefont
  {Guerin}}, \bibinfo {author} {\bibfnamefont {M.~O.}\ \bibnamefont
  {Ara\'ujo}}, \ and\ \bibinfo {author} {\bibfnamefont {R.}~\bibnamefont
  {Kaiser}},\ }\href {\doibase 10.1103/PhysRevLett.116.083601} {\bibfield
  {journal} {\bibinfo  {journal} {Phys. Rev. Lett.}\ }\textbf {\bibinfo
  {volume} {116}},\ \bibinfo {pages} {083601} (\bibinfo {year}
  {2016})}\BibitemShut {NoStop}%
\bibitem [{\citenamefont {Antezza}\ and\ \citenamefont
  {Castin}(2009{\natexlab{b}})}]{antezzaPRL2009}%
  \BibitemOpen
  \bibfield  {author} {\bibinfo {author} {\bibfnamefont {M.}~\bibnamefont
  {Antezza}}\ and\ \bibinfo {author} {\bibfnamefont {Y.}~\bibnamefont
  {Castin}},\ }\href {\doibase 10.1103/PhysRevLett.103.123903} {\bibfield
  {journal} {\bibinfo  {journal} {Phys. Rev. Lett.}\ }\textbf {\bibinfo
  {volume} {103}},\ \bibinfo {pages} {123903} (\bibinfo {year}
  {2009}{\natexlab{b}})}\BibitemShut {NoStop}%
\bibitem [{\citenamefont {Svidzinsky}\ \emph {et~al.}(2010)\citenamefont
  {Svidzinsky}, \citenamefont {Chang},\ and\ \citenamefont
  {Scully}}]{svidzinskyPRA2010}%
  \BibitemOpen
  \bibfield  {author} {\bibinfo {author} {\bibfnamefont {A.~A.}\ \bibnamefont
  {Svidzinsky}}, \bibinfo {author} {\bibfnamefont {J.-T.}\ \bibnamefont
  {Chang}}, \ and\ \bibinfo {author} {\bibfnamefont {M.~O.}\ \bibnamefont
  {Scully}},\ }\href {\doibase 10.1103/PhysRevA.81.053821} {\bibfield
  {journal} {\bibinfo  {journal} {Phys. Rev. A}\ }\textbf {\bibinfo {volume}
  {81}},\ \bibinfo {pages} {053821} (\bibinfo {year} {2010})}\BibitemShut
  {NoStop}%
\bibitem [{\citenamefont {Weber}\ and\ \citenamefont
  {Ford}(2004)}]{weberPRB2004}%
  \BibitemOpen
  \bibfield  {author} {\bibinfo {author} {\bibfnamefont {W.~H.}\ \bibnamefont
  {Weber}}\ and\ \bibinfo {author} {\bibfnamefont {G.~W.}\ \bibnamefont
  {Ford}},\ }\href {\doibase 10.1103/PhysRevB.70.125429} {\bibfield  {journal}
  {\bibinfo  {journal} {Phys. Rev. B}\ }\textbf {\bibinfo {volume} {70}},\
  \bibinfo {pages} {125429} (\bibinfo {year} {2004})}\BibitemShut {NoStop}%
\bibitem [{\citenamefont {Atala}\ \emph {et~al.}(2013)\citenamefont {Atala},
  \citenamefont {Aidelsburger}, \citenamefont {Barreiro}, \citenamefont
  {Abanin}, \citenamefont {Kitagawa}, \citenamefont {Demler},\ and\
  \citenamefont {Bloch}}]{atalaNaturephys2013}%
  \BibitemOpen
  \bibfield  {author} {\bibinfo {author} {\bibfnamefont {M.}~\bibnamefont
  {Atala}}, \bibinfo {author} {\bibfnamefont {M.}~\bibnamefont {Aidelsburger}},
  \bibinfo {author} {\bibfnamefont {J.~T.}\ \bibnamefont {Barreiro}}, \bibinfo
  {author} {\bibfnamefont {D.}~\bibnamefont {Abanin}}, \bibinfo {author}
  {\bibfnamefont {T.}~\bibnamefont {Kitagawa}}, \bibinfo {author}
  {\bibfnamefont {E.}~\bibnamefont {Demler}}, \ and\ \bibinfo {author}
  {\bibfnamefont {I.}~\bibnamefont {Bloch}},\ }\href@noop {} {\bibfield
  {journal} {\bibinfo  {journal} {Nature Physics}\ }\textbf {\bibinfo {volume}
  {9}},\ \bibinfo {pages} {795} (\bibinfo {year} {2013})}\BibitemShut {NoStop}%
\bibitem [{\citenamefont {Zhang}\ \emph {et~al.}(2018)\citenamefont {Zhang},
  \citenamefont {Wu}, \citenamefont {Kumar}, \citenamefont {Si},\ and\
  \citenamefont {Fung}}]{zhangPRB2018}%
  \BibitemOpen
  \bibfield  {author} {\bibinfo {author} {\bibfnamefont {Y.-L.}\ \bibnamefont
  {Zhang}}, \bibinfo {author} {\bibfnamefont {R.~P.~H.}\ \bibnamefont {Wu}},
  \bibinfo {author} {\bibfnamefont {A.}~\bibnamefont {Kumar}}, \bibinfo
  {author} {\bibfnamefont {T.}~\bibnamefont {Si}}, \ and\ \bibinfo {author}
  {\bibfnamefont {K.~H.}\ \bibnamefont {Fung}},\ }\href {\doibase
  10.1103/PhysRevB.97.144203} {\bibfield  {journal} {\bibinfo  {journal} {Phys.
  Rev. B}\ }\textbf {\bibinfo {volume} {97}},\ \bibinfo {pages} {144203}
  (\bibinfo {year} {2018})}\BibitemShut {NoStop}%
\bibitem [{\citenamefont {Feng}\ \emph {et~al.}(2017)\citenamefont {Feng},
  \citenamefont {El-Ganainy},\ and\ \citenamefont {Ge}}]{fengNaturephton2017}%
  \BibitemOpen
  \bibfield  {author} {\bibinfo {author} {\bibfnamefont {L.}~\bibnamefont
  {Feng}}, \bibinfo {author} {\bibfnamefont {R.}~\bibnamefont {El-Ganainy}}, \
  and\ \bibinfo {author} {\bibfnamefont {L.}~\bibnamefont {Ge}},\ }\href
  {\doibase 10.1038/s41566-017-0031-1} {\bibfield  {journal} {\bibinfo
  {journal} {Nature Photonics}\ }\textbf {\bibinfo {volume} {11}},\ \bibinfo
  {pages} {752} (\bibinfo {year} {2017})}\BibitemShut {NoStop}%
\bibitem [{\citenamefont {El-Ganainy}\ \emph {et~al.}(2018)\citenamefont
  {El-Ganainy}, \citenamefont {Makris}, \citenamefont {Khajavikhan},
  \citenamefont {Musslimani}, \citenamefont {Rotter},\ and\ \citenamefont
  {Christodoulides}}]{elganainyNaturephys2018}%
  \BibitemOpen
  \bibfield  {author} {\bibinfo {author} {\bibfnamefont {R.}~\bibnamefont
  {El-Ganainy}}, \bibinfo {author} {\bibfnamefont {K.~G.}\ \bibnamefont
  {Makris}}, \bibinfo {author} {\bibfnamefont {M.}~\bibnamefont {Khajavikhan}},
  \bibinfo {author} {\bibfnamefont {Z.~H.}\ \bibnamefont {Musslimani}},
  \bibinfo {author} {\bibfnamefont {S.}~\bibnamefont {Rotter}}, \ and\ \bibinfo
  {author} {\bibfnamefont {D.~N.}\ \bibnamefont {Christodoulides}},\ }\href
  {\doibase 10.1038/nphys4323} {\bibfield  {journal} {\bibinfo  {journal}
  {Nature Physics}\ }\textbf {\bibinfo {volume} {14}},\ \bibinfo {pages} {11}
  (\bibinfo {year} {2018})}\BibitemShut {NoStop}%
\bibitem [{\citenamefont {Heiss}(2012)}]{heissJPA2012}%
  \BibitemOpen
  \bibfield  {author} {\bibinfo {author} {\bibfnamefont {W.~D.}\ \bibnamefont
  {Heiss}},\ }\href {http://stacks.iop.org/1751-8121/45/i=44/a=444016}
  {\bibfield  {journal} {\bibinfo  {journal} {Journal of Physics A:
  Mathematical and Theoretical}\ }\textbf {\bibinfo {volume} {45}},\ \bibinfo
  {pages} {444016} (\bibinfo {year} {2012})}\BibitemShut {NoStop}%
\bibitem [{\citenamefont {Schnyder}\ \emph {et~al.}(2008)\citenamefont
  {Schnyder}, \citenamefont {Ryu}, \citenamefont {Furusaki},\ and\
  \citenamefont {Ludwig}}]{schnyderPRB2008}%
  \BibitemOpen
  \bibfield  {author} {\bibinfo {author} {\bibfnamefont {A.~P.}\ \bibnamefont
  {Schnyder}}, \bibinfo {author} {\bibfnamefont {S.}~\bibnamefont {Ryu}},
  \bibinfo {author} {\bibfnamefont {A.}~\bibnamefont {Furusaki}}, \ and\
  \bibinfo {author} {\bibfnamefont {A.~W.~W.}\ \bibnamefont {Ludwig}},\ }\href
  {\doibase 10.1103/PhysRevB.78.195125} {\bibfield  {journal} {\bibinfo
  {journal} {Phys. Rev. B}\ }\textbf {\bibinfo {volume} {78}},\ \bibinfo
  {pages} {195125} (\bibinfo {year} {2008})}\BibitemShut {NoStop}%
\bibitem [{\citenamefont {Ding}\ \emph {et~al.}(2015)\citenamefont {Ding},
  \citenamefont {Zhang},\ and\ \citenamefont {Chan}}]{dingPRB2015}%
  \BibitemOpen
  \bibfield  {author} {\bibinfo {author} {\bibfnamefont {K.}~\bibnamefont
  {Ding}}, \bibinfo {author} {\bibfnamefont {Z.~Q.}\ \bibnamefont {Zhang}}, \
  and\ \bibinfo {author} {\bibfnamefont {C.~T.}\ \bibnamefont {Chan}},\ }\href
  {\doibase 10.1103/PhysRevB.92.235310} {\bibfield  {journal} {\bibinfo
  {journal} {Phys. Rev. B}\ }\textbf {\bibinfo {volume} {92}},\ \bibinfo
  {pages} {235310} (\bibinfo {year} {2015})}\BibitemShut {NoStop}%
\bibitem [{\citenamefont {Yuce}(2015)}]{yucePLA2015}%
  \BibitemOpen
  \bibfield  {author} {\bibinfo {author} {\bibfnamefont {C.}~\bibnamefont
  {Yuce}},\ }\href {\doibase https://doi.org/10.1016/j.physleta.2015.02.011}
  {\bibfield  {journal} {\bibinfo  {journal} {Physics Letters A}\ }\textbf
  {\bibinfo {volume} {379}},\ \bibinfo {pages} {1213 } (\bibinfo {year}
  {2015})}\BibitemShut {NoStop}%
\bibitem [{\citenamefont {Wagner}(2017)}]{wagnerAP2017}%
  \BibitemOpen
  \bibfield  {author} {\bibinfo {author} {\bibfnamefont {M.}~\bibnamefont
  {Wagner}},\ }\href {\doibase 10.14311/AP.2017.57.0470} {\bibfield  {journal}
  {\bibinfo  {journal} {Acta Polytechnica}\ }\textbf {\bibinfo {volume} {57}},\
  \bibinfo {pages} {470} (\bibinfo {year} {2017})}\BibitemShut {NoStop}%
\bibitem [{\citenamefont {Martinez~Alvarez}\ \emph {et~al.}(2018)\citenamefont
  {Martinez~Alvarez}, \citenamefont {Barrios~Vargas},\ and\ \citenamefont
  {Foa~Torres}}]{alvarezPRB2018}%
  \BibitemOpen
  \bibfield  {author} {\bibinfo {author} {\bibfnamefont {V.~M.}\ \bibnamefont
  {Martinez~Alvarez}}, \bibinfo {author} {\bibfnamefont {J.~E.}\ \bibnamefont
  {Barrios~Vargas}}, \ and\ \bibinfo {author} {\bibfnamefont {L.~E.~F.}\
  \bibnamefont {Foa~Torres}},\ }\href {\doibase 10.1103/PhysRevB.97.121401}
  {\bibfield  {journal} {\bibinfo  {journal} {Phys. Rev. B}\ }\textbf {\bibinfo
  {volume} {97}},\ \bibinfo {pages} {121401} (\bibinfo {year}
  {2018})}\BibitemShut {NoStop}%
\bibitem [{\citenamefont {P{\'e}rez-Gonz{\'a}lez}\ \emph
  {et~al.}(2018)\citenamefont {P{\'e}rez-Gonz{\'a}lez}, \citenamefont {Bello},
  \citenamefont {G{\'o}mez-Le{\'o}n},\ and\ \citenamefont
  {Platero}}]{perezArxiv2018}%
  \BibitemOpen
  \bibfield  {author} {\bibinfo {author} {\bibfnamefont {B.}~\bibnamefont
  {P{\'e}rez-Gonz{\'a}lez}}, \bibinfo {author} {\bibfnamefont {M.}~\bibnamefont
  {Bello}}, \bibinfo {author} {\bibfnamefont {{\'A}.}~\bibnamefont
  {G{\'o}mez-Le{\'o}n}}, \ and\ \bibinfo {author} {\bibfnamefont
  {G.}~\bibnamefont {Platero}},\ }\href@noop {} {\bibfield  {journal} {\bibinfo
   {journal} {arXiv preprint arXiv:1802.03973}\ } (\bibinfo {year}
  {2018})}\BibitemShut {NoStop}%
\bibitem [{\citenamefont {Rhim}\ \emph {et~al.}(2017)\citenamefont {Rhim},
  \citenamefont {Behrends},\ and\ \citenamefont {Bardarson}}]{rhimPRB2017}%
  \BibitemOpen
  \bibfield  {author} {\bibinfo {author} {\bibfnamefont {J.-W.}\ \bibnamefont
  {Rhim}}, \bibinfo {author} {\bibfnamefont {J.}~\bibnamefont {Behrends}}, \
  and\ \bibinfo {author} {\bibfnamefont {J.~H.}\ \bibnamefont {Bardarson}},\
  }\href {\doibase 10.1103/PhysRevB.95.035421} {\bibfield  {journal} {\bibinfo
  {journal} {Phys. Rev. B}\ }\textbf {\bibinfo {volume} {95}},\ \bibinfo
  {pages} {035421} (\bibinfo {year} {2017})}\BibitemShut {NoStop}%
\bibitem [{\citenamefont {Wang}\ \emph
  {et~al.}(2018{\natexlab{a}})\citenamefont {Wang}, \citenamefont
  {R\"{o}ntgen}, \citenamefont {Morfonios}, \citenamefont {Pinheiro},
  \citenamefont {Schmelcher},\ and\ \citenamefont {Negro}}]{wangOL2018}%
  \BibitemOpen
  \bibfield  {author} {\bibinfo {author} {\bibfnamefont {R.}~\bibnamefont
  {Wang}}, \bibinfo {author} {\bibfnamefont {M.}~\bibnamefont {R\"{o}ntgen}},
  \bibinfo {author} {\bibfnamefont {C.~V.}\ \bibnamefont {Morfonios}}, \bibinfo
  {author} {\bibfnamefont {F.~A.}\ \bibnamefont {Pinheiro}}, \bibinfo {author}
  {\bibfnamefont {P.}~\bibnamefont {Schmelcher}}, \ and\ \bibinfo {author}
  {\bibfnamefont {L.~D.}\ \bibnamefont {Negro}},\ }\href {\doibase
  10.1364/OL.43.001986} {\bibfield  {journal} {\bibinfo  {journal} {Opt.
  Lett.}\ }\textbf {\bibinfo {volume} {43}},\ \bibinfo {pages} {1986} (\bibinfo
  {year} {2018}{\natexlab{a}})}\BibitemShut {NoStop}%
\bibitem [{\citenamefont {Skipetrov}\ and\ \citenamefont
  {Sokolov}(2014)}]{Skipetrov2014}%
  \BibitemOpen
  \bibfield  {author} {\bibinfo {author} {\bibfnamefont {S.~E.}\ \bibnamefont
  {Skipetrov}}\ and\ \bibinfo {author} {\bibfnamefont {I.~M.}\ \bibnamefont
  {Sokolov}},\ }\href {\doibase 10.1103/PhysRevLett.112.023905} {\bibfield
  {journal} {\bibinfo  {journal} {Phys. Rev. Lett.}\ }\textbf {\bibinfo
  {volume} {112}},\ \bibinfo {pages} {023905} (\bibinfo {year}
  {2014})}\BibitemShut {NoStop}%
\bibitem [{\citenamefont {St-Jean}\ \emph {et~al.}(2017)\citenamefont
  {St-Jean}, \citenamefont {Goblot}, \citenamefont {Galopin}, \citenamefont
  {Lema{\^\i}tre}, \citenamefont {Ozawa}, \citenamefont {Le~Gratiet},
  \citenamefont {Sagnes}, \citenamefont {Bloch},\ and\ \citenamefont
  {Amo}}]{stjeanNaturephoton2017}%
  \BibitemOpen
  \bibfield  {author} {\bibinfo {author} {\bibfnamefont {P.}~\bibnamefont
  {St-Jean}}, \bibinfo {author} {\bibfnamefont {V.}~\bibnamefont {Goblot}},
  \bibinfo {author} {\bibfnamefont {E.}~\bibnamefont {Galopin}}, \bibinfo
  {author} {\bibfnamefont {A.}~\bibnamefont {Lema{\^\i}tre}}, \bibinfo {author}
  {\bibfnamefont {T.}~\bibnamefont {Ozawa}}, \bibinfo {author} {\bibfnamefont
  {L.}~\bibnamefont {Le~Gratiet}}, \bibinfo {author} {\bibfnamefont
  {I.}~\bibnamefont {Sagnes}}, \bibinfo {author} {\bibfnamefont
  {J.}~\bibnamefont {Bloch}}, \ and\ \bibinfo {author} {\bibfnamefont
  {A.}~\bibnamefont {Amo}},\ }\href {\doibase 10.1038/s41566-017-0006-2}
  {\bibfield  {journal} {\bibinfo  {journal} {Nature Photonics}\ }\textbf
  {\bibinfo {volume} {11}},\ \bibinfo {pages} {651} (\bibinfo {year}
  {2017})}\BibitemShut {NoStop}%
\bibitem [{\citenamefont {Kraus}\ and\ \citenamefont
  {Zilberberg}(2012)}]{krausPRL2012}%
  \BibitemOpen
  \bibfield  {author} {\bibinfo {author} {\bibfnamefont {Y.~E.}\ \bibnamefont
  {Kraus}}\ and\ \bibinfo {author} {\bibfnamefont {O.}~\bibnamefont
  {Zilberberg}},\ }\href {\doibase 10.1103/PhysRevLett.109.116404} {\bibfield
  {journal} {\bibinfo  {journal} {Phys. Rev. Lett.}\ }\textbf {\bibinfo
  {volume} {109}},\ \bibinfo {pages} {116404} (\bibinfo {year}
  {2012})}\BibitemShut {NoStop}%
\bibitem [{\citenamefont {Eleuch}\ and\ \citenamefont
  {Rotter}(2016)}]{eleuchPRA2016}%
  \BibitemOpen
  \bibfield  {author} {\bibinfo {author} {\bibfnamefont {H.}~\bibnamefont
  {Eleuch}}\ and\ \bibinfo {author} {\bibfnamefont {I.}~\bibnamefont
  {Rotter}},\ }\href {\doibase 10.1103/PhysRevA.93.042116} {\bibfield
  {journal} {\bibinfo  {journal} {Phys. Rev. A}\ }\textbf {\bibinfo {volume}
  {93}},\ \bibinfo {pages} {042116} (\bibinfo {year} {2016})}\BibitemShut
  {NoStop}%
\bibitem [{\citenamefont {Syzranov}\ \emph {et~al.}(2016)\citenamefont
  {Syzranov}, \citenamefont {Wall}, \citenamefont {Zhu}, \citenamefont
  {Gurarie},\ and\ \citenamefont {Rey}}]{syzranovNaturecomms2016}%
  \BibitemOpen
  \bibfield  {author} {\bibinfo {author} {\bibfnamefont {S.~V.}\ \bibnamefont
  {Syzranov}}, \bibinfo {author} {\bibfnamefont {M.~L.}\ \bibnamefont {Wall}},
  \bibinfo {author} {\bibfnamefont {B.}~\bibnamefont {Zhu}}, \bibinfo {author}
  {\bibfnamefont {V.}~\bibnamefont {Gurarie}}, \ and\ \bibinfo {author}
  {\bibfnamefont {A.~M.}\ \bibnamefont {Rey}},\ }\href@noop {} {\bibfield
  {journal} {\bibinfo  {journal} {Nature communications}\ }\textbf {\bibinfo
  {volume} {7}},\ \bibinfo {pages} {13543} (\bibinfo {year}
  {2016})}\BibitemShut {NoStop}%
\bibitem [{\citenamefont {Olmos}\ \emph {et~al.}(2013)\citenamefont {Olmos},
  \citenamefont {Yu}, \citenamefont {Singh}, \citenamefont {Schreck},
  \citenamefont {Bongs},\ and\ \citenamefont {Lesanovsky}}]{olmosPRL2013}%
  \BibitemOpen
  \bibfield  {author} {\bibinfo {author} {\bibfnamefont {B.}~\bibnamefont
  {Olmos}}, \bibinfo {author} {\bibfnamefont {D.}~\bibnamefont {Yu}}, \bibinfo
  {author} {\bibfnamefont {Y.}~\bibnamefont {Singh}}, \bibinfo {author}
  {\bibfnamefont {F.}~\bibnamefont {Schreck}}, \bibinfo {author} {\bibfnamefont
  {K.}~\bibnamefont {Bongs}}, \ and\ \bibinfo {author} {\bibfnamefont
  {I.}~\bibnamefont {Lesanovsky}},\ }\href {\doibase
  10.1103/PhysRevLett.110.143602} {\bibfield  {journal} {\bibinfo  {journal}
  {Phys. Rev. Lett.}\ }\textbf {\bibinfo {volume} {110}},\ \bibinfo {pages}
  {143602} (\bibinfo {year} {2013})}\BibitemShut {NoStop}%
\bibitem [{\citenamefont {Wang}\ \emph
  {et~al.}(2018{\natexlab{b}})\citenamefont {Wang}, \citenamefont {Subhankar},
  \citenamefont {Bienias}, \citenamefont {\L{}acki}, \citenamefont {Tsui},
  \citenamefont {Baranov}, \citenamefont {Gorshkov}, \citenamefont {Zoller},
  \citenamefont {Porto},\ and\ \citenamefont {Rolston}}]{wangPRL2018}%
  \BibitemOpen
  \bibfield  {author} {\bibinfo {author} {\bibfnamefont {Y.}~\bibnamefont
  {Wang}}, \bibinfo {author} {\bibfnamefont {S.}~\bibnamefont {Subhankar}},
  \bibinfo {author} {\bibfnamefont {P.}~\bibnamefont {Bienias}}, \bibinfo
  {author} {\bibfnamefont {M.}~\bibnamefont {\L{}acki}}, \bibinfo {author}
  {\bibfnamefont {T.-C.}\ \bibnamefont {Tsui}}, \bibinfo {author}
  {\bibfnamefont {M.~A.}\ \bibnamefont {Baranov}}, \bibinfo {author}
  {\bibfnamefont {A.~V.}\ \bibnamefont {Gorshkov}}, \bibinfo {author}
  {\bibfnamefont {P.}~\bibnamefont {Zoller}}, \bibinfo {author} {\bibfnamefont
  {J.~V.}\ \bibnamefont {Porto}}, \ and\ \bibinfo {author} {\bibfnamefont
  {S.~L.}\ \bibnamefont {Rolston}},\ }\href {\doibase
  10.1103/PhysRevLett.120.083601} {\bibfield  {journal} {\bibinfo  {journal}
  {Phys. Rev. Lett.}\ }\textbf {\bibinfo {volume} {120}},\ \bibinfo {pages}
  {083601} (\bibinfo {year} {2018}{\natexlab{b}})}\BibitemShut {NoStop}%
\bibitem [{\citenamefont {F{\"o}lling}\ \emph {et~al.}(2007)\citenamefont
  {F{\"o}lling}, \citenamefont {Trotzky}, \citenamefont {Cheinet},
  \citenamefont {Feld}, \citenamefont {Saers}, \citenamefont {Widera},
  \citenamefont {M{\"u}ller},\ and\ \citenamefont
  {Bloch}}]{follingNaturephys2007}%
  \BibitemOpen
  \bibfield  {author} {\bibinfo {author} {\bibfnamefont {S.}~\bibnamefont
  {F{\"o}lling}}, \bibinfo {author} {\bibfnamefont {S.}~\bibnamefont
  {Trotzky}}, \bibinfo {author} {\bibfnamefont {P.}~\bibnamefont {Cheinet}},
  \bibinfo {author} {\bibfnamefont {M.}~\bibnamefont {Feld}}, \bibinfo {author}
  {\bibfnamefont {R.}~\bibnamefont {Saers}}, \bibinfo {author} {\bibfnamefont
  {A.}~\bibnamefont {Widera}}, \bibinfo {author} {\bibfnamefont
  {T.}~\bibnamefont {M{\"u}ller}}, \ and\ \bibinfo {author} {\bibfnamefont
  {I.}~\bibnamefont {Bloch}},\ }\href@noop {} {\bibfield  {journal} {\bibinfo
  {journal} {Nature}\ }\textbf {\bibinfo {volume} {448}},\ \bibinfo {pages}
  {1029} (\bibinfo {year} {2007})}\BibitemShut {NoStop}%
\bibitem [{\citenamefont {Wirth}\ \emph {et~al.}(2011)\citenamefont {Wirth},
  \citenamefont {{\"O}lschl{\"a}ger},\ and\ \citenamefont
  {Hemmerich}}]{wirthNaturephys2011}%
  \BibitemOpen
  \bibfield  {author} {\bibinfo {author} {\bibfnamefont {G.}~\bibnamefont
  {Wirth}}, \bibinfo {author} {\bibfnamefont {M.}~\bibnamefont
  {{\"O}lschl{\"a}ger}}, \ and\ \bibinfo {author} {\bibfnamefont
  {A.}~\bibnamefont {Hemmerich}},\ }\href@noop {} {\bibfield  {journal}
  {\bibinfo  {journal} {Nature Physics}\ }\textbf {\bibinfo {volume} {7}},\
  \bibinfo {pages} {147} (\bibinfo {year} {2011})}\BibitemShut {NoStop}%
\bibitem [{\citenamefont {Bernien}\ \emph {et~al.}(2017)\citenamefont
  {Bernien}, \citenamefont {Schwartz}, \citenamefont {Keesling}, \citenamefont
  {Levine}, \citenamefont {Omran}, \citenamefont {Pichler}, \citenamefont
  {Choi}, \citenamefont {Zibrov}, \citenamefont {Endres}, \citenamefont
  {Greiner} \emph {et~al.}}]{bernienNature2017}%
  \BibitemOpen
  \bibfield  {author} {\bibinfo {author} {\bibfnamefont {H.}~\bibnamefont
  {Bernien}}, \bibinfo {author} {\bibfnamefont {S.}~\bibnamefont {Schwartz}},
  \bibinfo {author} {\bibfnamefont {A.}~\bibnamefont {Keesling}}, \bibinfo
  {author} {\bibfnamefont {H.}~\bibnamefont {Levine}}, \bibinfo {author}
  {\bibfnamefont {A.}~\bibnamefont {Omran}}, \bibinfo {author} {\bibfnamefont
  {H.}~\bibnamefont {Pichler}}, \bibinfo {author} {\bibfnamefont
  {S.}~\bibnamefont {Choi}}, \bibinfo {author} {\bibfnamefont {A.~S.}\
  \bibnamefont {Zibrov}}, \bibinfo {author} {\bibfnamefont {M.}~\bibnamefont
  {Endres}}, \bibinfo {author} {\bibfnamefont {M.}~\bibnamefont {Greiner}},
  \emph {et~al.},\ }\href@noop {} {\bibfield  {journal} {\bibinfo  {journal}
  {Nature}\ }\textbf {\bibinfo {volume} {551}},\ \bibinfo {pages} {579}
  (\bibinfo {year} {2017})}\BibitemShut {NoStop}%
\bibitem [{\citenamefont {Nascimbene}\ \emph {et~al.}(2015)\citenamefont
  {Nascimbene}, \citenamefont {Goldman}, \citenamefont {Cooper},\ and\
  \citenamefont {Dalibard}}]{nascimbenePRL2015}%
  \BibitemOpen
  \bibfield  {author} {\bibinfo {author} {\bibfnamefont {S.}~\bibnamefont
  {Nascimbene}}, \bibinfo {author} {\bibfnamefont {N.}~\bibnamefont {Goldman}},
  \bibinfo {author} {\bibfnamefont {N.~R.}\ \bibnamefont {Cooper}}, \ and\
  \bibinfo {author} {\bibfnamefont {J.}~\bibnamefont {Dalibard}},\ }\href
  {\doibase 10.1103/PhysRevLett.115.140401} {\bibfield  {journal} {\bibinfo
  {journal} {Phys. Rev. Lett.}\ }\textbf {\bibinfo {volume} {115}},\ \bibinfo
  {pages} {140401} (\bibinfo {year} {2015})}\BibitemShut {NoStop}%
\bibitem [{\citenamefont {Gonz{\'a}lez-Tudela}\ \emph
  {et~al.}(2015)\citenamefont {Gonz{\'a}lez-Tudela}, \citenamefont {Hung},
  \citenamefont {Chang}, \citenamefont {Cirac},\ and\ \citenamefont
  {Kimble}}]{gonzalezNaturephoton2015}%
  \BibitemOpen
  \bibfield  {author} {\bibinfo {author} {\bibfnamefont {A.}~\bibnamefont
  {Gonz{\'a}lez-Tudela}}, \bibinfo {author} {\bibfnamefont {C.-L.}\
  \bibnamefont {Hung}}, \bibinfo {author} {\bibfnamefont {D.~E.}\ \bibnamefont
  {Chang}}, \bibinfo {author} {\bibfnamefont {J.~I.}\ \bibnamefont {Cirac}}, \
  and\ \bibinfo {author} {\bibfnamefont {H.}~\bibnamefont {Kimble}},\
  }\href@noop {} {\bibfield  {journal} {\bibinfo  {journal} {Nature Photonics}\
  }\textbf {\bibinfo {volume} {9}},\ \bibinfo {pages} {320} (\bibinfo {year}
  {2015})}\BibitemShut {NoStop}%
\bibitem [{\citenamefont {Gullans}\ \emph {et~al.}(2012)\citenamefont
  {Gullans}, \citenamefont {Tiecke}, \citenamefont {Chang}, \citenamefont
  {Feist}, \citenamefont {Thompson}, \citenamefont {Cirac}, \citenamefont
  {Zoller},\ and\ \citenamefont {Lukin}}]{gullansPRL2012}%
  \BibitemOpen
  \bibfield  {author} {\bibinfo {author} {\bibfnamefont {M.}~\bibnamefont
  {Gullans}}, \bibinfo {author} {\bibfnamefont {T.~G.}\ \bibnamefont {Tiecke}},
  \bibinfo {author} {\bibfnamefont {D.~E.}\ \bibnamefont {Chang}}, \bibinfo
  {author} {\bibfnamefont {J.}~\bibnamefont {Feist}}, \bibinfo {author}
  {\bibfnamefont {J.~D.}\ \bibnamefont {Thompson}}, \bibinfo {author}
  {\bibfnamefont {J.~I.}\ \bibnamefont {Cirac}}, \bibinfo {author}
  {\bibfnamefont {P.}~\bibnamefont {Zoller}}, \ and\ \bibinfo {author}
  {\bibfnamefont {M.~D.}\ \bibnamefont {Lukin}},\ }\href {\doibase
  10.1103/PhysRevLett.109.235309} {\bibfield  {journal} {\bibinfo  {journal}
  {Phys. Rev. Lett.}\ }\textbf {\bibinfo {volume} {109}},\ \bibinfo {pages}
  {235309} (\bibinfo {year} {2012})}\BibitemShut {NoStop}%
\end{thebibliography}%
\end{document}